\documentclass[11pt,preprint]{aastex}
\usepackage{graphicx}
\usepackage{amsmath,amssymb} 
\usepackage{subfigure}

\begin{document}

\newcommand{\rhat}{\hat{\boldsymbol r}}
\newcommand{\that}{\hat{\boldsymbol \theta}}
\newcommand{\phat}{\hat{\boldsymbol \phi}}
\newcommand{\bfnabla}{\boldsymbol \nabla}
\newcommand{\bfv}{\boldsymbol v}
\newcommand{\bfomega}{\boldsymbol \omega}
\newcommand{\ba}{\begin{eqnarray}}
\newcommand{\ea}{\end{eqnarray}}
\newcommand{\eh}{\hat{\boldsymbol e}}
\newcommand{\cp}{\varpi}
\newcommand{\ol}{\overline}
\newcommand{\sgn}{\mathop{\rm sgn}}
\newcommand{\ephi}{{\bf e}_\varphi}
\newcommand{\Fdis}{F_{\rm w,dis}}


\title{Angular momentum transport and variability in boundary 
layers of accretion disks driven by global acoustic modes.}

\author{Mikhail A. Belyaev\altaffilmark{1} \& Roman
  R. Rafikov\altaffilmark{1,2} \& James M. Stone\altaffilmark{1}}
\altaffiltext{1}{Department of Astrophysical Sciences, 
Princeton University, Ivy Lane, Princeton, NJ 08540; 
rrr@astro.princeton.edu}
\altaffiltext{2}{Sloan Fellow}


\begin{abstract}
Disk accretion onto a weakly magnetized central object, e.g. a star,
is inevitably accompanied by the 
formation of a boundary layer
near the surface, in which matter slows
down from the highly supersonic orbital velocity of the
disk to the rotational velocity of the star. We perform 
high resolution 2D hydrodynamical simulations in the equatorial 
plane of an astrophysical boundary layer with the goal of 
exploring the dynamics of non-axisymmetric structures that 
form there. We generically find that the supersonic 
shear in the boundary layer excites non-axisymmetric 
quasi-stationary acoustic modes that are trapped between 
the surface of the star and a Lindblad resonance in the disk.
These modes rotate in a prograde fashion, are stable for hundreds of
orbital periods, and have a pattern speed that is less than and of
order the rotational velocity at the inner edge of the disk. 
The origin of these intrinsically global modes is intimately
related to the operation of a corotation amplifier in the 
system. Dissipation of acoustic modes in weak shocks provides a 
universal mechanism for angular momentum and mass transport 
even in purely hydrodynamic (i.e. non-magnetized) boundary 
layers. We discuss the possible implications of these trapped 
modes for explaining the variability seen in accreting compact 
objects.
\end{abstract}



\section{Introduction}
\label{sect:intro}

One of the outstanding problems in astrophysical accretion disk theory 
is the physics of the boundary layer (hereafter BL). The BL is the innermost region 
of the accretion disk in which the rotation profile of
the star attaches smoothly to that of the disk \citep{LyndenBellPringle}. 
In the BL, the rotation velocity of the disk fluid necessarily rises 
with increasing radius, which has important implications for 
angular momentum transport. BLs occur in a variety of systems 
--- young stars, white dwarfs, neutron stars --- whenever 
the accretion rate $\dot M$ is high enough for the disk to extend all the way 
down to the surface of the central object (hereafter referred to as ``star'' 
for simplicity), without being disrupted by the stellar magnetic field. 
In such cases, up to half of the accretion energy is dissipated in the 
BL \citep{Kluzniak}, which leads to intense local heating and gives 
rise to X-ray and ultraviolet emission in accreting neutron stars 
\citep{InogamovSunyaev} and white dwarfs 
\citep{Pringle, PringleSavonije, NarayanPopham,PophamNarayan, 
PiroBildsten}. 

Since the rotational velocity of the gas passing through the
BL changes with radius, some mechanism of angular momentum 
transport must operate there, ultimately leading to the 
radial mass transport. The nature of this mechanism is not understood 
at the moment. The magnetorotational instability (MRI;
\citet{Velikhov,Chandra,BH91a,BH91b}) usually invoked to 
explain mass transport in accretion disks does not operate in 
the BL because $d\Omega^2/d\cp>0$ there ($\cp$ is the cylindrical 
radius). Magnetic field carried into the BL by accreting gas 
is sheared by differential rotation 
\citep{Armitage,SteinackerPapaloizou} but whether this 
can lead to sustained mass accretion has not been conclusively 
demonstrated. Other transport mechanisms --- Kelvin-Helmholtz 
instability \citep{KT78}, baroclinic instability \citep{Fujimoto},
Tayler-Spruit dynamo \citep{Tayler,Spruit,PiroBildsten} have also 
been proposed but whether they can operate in the BL and give rise 
to efficient momentum transport there is far from clear. In the 
absence of a good understanding of the momentum and mass transport, 
even the morphology of the BL has been a matter of debate,
with the proposed geometries ranging from a BL 
contained inside the star\citep{KT78} to a spreading layer
extending to significant latitudes over the stellar surface 
\citep{InogamovSunyaev,InogamovSunyaev10}.

The BL has also been associated with variability in accreting systems
that occurs on timescales comparable to the dynamical timescale at the
surface of the star. One example of such variability are dwarf nova
oscillations (DNOs) in accreting white dwarf systems \citep{Warner}, 
which do not have a definitive explanation. DNOs are 
characterized by an oscillation period of
$P_{DNO} \gtrsim P_K(R_\star)$, where
$P_{DNO}$ is the DNO period and $P_K(R_\star)$ is the Keplerian 
rotational period at the surface of the star \citep{Patterson}. Since
the DNO period is comparable to the Keplerian rotational period at the
surface of the star, it is natural to associate DNOs with the boundary
layer or with the region of the disk directly adjacent to it. 

In this work we describe a set of first principles, two-dimensional 
(2D) hydrodynamical simulations of the equatorial plane of the 
disk-star system carried out in cylindrical geometry. The main 
result of our work is that we 
generically observe the excitation of a non-axisymmetric 
mode in the BL with a unique 
wavelength and pattern speed that is stable for many hundreds 
of orbital periods. 

The mechanism of excitation for this mode is the sonic
instability \citep{Glatzel, BR}, which is a type of shear instability
that occurs for supersonic, highly compressible flows. The sonic
instability is closely related to the Papaloizou-Pringle instability
\citep{PPI, Goodman} and operates very differently from the more
familiar KH instability, which occurs in the subsonic
regime. 

Dissipation of the non-axisymmetric BL mode by virtue of weak
shocks results in angular momentum and mass transport in 
the BL. It has also been argued that the presence of a
non-axisymmetric mode in the BL or on the surface of the star can
modulate stellar emission, providing an explanation for DNOs 
\citep{PapaloizouPringle,Popham,PiroBildsten1}. Previously, however,
there was no acknowledged mechanism for exciting non-axisymmetric
modes on the surface of the star. Since such modes emerge naturally in
our simulations, our work may help to understand the mechanism of
DNOs, or more generically, any periodic phenomena having a period $P
\sim P_K(R_\star)$.

The paper is organized as follows: In \S \ref{nummod} we present the
model we use in our simulations, and in \S \ref{results_sec} we
present our results. We find that the boundary
layer is unstable to sonic instabilities (\S \ref{sonic_sec}), which tend to
excite a single surface
mode. This mode is characterized by vortices at the base of the BL and
by shocks launched from the top of the BL that propagate through the
disk and are reflected back
towards the BL at a Lindblad resonance. The shocks are weak with a small
compression ratio, and we
find that their properties can be easily understood in terms of the WKB
theory for weak disturbances in the disk (\S \ref{mode_sec}). We also
study the stresses in the BL caused by the instabilities
that operate there (\S \ref{angmom_sec}-\ref{highlow_sec}), transport 
of mass (\S \ref{sect:mass_accretion}+), how the
pattern speed of the modes is affected by Mach number
(\S \ref{pattern_sec}), and how the BL thickness varies as a function of
both time and Mach number (\S \ref{BLwidth_sec}). In \S \ref{fulldisk_sec},
we simulate the full range of azimuthal angle from $0-2\pi$ and
observe long wavelength non-axisymmetric modes. 


\section{Numerical Model}
\label{nummod}

We perform a set of 2D hydrodynamical simulations in the 
equatorial ($r-\phi$) plane of a star with an adjacent 
accretion disk. This setup ignores the vertical structure 
of the disk and star but allows us to capture the formation of the 
long-lived, non-axisymmetric structures in the BL at 
reasonable numerical cost. Similar to \citet{Armitage} and 
\citet{SteinackerPapaloizou} we disregard thermodynamical
evolution of the system by adopting the isothermal equation 
of state in our simulations, so the pressure is given by 
\ba
P = \Sigma s^2. 
\label{eq:EOS}
\ea
Here $\Sigma$ is the density, and $s$ is the isothermal sound 
speed, which we assume to be the same throughout the simulation 
domain. Adoption of isothermal equation of state is equivalent 
to assuming 
fast cooling in the BL. This may not be realistic for all systems,
but we still adopt this approximation since we are primarily 
interested in providing a proof of concept for the trapped acoustic
modes, which are the main focus of our paper.

Previously, \citet{Armitage, SteinackerPapaloizou, Romanova} 
have performed 3D MHD simulations of the BL. We plan to
include magnetic fields in future 3D simulations, but
first it is useful to investigate and understand the restricted 2D
hydro case, since it allows us to study in detail trapped 
modes excited by sonic instabilities, which are
the main focus of our paper.

We numerically evolve the following set of equations 
\ba
\label{eq:cont}
\frac{\partial \Sigma}{\partial t} + \bfnabla \cdot (\Sigma \bfv) = 0,
\\
\label{eq:mom}
\frac{\partial (\Sigma \bfv)}{\partial t} + \bfnabla \cdot (\Sigma \bfv
\bfv) + \bfnabla P + \Sigma \bfnabla \Phi = 0,
\ea
in the equatorial plane of the disk+star system, where $\Phi$ is a fixed,
time-independent potential. The system of equations is closed by the
isothermal equation of state, equation
(\ref{eq:EOS}).  

For the numerical evolution of equations
(\ref{eq:cont}) and (\ref{eq:mom}), we use
the Godunov code Athena \citep{Stone} in cylindrical coordinates
\citep{SkinnerOstriker}. Athena is highly-parallelized and exhibits 
very low levels of numerical viscosity  (\S
\ref{sect:mass_accretion}) making it well-suited for studying the
long-term evolution of the boundary layer.

We use a significantly higher resolution than previous 
authors, which lets us properly resolve the scale height of the
star. This is important for studying
modes excited in the BL, since the scale height
represents a natural length scale in the problem. \citet{Balsaraetal}
have also run high resolution 2D simulations of
the BL in the meridional ($r - \theta$) plane, but the axisymmetric 
setup of their simulations precluded them from seeing the sonic
instability. Moreover, their simulations were run for $\sim 1$
orbital period, whereas we run our simulations for {\it hundreds} of
orbital periods, as measured at the inner radius of the disk.
This allows us to investigate the long-term evolution of the BL and
observe the stability of modes excited on the star on very long 
timescales.


\subsection{Physical Setup}
\label{sect:phys_setup}

In our initial setup, we consider a non-rotating star in hydrostatic
equilibrium which is
surrounded by a rotationally supported disk of constant density. We
nondimensionalize quantities by setting the
radius of the star to $R_\star=1$, the Keplerian orbital velocity at
the surface of the star to $v_K(R_\star) = 1$, and the surface density in the disk
to $\Sigma = 1$. Thus, time is scaled such that the Keplerian orbital period is
$P_K=2\pi$ at $\cp = 1$, where $\cp$ is the cylindrical radius. We take
the potential in the system to be given by the fixed
cylindrically-symmetric potential
\ba
\label{poteq}
\Phi(\cp) = -1/\cp,
\ea
and we ignore self-gravity in the simulations. 
Since the potential given by Equation \ref{poteq} is
cylindrically symmetric, our calculations do not include the vertical
stratification of the disk. 

\begin{figure}[!h]
\centering
  \includegraphics[width=0.7\textwidth]{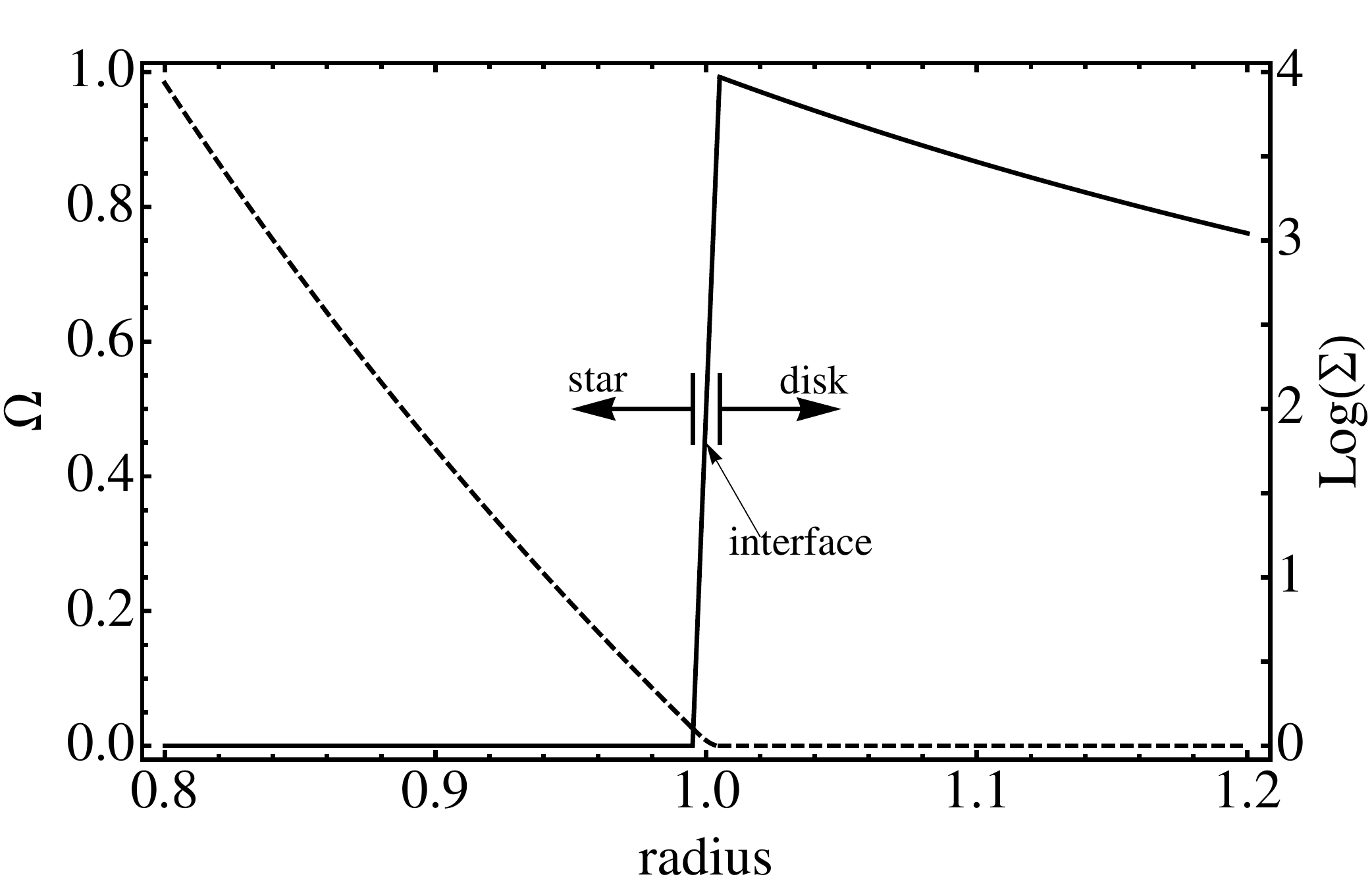}
\caption{Initial rotation profile $\Omega(\cp)$ (equation 
[\ref{regions}], {\it solid line}), 
and the logarithm of the initial density (equation 
[\ref{hydrostateq}], {\it dashed line}) for 
a simulation with $M=6$. Note that the jump in $\Omega$ at $\cp = 1$ 
is not discontinuous and is resolved in the simulations. Various 
regions of the simulation domain are indicated.}
\label{densrot}
\end{figure}

At the start of the simulation, the star is joined to the disk via
a thin region of width $\delta_{BL,0} \ll 1$, inside of which the
rotational velocity $v_\phi$ rises very nearly linearly 
from zero (the velocity in the
star) to the Keplerian orbital velocity (the velocity in the
disk). The initial rotation profile is given by
\ba
\label{regions}
\Omega(\cp) = \left\{
     \begin{array}{lc}
       0 & \cp < 1-\frac{\delta_{BL,0}}{2} \ \ \ \ \ \text{``star''} \\
       \cp^{-3/2}\left(\frac{\cp-1}{\delta_{BL,0}} + \frac{1}{2}\right) & 1-\frac{\delta_{BL,0}}{2} \le \cp \le 1 +
       \frac{\delta_{BL,0}}{2} \ \ \ \ \ \text{``interface''}\\
       \cp^{-3/2} & \cp > 1 + \frac{\delta_{BL,0}}{2} \ \ \ \ \ \text{``disk''}
     \end{array}
   \right. ,
\ea 
and from now on, when speaking about the initial conditions of the
simulations, we will refer to the region $\cp \le 1 - \delta_{BL,0}/2$ as
the ``star'', the region $1 - \delta_{BL,0}/2 < \cp \le 1 + \delta_{BL,0}/2$
as the ``interface'', and the region $\cp > 1 + \delta_{BL,0}/2$ as the
``disk'', see Figure \ref{densrot}. It is important to bear in 
mind, though, that the rotation profile will evolve in time as 
a result of sonic instabilities (discussed below) that induce angular 
momentum transport. These instabilities will rapidly cause the 
initial ``interface'' region to thicken into a self-consistent BL, 
see \S \ref{sonic_sec}.

The initial density profile is specified everywhere throughout the 
domain through the equation of hydrostatic equilibrium
\ba
\label{hydrostateq}
\frac{1}{\Sigma}\frac{dP}{d\cp} = -\frac{d \Phi}{d \cp} + \Omega^2 \cp.
\ea
This results in $\Sigma = const$ within the disk, and $\Sigma \propto
\exp[-\Phi(\cp)/s^2]$ within the star. 
Figure \ref{densrot} shows the initial rotation profile from equation
(\ref{regions}), as well as the initial density profile which is
determined by hydrostatic equilibrium.

For each of our simulations we
define a characteristic Mach number $M = 1/s$, where $s$ is the
isothermal sound speed. In our units the gravitational acceleration is
$g=1$ at $\cp = 1$ (equation \ref{poteq}), which means that $M$
corresponds to the
actual Mach number at $\cp = 1$ for a Keplerian disk, since the
Keplerian velocity is $v_K = 1$ at $\cp = 1$ in our simulations. We also
define a characteristic radial scale in the star using
\ba
\label{scheight}
h_s = M^{-2}R_\star = M^{-2}.
\ea
This is the exact pressure and density scale height at $\cp = 1$ for
an unrotating flow since
\ba
\label{scheight2}
h_s = \frac{s^2}{g},
\ea
and $g$ can be written in terms of the Keplerian velocity as $g =
v_K^2/\cp$. Plugging this into equation (\ref{scheight2}) and using
$\cp = 1$, we recover equation (\ref{scheight}).
Note that $h_s$ is different from the scale height $h_d$ of the 
accretion disk in the vertical direction (which is not considered 
in our simulations). The latter is given by $h_d=s/\Omega_K$
so that $h_s=h_d^2(R_\star)/R_\star = h_d(R_\star) M^{-1} \ll h_d(R_\star)$. 

At this point it is clear that if the isothermal equation of state is
assumed, the only free parameter in the problem is the Mach
number. However, we emphasize this point by explicitly showing that
this is the case. To
do so, we non-dimensionalize the isothermal fluid dynamics equations by
scaling the unit of length according to $R_\star$, the unit of time
according to $\Omega(R_\star)^{-1}$, and the density according to
$\Sigma_{0}$, which is some arbitrary density, such as the density of
the disk which is a constant in our model. Assuming
a Keplerian rotation profile ($1/\cp$
form of the potential), equations (\ref{eq:cont}) and (\ref{eq:mom}) become
\ba
\frac{\partial \Sigma'}{\partial t'} + \bfnabla' \cdot (\Sigma' \bfv') = 0,
\\
\frac{\partial (\Sigma' \bfv')}{\partial t'} + \bfnabla' \cdot (\Sigma' \bfv'
\bfv') + M^{-2}\bfnabla' \Sigma' + \Sigma' \xi^{-2}  = 0,
\ea
where $\xi \equiv \cp/R_\star$, and the primes denote the
non-dimensionalized forms of the variables and operators,
i.e. $\Sigma' = \Sigma/\Sigma_0$, $\bfnabla' = R_\star \bfnabla$, etc.
When the equations are written in non-dimensionalized form, it is
clear that the Mach number is the only free parameter in the problem
for a fixed geometry. Thus, the solutions for the
isothermal boundary layer form a one-parameter family in the Mach
number, and the Mach number itself plays a role which is analogous
to the Reynolds number for the incompressible Navier-Stokes
equations. Of course, if the isothermal assumption is relaxed,
the solution will also depend on other dimensionless parameters. However,
since this paper is a proof of concept, we do not concern ourselves
with those parameters here.


\subsection{Numerical Details}
\label{secty:num_details}

We now summarize the numerical details which pertain to all of our
simulations, and we begin by
discussing the condition of hydrostatic equilibrium. \citet{Zingale}
discuss how to set the initial conditions to achieve numerical
hydrostatic equilibrium in a Godunov code. However, we opt for a
simpler approach in which we initialize
the simulation according to a numerical integration of
equation (\ref{hydrostateq}) and wait for the fluctuations from
the initial transient to damp out.

The amplitude of the initial fluctuations depends on the
resolution of the simulation, but even small fluctuations close to the
inner radial boundary can amplify to become shocks as they propagate
into the disk. This is due to the fact that the amplitude of a
wave propagating in an isothermal atmosphere is $\propto \rho^{-1/2}$
\citep{LL} and there is a large density contrast between the density
in the disk and the density at the inner radius of the simulation domain
(factor of $\sim 10^6$). For this
reason, we burn in all of our simulations by running the analytic
hydrostatic equilibrium solution for a time of $t \sim 100$ with no
perturbations, which allows the simulation to settle
down to a numerical hydrostatic
equilibrium. 

\begin{table}[!htbp]
\centering
\begin{tabular}{|c|c|c|c|c|c|c|}
\hline
label & $N_\cp \times N_\phi \times N_z$ & $\cp$-range & $\phi$-range &
$M$ \\
\hline
A1 & $1024 \times 256 \times 1$ & (0.7, 2.5) & (-0.4, 0.4) &  6 \\
A2 & $2048 \times 512 \times 1$ & (0.7, 2.5) & (-0.4, 0.4) &  6  \\
A3 & $8192 \times 1024 \times 1$ & (0.7, 2.5) & (-0.4, 0.4) &  6  \\
A4 & $2048 \times 2048 \times 1$ & (0.7, 2.5) & (0, 2$\pi$) &  6  \\
A5 & $4096 \times 512 \times 1$ & (0.7, 4.3) & (-0.4, 0.4) &  6  \\
B1 & $2048 \times 512 \times 1$ & (0.85, 2.5) & (-0.4, 0.4) &  9  \\
B2 & $4096 \times 4096 \times 1$ & (0.85, 2.5) & (0, 2$\pi$) &  9  \\
C1 & $6144 \times 1024 \times 1$ & (0.9, 2.5) & (-0.4, 0.4) &  12  \\
C2 & $6144 \times 6144 \times 1$ & (0.9, 2.5) & (0, 2$\pi$) &  12  \\
D1 & $6144 \times 1024 \times 1$ & (0.94, 2.5) & (-0.4, 0.4) &  15  \\
D2 & $6144 \times 6144 \times 1$ & (0.94, 2.5) & (0, 2$\pi$) &  15  \\
\hline
\end{tabular}
\caption{Simulation label, dimension of the box in cells, $\cp$ \& $\phi$
  extent of the box, Mach number.}
\label{simdetails}
\end{table}

We then restart the simulation by adding a
random perturbation to the radial velocity, $\delta v_\cp \sim 0.001$, at
each grid point for $\cp > 1$, and for the remainder of the paper we
take $t=0$ to be the time when the perturbations are added.
The reason not to create perturbations
to the velocity within the star stems from the fact
that the density rises exponentially inside the star, and any
disturbance propagating from the star towards
the disk amplifies exponentially as well, with the relative amplitude going as
the inverse square root of the density \citep{LL}. 

The boundary conditions we use for the simulations are periodic 
in $\phi$ and ``do-nothing'' at the inner and outer radial 
boundaries. The do-nothing boundary condition simply means that 
the fluid variables on the boundary retain their initial values 
for the duration of the simulation. This is a convenient boundary 
condition to use, since essentially no action takes place near 
the inner or outer boundaries: there is no mass inflow or outflow 
through these boundaries, and any incident waves are largely 
absorbed with minimal reflection.

In order to accurately capture the physics of modes excited on the
surface of the star
(\S \ref{mode_sec}), it is both necessary that the simulation extend to several
stellar radii, and that we
resolve the radial scale height (equation [\ref{scheight}]) inside the
star. To satisfy the second of these conditions, we use $\sim 15-30$
cells per radial scale
height $h_s$ within the star, depending on the simulation. However, the
first condition that the simulation domain extend to several stellar
radii makes it
computationally prohibitive to run the simulations at realistic Mach
numbers, since the scale height goes as $h_s = M^{-2}$. As a compromise, we
use $M=6-15$ for our simulations, even though this is
somewhat low in an
astrophysical context. For example, a typical accreting white dwarf 
has an inner disk temperature of $T\sim 10^5 \ \text{K}$, a stellar
radius of $R_\star\sim 10^9 \ \text{cm}$, and a mass of 
$M_\star\sim 0.6 \ M_\sun$, meaning that 
\ba
M=\left(\frac{GM_\star}{R_\star}\frac{\mu}{kT}\right)^{1/2} 
\approx 100\left(\frac{M_\star}{0.6M_\odot}\frac{10^9\mbox{cm}}
{R_\star}\frac{10^5\mbox{K}}{T}\right)^{1/2} 
\label{eq:M_est}
\ea
for such a system.

In all of our simulations we use $\delta_{BL,0} = 0.01$ for the
width of the interface, so that $\delta_{BL,0} \sim h_s$. This is as
thin as we can make the interface region, while still being able to
resolve the physics that goes on there.

Table \ref{simdetails} summarizes the parameters of our
simulations. In all of our simulations, we only vary one physical parameter
which is the Mach number. Simulations that have different Mach numbers
are labeled by a
different letter, whereas simulations that differ only in their numerical
parameters, such as resolution or box size, are labeled by
the same letter, but by a different number.


\section{Results}
\label{results_sec}


\subsection{Sonic Instability}
\label{sonic_sec}

Although hydrodynamical disks with a
Keplerian rotation profile are stable to axisymmetric perturbations by
the Rayleigh criterion, the large shear initially present  
in the interface (equation [\ref{regions}]) makes that region
unstable to non-axisymmetric shear instabilities. Since the initial flow has a
supersonic drop in the velocity across the interface, the instability
that sets in is not the classical KH instability for an incompressible
fluid, but rather the sonic instability \citep{Glatzel, BR}. This is a
global instability that cannot be derived from a local analysis and is
similar to the Papaloizou-Pringle instability \citep{PPI, Goodman,
  Glatzel}. 

The sonic instability operates in one of two ways, either by 
{\it overreflection} of sound waves from a critical layer 
(where the radial wavenumber becomes equal to zero) or 
by {\it radiation} of energy away from the BL, see 
\citet{Goodman, Glatzel, BR}. For a rotating flow, the critical layer
of a mode is equivalent to the corotation resonance, which is located
at the radius $\cp_{CR}$ where $\Omega(\cp_{CR})$ equals the 
pattern frequency, $\Omega_P$. Modes, which have a corotation
resonance (critical layer) inside the fluid domain 
are susceptible to instability. 

The origin of the sonic instability 
is closely related to the existence of a conserved action 
(or pseudo-energy) associated with the mode, which
is quadratic in perturbed fluid variables and changes sign at
corotation \citep{Goodman}. A wave incident upon one side of the 
corotation region with a positive sign of the action
is predominantly reflected. However, there is partial tunneling through the
evanescent region around corotation (see \S \ref{mode_sec}), and the
wave emerging on the other side of $\cp_{CR}$ has the opposite sign of
the action (i.e. negative).
Global conservation of action then requires the reflected wave to 
have greater positive action than that of the incident wave implying
amplification of the wave on that side of corotation upon 
reflection. If a wave is trapped between two corotation resonances or
a corotation resonance and solid wall, this amplification mechanism,
known as a corotation amplifier \citep{Mark,GT78}, ultimately results
in instability. 

An important difference between the sonic and KH instabilities is that
the sonic instability operates regardless of the density contrast
between the star and the disk. \citet{BR} have investigated
the operation of the sonic instability in the presence of a density
discontinuity for a linear shear profile. They found that the growth
rate for the sonic instability was almost independent of the density
contrast across the discontinuity. Indeed, even in the case when the shear
flow occured over a plane reflecting wall, which can be thought of as the
limit of infinite density contrast, the growth rate was comparable to
the case of no density contrast at all.

\begin{figure}[!h]
\centering
\includegraphics[width=0.49\textwidth]{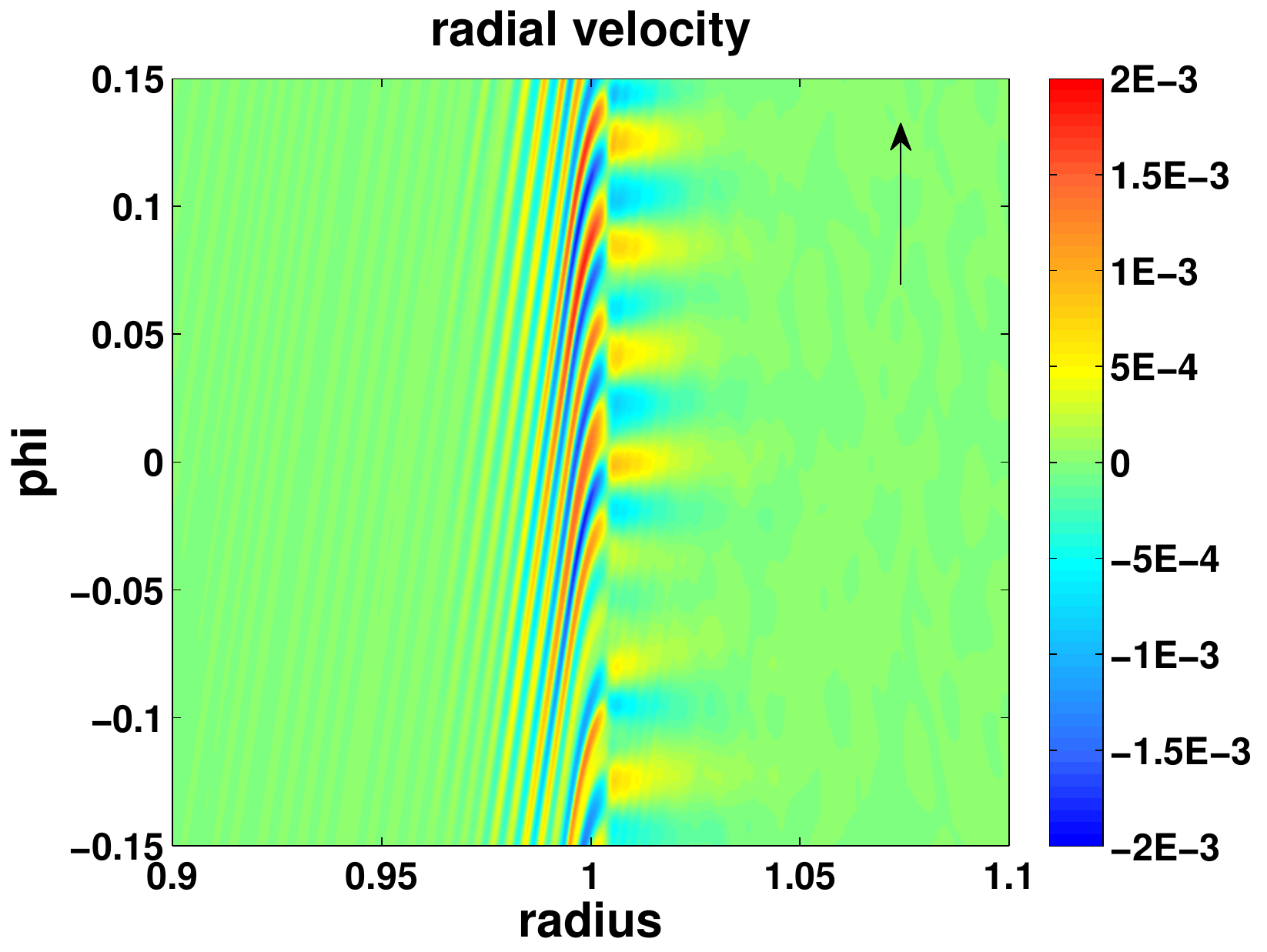}
\caption{a) $\cp - \phi$ plot of the radial velocity at $t=6$ for
simulation C1 ($M=12$) in the vicinity of the interface, showing 
the development of sonic instability. Sound waves propagating in 
the flow both into the star (for $\cp<1$) and into the disk 
(for $\cp>1$) are clearly visible. The black arrow
shows the direction of the unperturbed flow.}
\label{linfig}
\end{figure}

In our simulations, the signature of the sonic
instability during the linear stage of growth is the presence of sound
waves that are generated within the shear layer. In our initial setup,
the region that has the greatest shear is the interface, and Fig. \ref{linfig}
shows the radial velocity, $v_\cp$, at $t=6$ for
simulation C1. One can compare this with Fig. 5a of
\citet{BR}, which
shows the linear stage of the sonic instability in a uniform density
medium for a shear
layer with a piecewise linear velocity profile\footnote{Fig. 5a
  of \citet{BR} is rotated by 90 degrees relative to
  Figure \ref{linfig}.}. The similarities between the two figures
confirm that our initial setup is unstable to the sonic
instability. In particular, a critical layer where the pattern speed 
of the mode matches the azimuthal speed of the flow and the perturbation
undergoes a phase shift in the azimuthal direction can be
seen in Figure \ref{linfig} close to the top of the interface region, at $\cp
\approx 1.005$. Such a critical layer is a clear signature of sonic
instabilities and is necessary for them to operate. We note that the
sound waves emitted from the interface appear to vanish in
Fig. \ref{linfig} as they propagate inwards. This is due to
the exponentially rising density towards the interior of the star,
which causes their amplitude to decay as $\Sigma^{-1/2}$. On
the contrary, there is no stratification in Fig. 5a of \citet{BR} so
sound waves propagating away from the shear layer do not diminish
as quickly. However there is still some reduction in amplitude due to
the fact that sound waves emitted at later times have a larger
amplitude than sound waves emitted at earlier times, due to the
presence of an instability (see \S 4.2 of \citet{BR}). 

\citet{BR} have also shown that a high numerical
resolution, of order several hundred cells across the shear layer for
$M=10$, is
necessary during the initial stages of evolution to correctly
capture the linear growth rate of the sonic instability. Because it would be
prohibitive to use such a high resolution in our simulations (we use
$\sim 30$ cells across the interface), we do
not attempt to accurately resolve the growth rate of the sonic
instabilities during the linear regime. Despite this, we show in \S
\ref{BLwidth_sec}
that simulations with different numerical resolutions converge at late
times, so the resolution does not affect the long term evolution of
the BL. Moreover, using a lower resolution underestimates the growth
rate, so going to higher resolution would simply make the initial
instabilities proceed faster, rather than stabilizing the instability.

After $t \sim 10$ the sonic instabilities saturate and
after $t \sim 200$ the simulations settle down to a
quasi-steady state. In this quasi-steady state, the original interface
has thickened substantially forming a BL. For simplicity, we define
the self-consistent BL to be the region in which 
\ba
\label{BLwidth}
0.1 < \langle v_\phi(\cp)\rangle/v_K(R_\star) < 0.9, 
\ea
where $v_K(R_\star) = 1$ is the Keplerian velocity at the surface of the
star and $\langle v_\phi(\cp)\rangle$ is the azimuthally-averaged
azimuthal velocity.


\subsection{Trapped Modes}
\label{mode_sec}

\begin{figure}[!h]
\centering
  \includegraphics[width=0.5\textwidth]{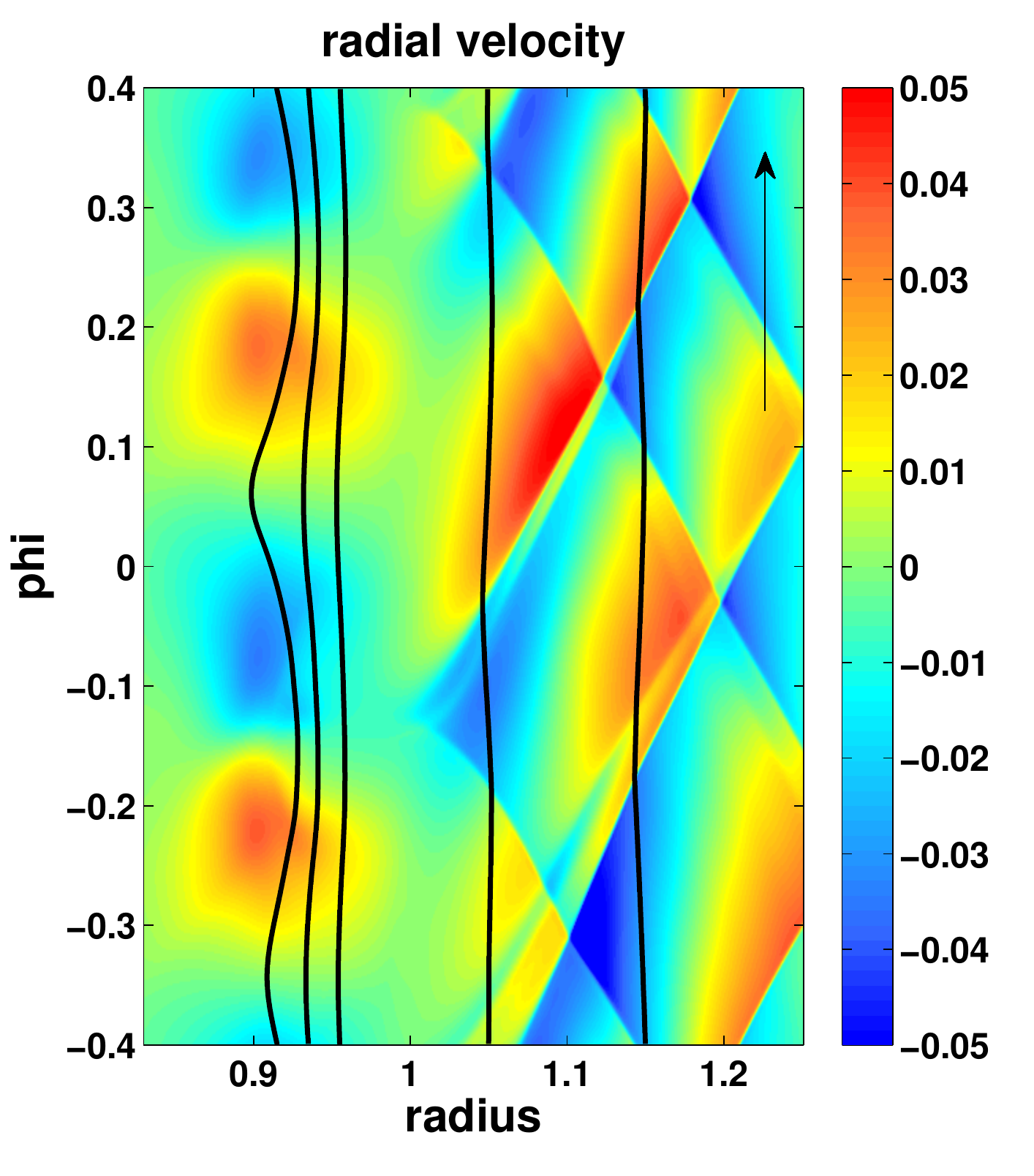}
\caption{$r - \phi$ plot of the radial velocity $v_\cp$ in the vicinity 
of the BL at $t=298$ for simulation A2 ($M=6$). The general sense of 
the flow is from
  bottom to top, and the black lines are streamlines which trace
  the deformation of the interface. A pair of vortices are seen at the
  base of the BL
  at $\cp \approx 0.9$, and two pairs of weak shocks, each consisting of
  a leading and a trailing shock, are seen originating from $\cp \approx 1$.}
\label{bump}
\end{figure}

By the time a quasi-steady state has been
established and the sonic instability has saturated, large scale
vortices are present at the base of the
BL.  Fig. \ref{bump} shows an image of $v_\cp$ at $t=298$ for
simulation A2, and the vortices are clearly visible at the base of the
BL. We mention that in this particular case, the number of vortices is
set by the box size in the azimuthal direction, but later in \S
\ref{fulldisk_sec}, we discuss simulations 
that span the full $2 \pi$ in azimuthal angle for which this is not
the case. Associated with the vortices is a deformation of the interface,
which is shown in Fig. \ref{bump} by the curved streamline. Since the
flow in the disk is supersonic over the surface of the BL, information
about the interface deformation cannot propagate upstream. As a
result, oblique shocks are created that guide the
supersonic flow in the disk around the deformed surface of the
BL. These shocks are clearly visible in Fig. \ref{bump} at radii $\cp
> 1$. 

\begin{figure}[!h]
\centering
\subfigure[]{\includegraphics[width=0.85\textwidth]{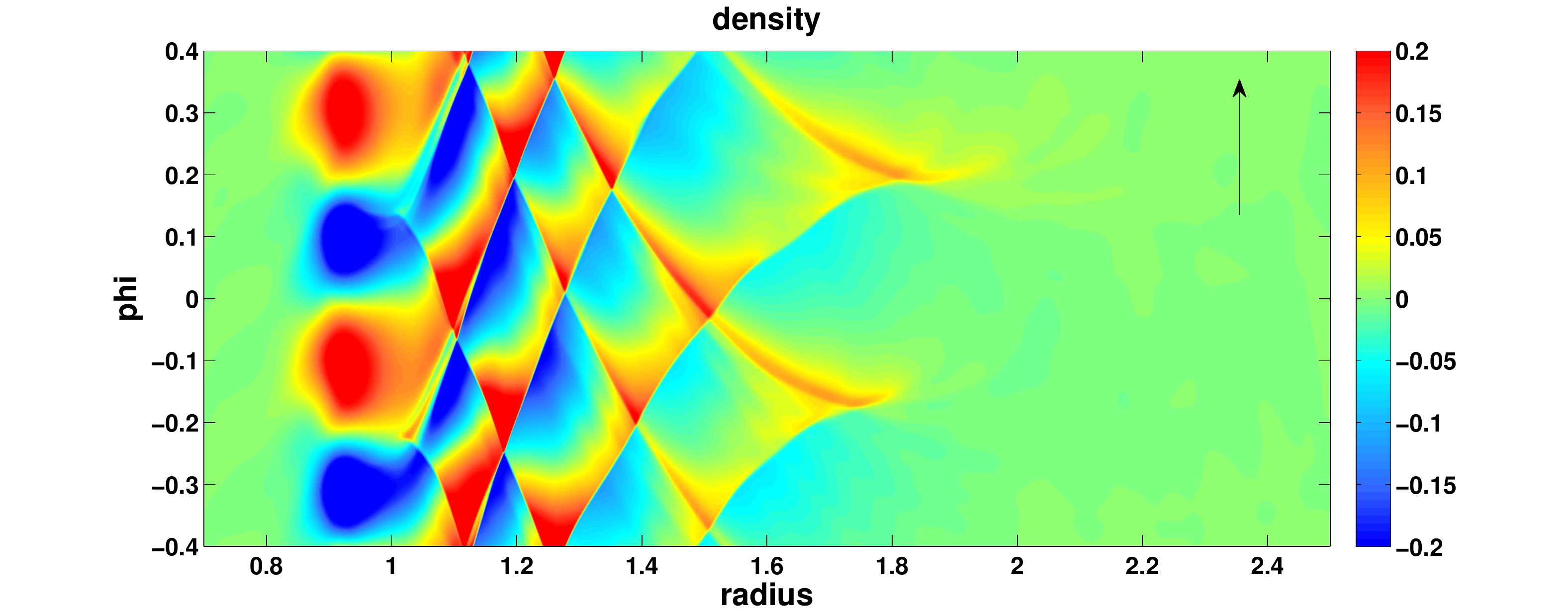}}
\subfigure[]{\includegraphics[width=0.85\textwidth]{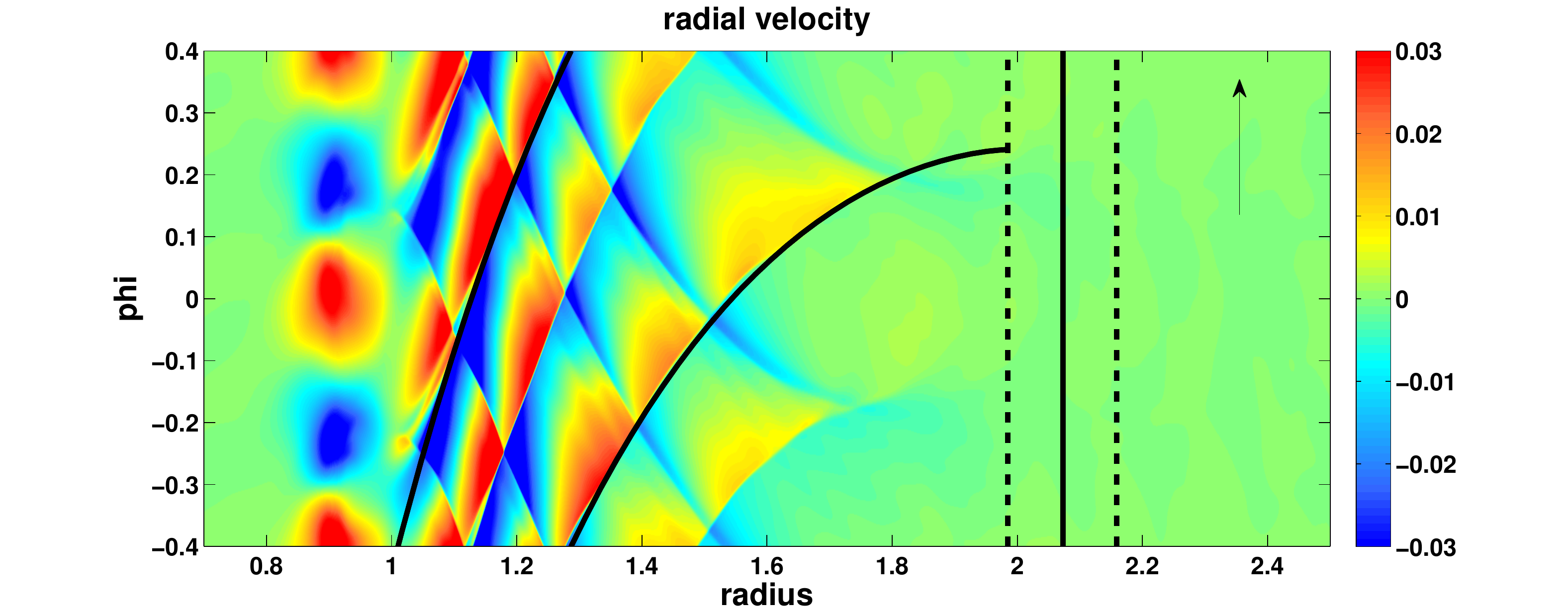}}
\caption{(a) Plot of relative density perturbation $\delta 
\Sigma/\langle\Sigma\rangle$ at $t = 284$ for
  simulation A2. The density is enhanced at shock crossings, but
  otherwise $\delta \Sigma/\langle\Sigma\rangle \sim 0.1-0.2$. (b) Plot of
radial velocity   $v_\cp$
  at $t=284$ for simulation A2 ($M=6$). The
  black curve shows the analytic solution for the shock front from the
  dispersion relation (\ref{keq}) using $\Omega_P = 0.335$. The dashed
  vertical lines show the
  locations of the inner and outer Lindblad resonances, and the solid
  vertical line shows the corotation radius. The black arrows show the
  direction of the unperturbed flow.}
\label{shockfig}
\end{figure}

For each bump in the BL, there are two weak shocks having $\delta \Sigma /
\langle\Sigma\rangle\sim 0.1-0.2 \lesssim 1$, where 
\ba
\label{sigma0}
\langle\Sigma(\cp)\rangle = \frac{1}{2\pi} 
\int_0^{2\pi} d \phi \Sigma(\cp,\phi)
\ea
is the azimuthally-averaged density, and
\ba
\label{deltasigmaeq}
\delta \Sigma = \Sigma - \langle\Sigma\rangle
\ea
is the density
perturbation. One of the shocks is an outgoing shock, and the other is
an incoming shock, which is simply an outgoing shock that has been
reflected from a Lindblad resonance within the disk (we discuss the
details of this
process below). The incoming and outgoing shocks intersect within the
disk, which leads to shock crossings that result in a local
enhancement of the density perturbation (Fig. \ref{shockfig}a). The
whole structure consisting of the vortices,
deformed interface, and shocks rotating in a prograde fashion with a
pattern speed $0 < \Omega_P < 1$.

We now consider in more detail the structure of intersecting shocks caused by
supersonic flow of disk material over the surface of the star. Since
the shocks are weak, we can apply linear theory to study their properties. 
In the linear approximation, the dispersion relation for a normal mode
perturbation of the form 
\ba
\delta \Sigma \propto \exp\left[i\left(\int^\cp k(\cp') d\cp' + m (\phi -
  \Omega_P t)\right)\right]
\label{eq:WKB}
\ea 
is given by
\ba
\label{keq}
k(\cp) = \frac{1}{s}\sqrt{m^2[\Omega(\cp) - \Omega_P]^2 - \kappa^2(\cp)},
\ea
which is simply the WKB dispersion relation for a fluid disk in
the absence of self-gravity \citep{BinneyTremaine}. Here, $k(\cp)$ is
the radial  wavenumber, $m$ is the azimuthal wavenumber, $\Omega_P$ is
the pattern speed, and $\kappa(\cp)\equiv \left(4\Omega^2+\cp 
d\Omega^2/d\cp \right)^{1/2}$ is the epicyclic frequency.

For a given pattern speed $\Omega_P$, the right hand side of equation
(\ref{keq}) consists entirely of known quantities so the
radial wavenumber as a function of radius is uniquely
determined. There are two corotation resonances in the system, one of
which is located inside the boundary
layer with the other one located inside the disk. Each of the two corotation
resonances is flanked by two Lindblad resonances which occur at
$\Omega = \Omega_P \pm \kappa/m$. In between a pair of Lindblad resonances,
$k$ is imaginary (equation [\ref{keq}]), and these regions
are forbidden to propagating waves, i.e. the waves are evanescent there. 

It should be noted that the WKB
approximation, which assumes $k \cp/m \gg 1$ breaks down near the
Lindblad resonances, since $k=0$ at a Lindblad
resonance. Nevertheless, we find that the WKB approximation works
well in the disk, even near a Lindblad resonance. On the other
hand, it performs poorly within the boundary layer, perhaps because
the WKB approximation ignores radial density gradients which are large within
the BL. For this reason, we will use a Keplerian rotation profile,
$\Omega(\cp) = \cp^{-3/2}$, rather than using
profiles for $\kappa$ and $\Omega$ determined from the
simulations. A Keplerian profile is accurate inside the disk, and
using it has the advantage of simplifying some of the
analysis without compromising on the physics. For
a Keplerian rotation profile, the
corotation resonance in the disk is located at
\ba
\label{CReq}
\cp_{CR} = \Omega_P^{-2/3},
\ea
and the Lindblad resonances are located at
\ba
\label{LReq}
\cp_{LR} = \left(\frac{m}{m \pm 1}\Omega_P\right)^{-2/3}.
\ea

When a shock launched from the BL
encounters the forbidden region in the disk between the 
two Lindblad resonances, it is partially reflected back
towards the BL. Moreover, this reflection is nearly perfect, 
and the transmission coefficient through the forbidden
region is small - so small that it cannot be measured
in the simulations. The reflected shock then propagates inward 
and is reflected again within the BL. The reflection within
the BL could either be due to the presence of Lindblad and 
corotation resonances
within the BL or because the radial scale height within the BL
becomes smaller than the radial wavelength of the shock, i.e. $k h_s <
1$. This process of reflections repeats cyclically resulting in a trapped
mode between the BL and the forbidden region in the disk. 

We point out that although the trapped modes are acoustic in nature,
they are distinct from the sonic instabilities described in \S
\ref{sonic_sec}. First, the sonic instabilities saturate very early in the
simulations ($t \sim 10$), whereas the trapped modes only develop
much later around $t \sim 300$. The likely reason for this is that the
boundary layer must settle down from the violent initial state
that occurs when the sonic instabilities saturate before the trapped
modes can develop. Second, there is clear leakage of wave action past
the critical layer during the sonic instability phase (e.g. Figure
\ref{linfig}), which is unbalanced by the wave dissipation, resulting
in growth of the wave amplitude. At the same time, during the trapped 
mode phase wave action tunneling past the corotation region is not 
so noticeable (e.g. Figure \ref{shockfig}). As a result, a quasi-steady state
is established during this phase of evolution, and the trapped modes 
are stable for hundreds of orbital periods. 

The black line in Figure \ref{shockfig}b is the solution for an outgoing shock
according to the dispersion relation (\ref{keq}) assuming a Keplerian
rotation curve for $\Omega(\cp)$ and using $\Omega_P = 0.335$. Due to
the periodic boundary conditions in the $\phi$ direction, the shock
exits from the top of the box and comes out again from the
bottom. From Figure \ref{shockfig}b, it is clear that the analytic
solution given by the black line traces the numerical solution for an
outgoing shock front very well, confirming the validity of the WKB
approximation. As a sanity check we confirmed that the value of $\Omega_P =
0.335$ was consistent with a movie of the simulation.

The dashed vertical lines in Fig. \ref{shockfig}b show the location of
the inner and outer Lindblad resonances, which are located at $\cp =
1.99$ and $\cp = 2.16$ respectively, assuming $\Omega_P = 0.335$. The
solid vertical line is the corotation resonance, which is located at $\cp =
2.07$. In between the Lindblad
resonances, $k$ in equation (\ref{keq}) is imaginary. This region
is forbidden to propagating waves, and outgoing waves incident upon it
are reflected
back towards the boundary layer. The outgoing shock (yellow) traced by the
black line in Fig. \ref{shockfig}b is indeed seen to be reflected into an
incoming shock (blue) in the vicinity of the inner Lindblad resonance.


\subsection{Conservation of the Angular Momentum Flux}
\label{angmom_sec}

A prediction of linear theory that we confirm in our simulations
is conservation of the angular momentum flux. In its most general 
form, the angular momentum flux for a disk with no self-gravity 
is given by
\ba
\label{genangmom}
C_L(\cp) = \cp^2 \int_0^{2 \pi} d \phi \Sigma v_\phi v_\cp.
\ea
Proof that $C_L(\cp) = const$, i.e. that the angular
momentum flux is conserved for stellar disks was first 
given by \citet{Toomre} using results about action density 
conservation from \citet{Whitham, Bretherton}. 

For small perturbations away from axisymmetry (linear theory), the
angular momentum flux is a second order quantity, and equation
(\ref{genangmom}) becomes \citep{BinneyTremaine, BalbusPapaloizou,
  SteinackerPapaloizou}
\ba
C_L(\cp) = 2 \pi \cp^2 \langle\Sigma\rangle \ol{\delta v_\phi \delta v_\cp}.
\label{linangmom}
\ea
Here $\langle\Sigma\rangle$ is the average surface density (equation
[\ref{sigma0}]), and the overbar denotes a density-averaged quantity
\ba
\ol{f}(\cp,t) \equiv \frac{\int_0^{2 \pi} d \phi
\Sigma f}{\int_0^{2 \pi} d\phi \Sigma}.
\ea
The velocity perturbations $\delta v_\phi$ and $\delta v_\cp$ are given by
\ba
\delta v_\cp  &=& v_\cp - \ol{v_\cp} \\ 
\delta v_\phi &=& v_\phi - \ol{v_\phi}.
\ea

We now test the linear theory during the quasi-steady state when there
is a single mode rotating at a fixed pattern speed (\S
\ref{mode_sec}). During the quasisteady state,
we have outgoing shocks that undergo almost perfect reflection
into incoming shocks at the Lindblad radius in the disk, so that
$C_L(\cp) \approx 0$. Thus rather than measuring $C_L(\cp)$, we
instead measure the quantity
\ba
\label{tildeCL}
\widetilde{C_L}(\cp) = 2 \pi \cp^2 \langle\Sigma\rangle 
\ol{|\delta v_\phi \delta v_\cp|},
\ea
where the vertical lines denote taking the absolute value.
Clearly, $\widetilde{C_L}(\cp) > 0$ for nonzero perturbations. Moreover, away
from shock crossings, when the outgoing and incoming shocks at a given
radius are well-separated in azimuth, we have $\widetilde{C_L}(\cp) \approx
const$. This is because for well-separated shocks, $\widetilde{C_L}(\cp)$ measures
the sum of the {\it absolute values} of the angular
momentum flux carried by the individual outgoing and incoming
shocks; the angular momentum flux is
individually conserved for each of these shocks, implying that the sum
of the absolute values of the angular momentum flux over all the shocks is
also conserved.

In Appendix \ref{totalmom}, we demonstrate that sufficiently far from the 
Lindblad resonances (when $|\Omega(\cp)-\Omega_P|\gtrsim \kappa$) and
away from shock crossings, $\widetilde{C_L}$ is related to the surface density
perturbation $\delta\Sigma$ (equation[\ref{deltasigmaeq}]) via
\ba
\label{finaleq}
\widetilde{C_L}(\cp) = \frac{\cp s^3}{\langle\Sigma\rangle
  (\Omega -\Omega_P)} \int_0^{2 \pi} d \phi \left(\delta \Sigma\right)^2.
\ea
A similar result was previously obtained in the shearing-sheet
approximation by \citet{GoodmanRafikov}.

\begin{figure}[!h]
\centering
  \includegraphics[width=0.7\textwidth]{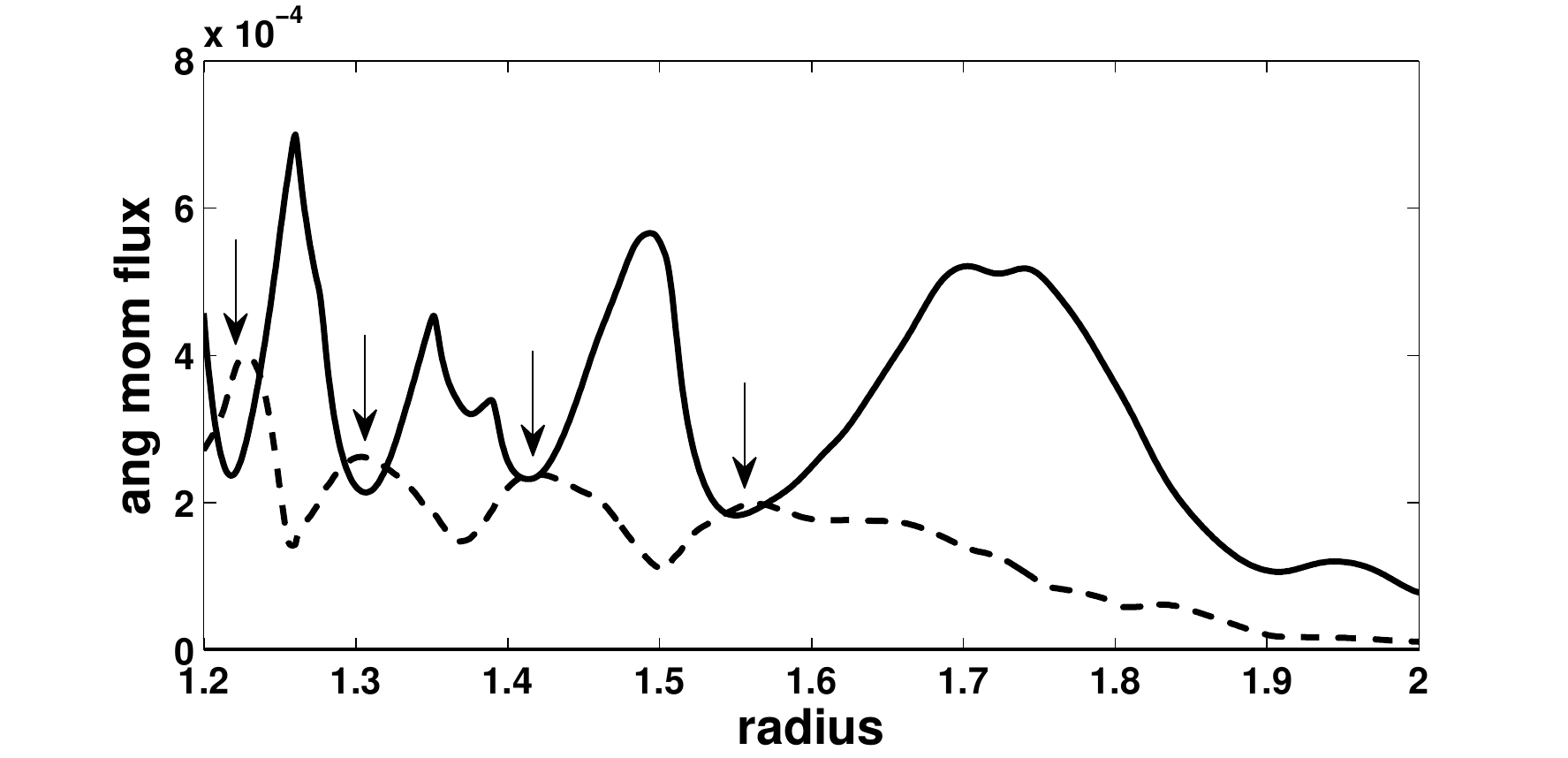}
\caption{Plot of $\widetilde{C_L}(\cp)$ at time
  $t=284$ for simulation A2 ($M=6$) computed using equation
  (\ref{finaleq}) (solid line) and equation (\ref{tildeCL}) (dashed
  line). The arrows indicate the radii which are in between shock
  crossings and where there is good agreement between the two forms
  for $\widetilde{C_L}$.}
\label{constfig}
\end{figure}

Figure \ref{constfig} shows $\widetilde{C_L}$ from simulation A2 at
$t=284$, computed using equation
(\ref{finaleq}) (solid line) and equation (\ref{tildeCL}) (dashed
line). It is not a problem that the simulation domain extends only
from $-0.4 < \phi < 0.4$ rather than the full $2 \pi$ in azimuth, since
the boundary conditions are periodic and we can stack domains
azimuthally. The arrows in the figure indicate the radii at which the
two forms of
$\widetilde{C_L}$ are approximately equal. We have checked using
Figure \ref{shockfig}a that
these radii correspond to regions which are in between shock
crossings. In these regions,
the individual shocks are well-separated in azimuth, which is precisely
the regime under which equations (\ref{tildeCL}) and (\ref{finaleq})
must give the same answer. When the shocks are not well-separated,
equations (\ref{tildeCL}) and (\ref{finaleq}) no longer agree, and the
curves exhibit oscillations due to the presence of shock
crossings. However, the two most important points are that away
from the shock crossings: (1) equations (\ref{tildeCL}) and (\ref{finaleq})
give similar values for $\widetilde{C_L}$, and (2)
$\widetilde{C_L}$ is roughly constant in radius with perhaps a slight,
decreasing trend. 

We note that the angular momentum flux is not expected to be 
exactly conserved as the waves propagate since the shocks 
are not adiabatic and lose energy due to dissipation. 
However, accounting for the
effect of dissipation on the angular momentum flux as a function of
radius is made complicated by the fact that we have both
outgoing and incoming (reflected) shocks in the simulation.


\subsection{Effective Value of $\alpha$}
\label{alpha_sec}

It is common practice in semianalytic studies of the boundary layer to
parametrize the angular momentum transport using an {\it ad hoc} prescription.
\citet{Kley, PophamNarayan, InogamovSunyaev, PiroBildsten,
  Balsaraetal} all used modified forms of the standard \citet{ShakuraSunyaev}
$\alpha$-viscosity prescription for disks, but the details of the
prescription vary from one author to the next. The reason for such a lack of
consensus comes from the fact that a simple
$\alpha$-viscosity prescription leads to
supersonic infall velocities in the BL, which \citet{Pringle} has
argued are unphysical on the basis that the disk should remain
causally connected to the star. Consequently, it is not clear
how the angular momentum transport should be parametrized inside the
BL, or whether it
can be parametrized at all with a simple prescription, given the
complicated shock structures that can develop.

In the purely hydrodynamical case, the starting
point of any prescription for the angular momentum transport should be
the Reynolds stress. We define the Reynolds
stress as \citep{BalbusPapaloizou,SteinackerPapaloizou}
\ba
\tau_{Re}(\cp,t) \equiv  \langle\Sigma\rangle \ol{\delta v_\cp \delta v_\phi}.
\ea
From this, it is possible to define a dimensionless parameter
\ba
\label{alphaeq}
\alpha_{Re}(\cp,t) \equiv \frac{\tau_{Re}}{\langle\Sigma\rangle s^2}.
\ea
It is clear that $\alpha_{Re}$ is a dimensionless
Reynolds stress, and this definition is consistent with the original definition
of \citet{ShakuraSunyaev}. However, one major point to keep in mind
is that one should expect $\alpha_{Re}$ to be {\it negative} inside the
BL. This is because the rotation profile
  rises in the BL ($d \ol{\Omega}/d \cp > 0$) so angular momentum is
  transported inwards, spinning up the star. An inward transport of
  angular momentum, leads to a negative value of the Reynolds stress,
  and hence a negative value of $\alpha_{Re}$.

\begin{figure}[!h]
\centering
  \subfigure[]{\includegraphics[width=0.49\textwidth]{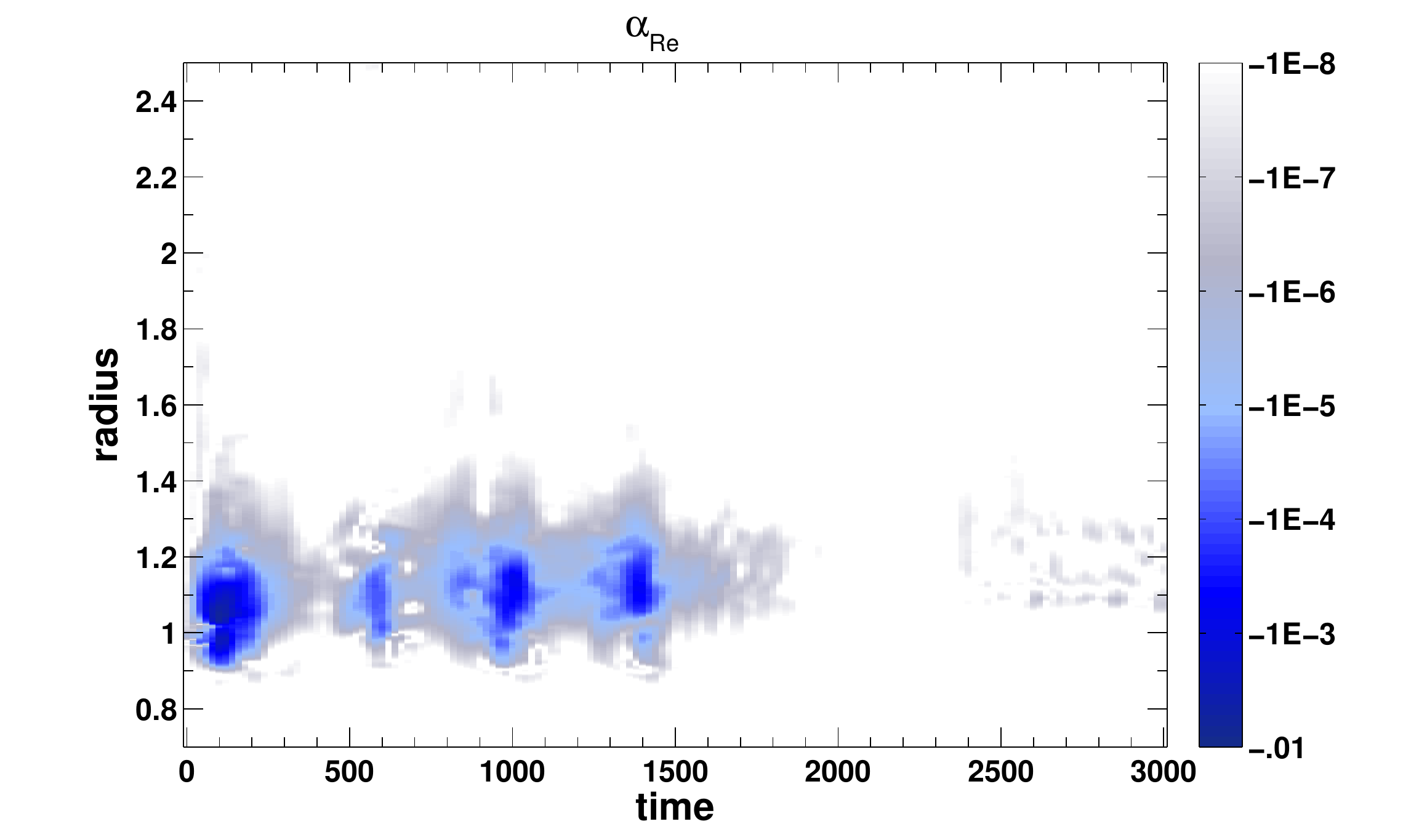}}
  \subfigure[]{\includegraphics[width=0.49\textwidth]{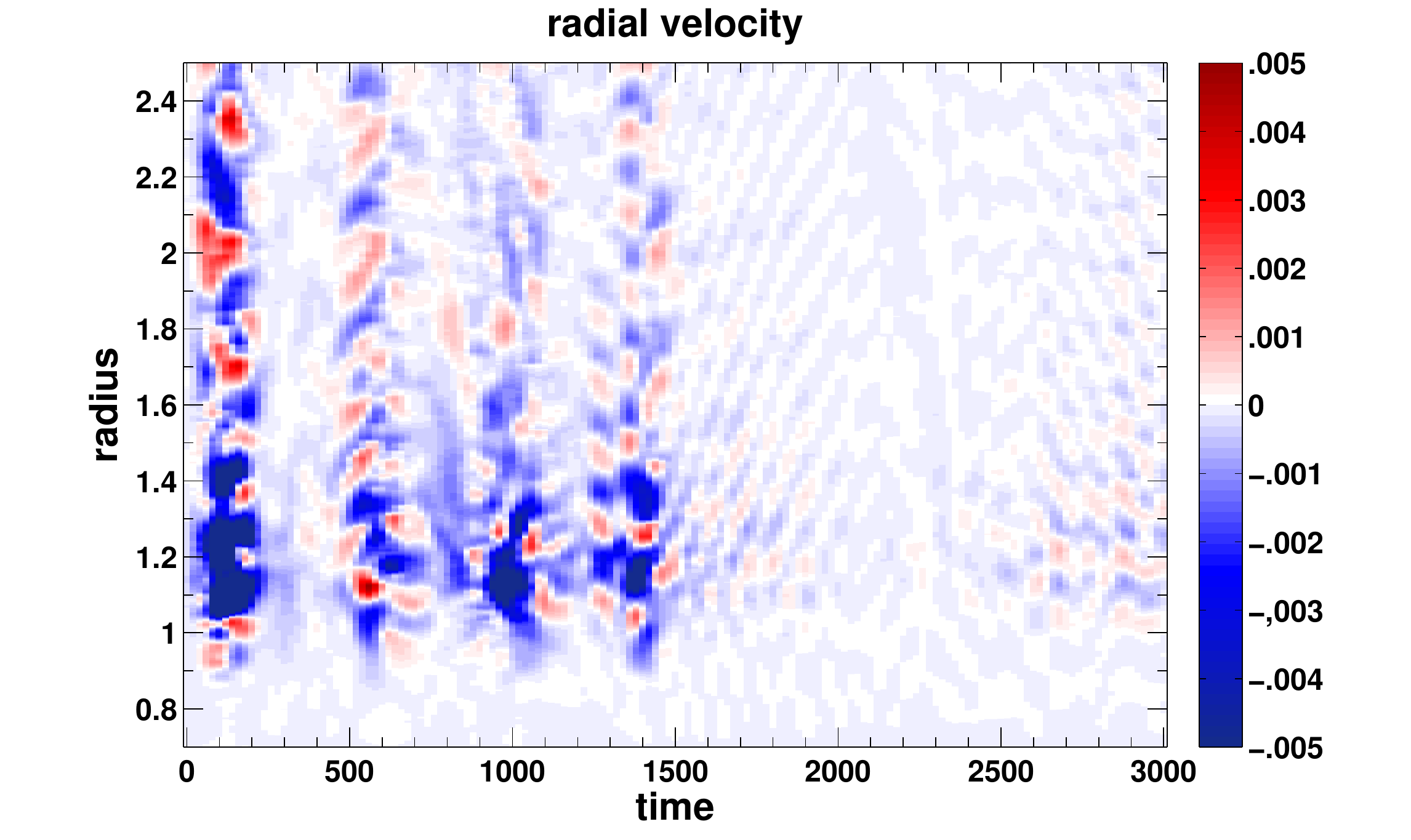}}
  \subfigure[]{\includegraphics[width=0.49\textwidth]{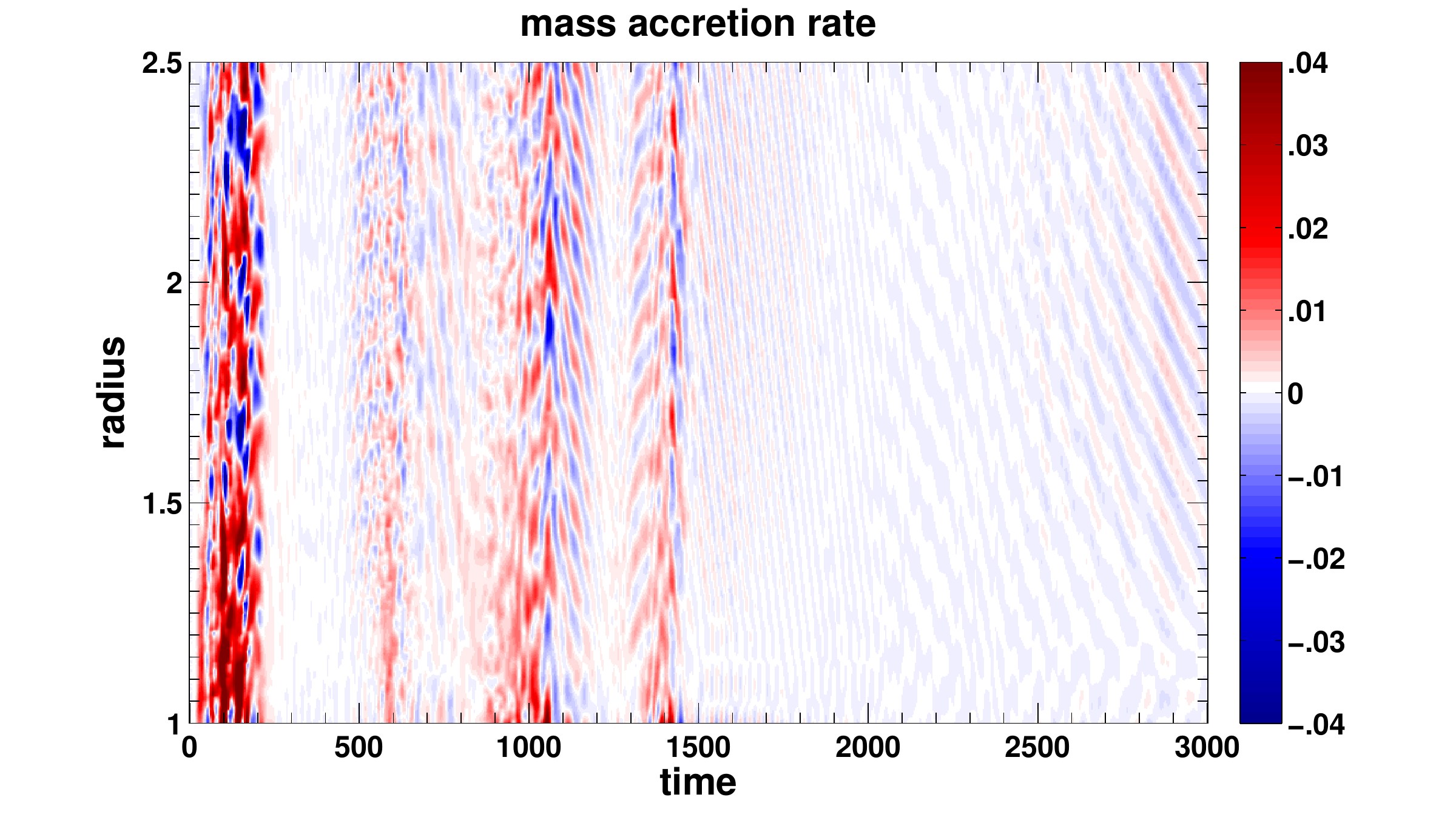}}
\caption{(a) Time-radius image of the dimensionless Reynold's stress
  parameter $\alpha_{Re}$ defined in
equation (\ref{alphaeq}) for run A2 ($M=6$). 
(b) The radial velocity, $\ol{v_\cp}$, for the
  same run. c) The mass accretion rate, $\dot{M}$, as defined in equation
  (\ref{massaccretobs})
  for the same run. The units for b) and c) are the code units
  discussed in \S \ref{nummod}.}
\label{M6spacetime}
\end{figure}

Figure \ref{M6spacetime}a shows a time-radius image of
$\alpha_{Re}$ from simulation A2. It is evident that $\alpha_{Re}$ is negative
and is spatially
nonzero only in the vicinity of the BL ($\cp \sim 1$). The latter is due to
the fact that
we have no MRI in the disk and that shear instabilities only operate in
the vicinity of the BL. It is also clear that $\alpha_{Re}$ is not constant
in time, exhibiting
temporal spikes at $t=150$, $t=600$, $t=1000$, and $t=1400$. We
explain the reason for these spikes in \S \ref{highlow_sec}. Typical
maximum values of $\alpha_{Re}$ as a function of radius
during the low and high accretion states (see \S \ref{highlow_sec} for
the definition of low and high accretion state) are
$\alpha_{Re} \sim -2 \times
10^{-3}$ and $\alpha_{Re} \sim -0.05$ respectively. 

Figures \ref{M6spacetime}b,c show time-radius images of the radial
velocity, $\ol{v_\cp}$, and the mass accretion rate $\dot{M}$
(equation [\ref{massaccretobs}]) within the disk. The radial velocity in the
vicinity of the
BL is predominantly negative and the mass accretion rate positive,
indicating accretion onto the central
object, and there are temporal spikes in the radial velocity and mass
accretion rate that
are coincident with the spikes in $\alpha_{Re}$. In the disk proper
outside the BL, the radial velocity exhibits alternating negative and
positive regions, which are the imprints of the inward- and 
outward-propagating modes.
In code units (\S \ref{nummod}), the typical values for the mean
radial infall velocity
during the low and high accretion states are $\ol{v_\cp} \sim -2
\times 10^{-4}$ and $\ol{v_\cp} \sim -0.01$ respectively, and typical
values for the mass accretion rate during the low and high accretion
states are $\dot{M} \sim 10^{-3}$ and $\dot{M} \sim .05$, respectively.

The negative value of the Reynolds stress in the BL yields a negative
value of $\alpha_{Re}$. It is straightforward, however, to define a
parameter, $\alpha_\nu$, which one might expect to be positive in both the
BL and the disk. We can do this through a prescription for the
kinematic viscosity
\ba
\nu(\cp,t) \equiv \alpha_\nu s l_t,
\ea 
where $l_t$ is the characteristic length scale for turbulent motions.
The viscous stress is given by
\ba
\tau(\cp,t) = - \langle \Sigma \rangle \nu \cp d \ol{\Omega}/d \cp,
\ea
and in the purely hydrodynamical case that we consider here, $\tau =
\tau_{Re}$, since there are no magnetic stresses. Thus, we can write
$\alpha_\nu$ as
\ba
\label{alphanueq}
\alpha_\nu(\cp,t) \equiv -\frac{\tau_{Re}}{\langle \Sigma \rangle s l_t \cp d
   \ol{\Omega}/ d \cp}.
\ea
It is clear, then, that the condition for $\alpha_\nu$ to be positive
   is that 
\ba
\label{alphacond}
\sgn(\tau_{Re}) = -\sgn(d\ol{\Omega}/d\cp),
\ea
and the stress should vanish at the location where $d\ol{\Omega}/d\cp = 0$.

For the value of $l_t$ in equation (\ref{alphanueq}), we adopt the
prescription of \citet{PophamNarayan}
\ba
l_t(\cp,t) = \left[ \left(\frac{\ol{\Omega}}{s}\right)^2 +
  \left(\frac{dP/d\cp}{P}\right)^2 \right]^{-1/2},
\ea 
and we define $(dP/d\cp)/P \equiv (d\langle P \rangle /d\cp)/\langle P
\rangle$, for purposes of readability.
This prescription essentially sets the turbulent scale height to be
the smaller of the vertical disk scale height, $s/\ol{\Omega}$, and the
radial pressure scale height, $|P/(dP/d\cp)|$. This generalization is
necessary for the BL, since the disk scale height defines the
characteristic length scale inside the disk, whereas the pressure scale height
defines it inside the star.

\begin{figure}[!h]
\centering
  \subfigure[]{\includegraphics[width=0.49\textwidth]{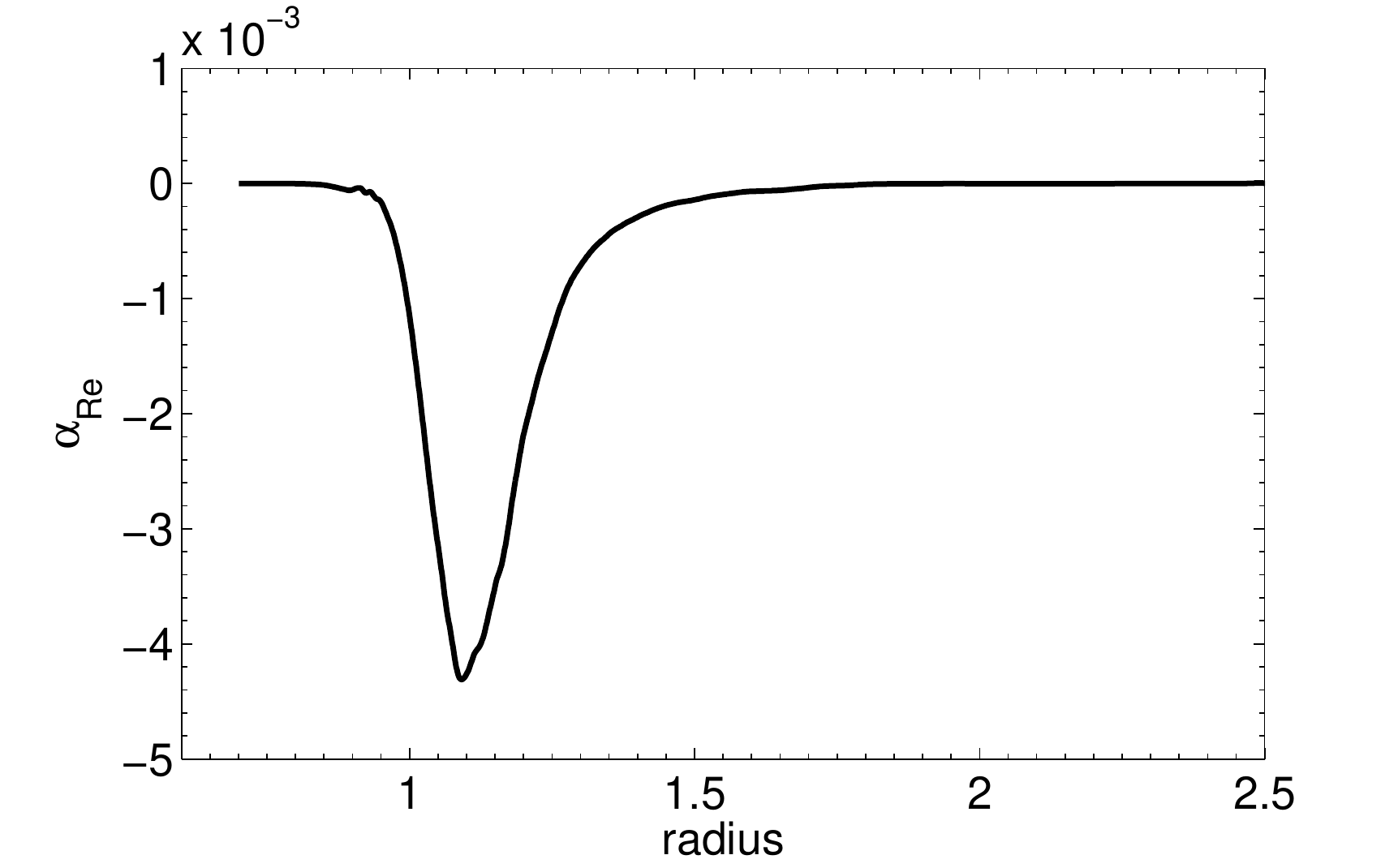}}
  \subfigure[]{\includegraphics[width=0.49\textwidth]{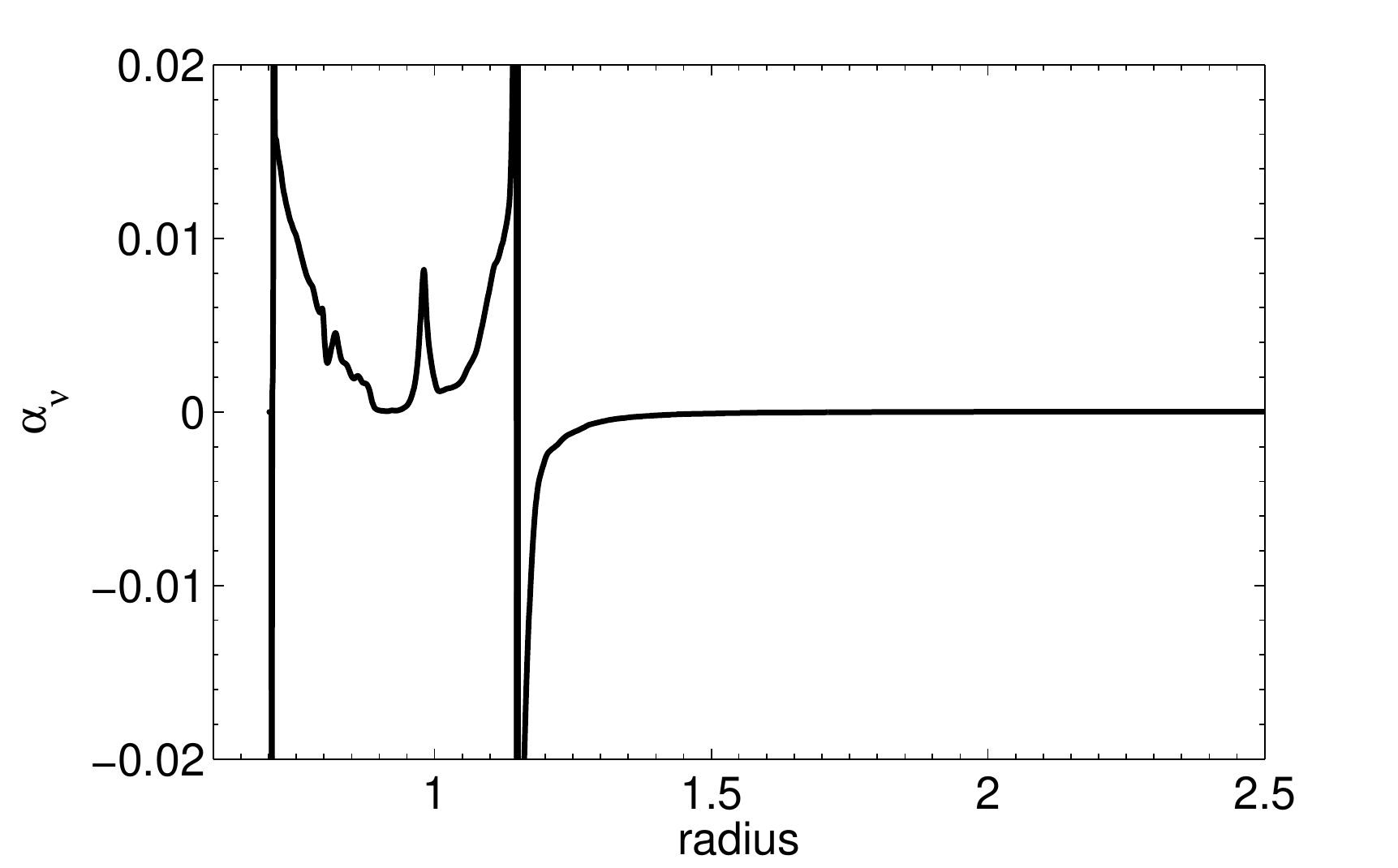}}
  \subfigure[]{\includegraphics[width=0.49\textwidth]{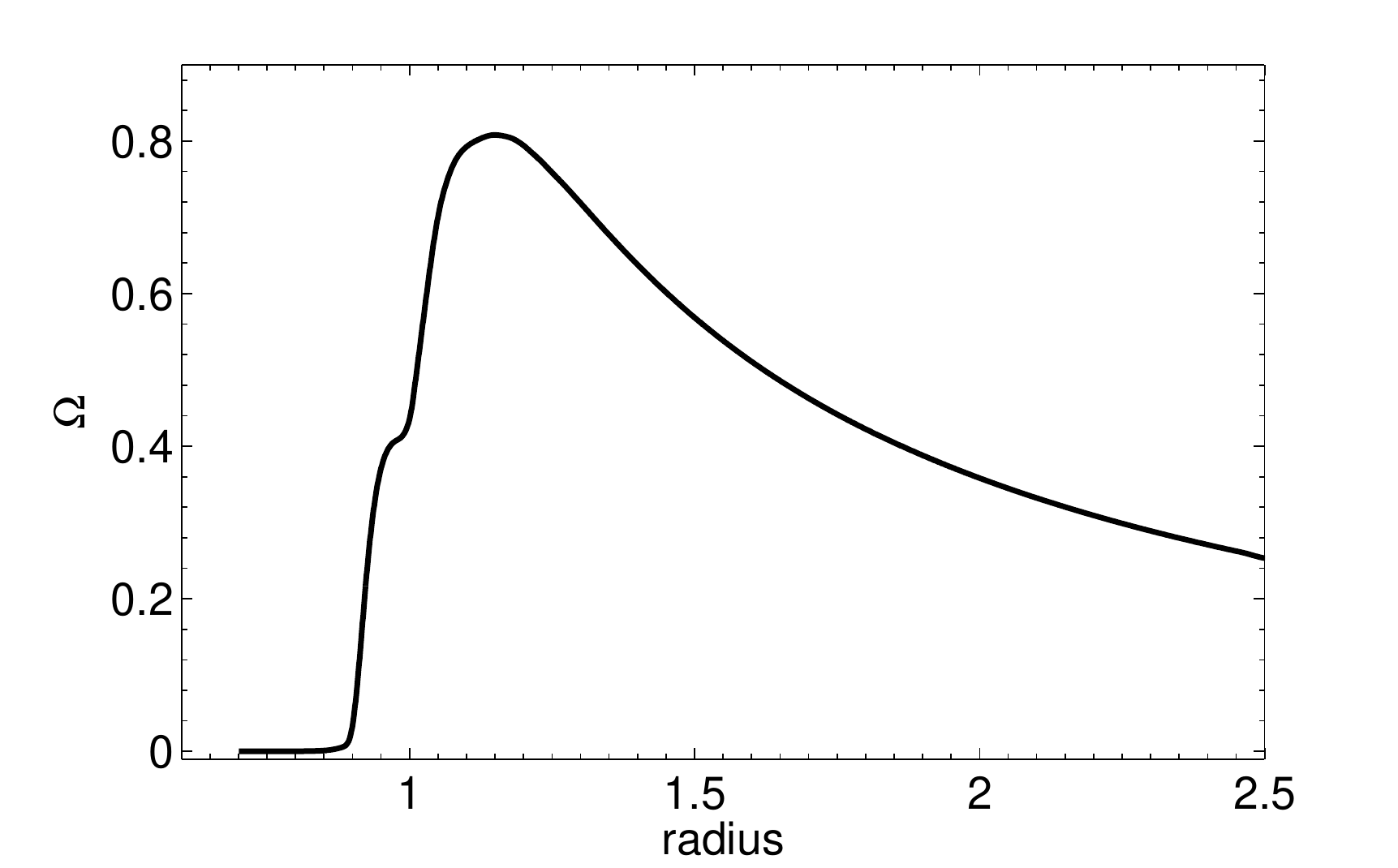}}
  \subfigure[]{\includegraphics[width=0.49\textwidth]{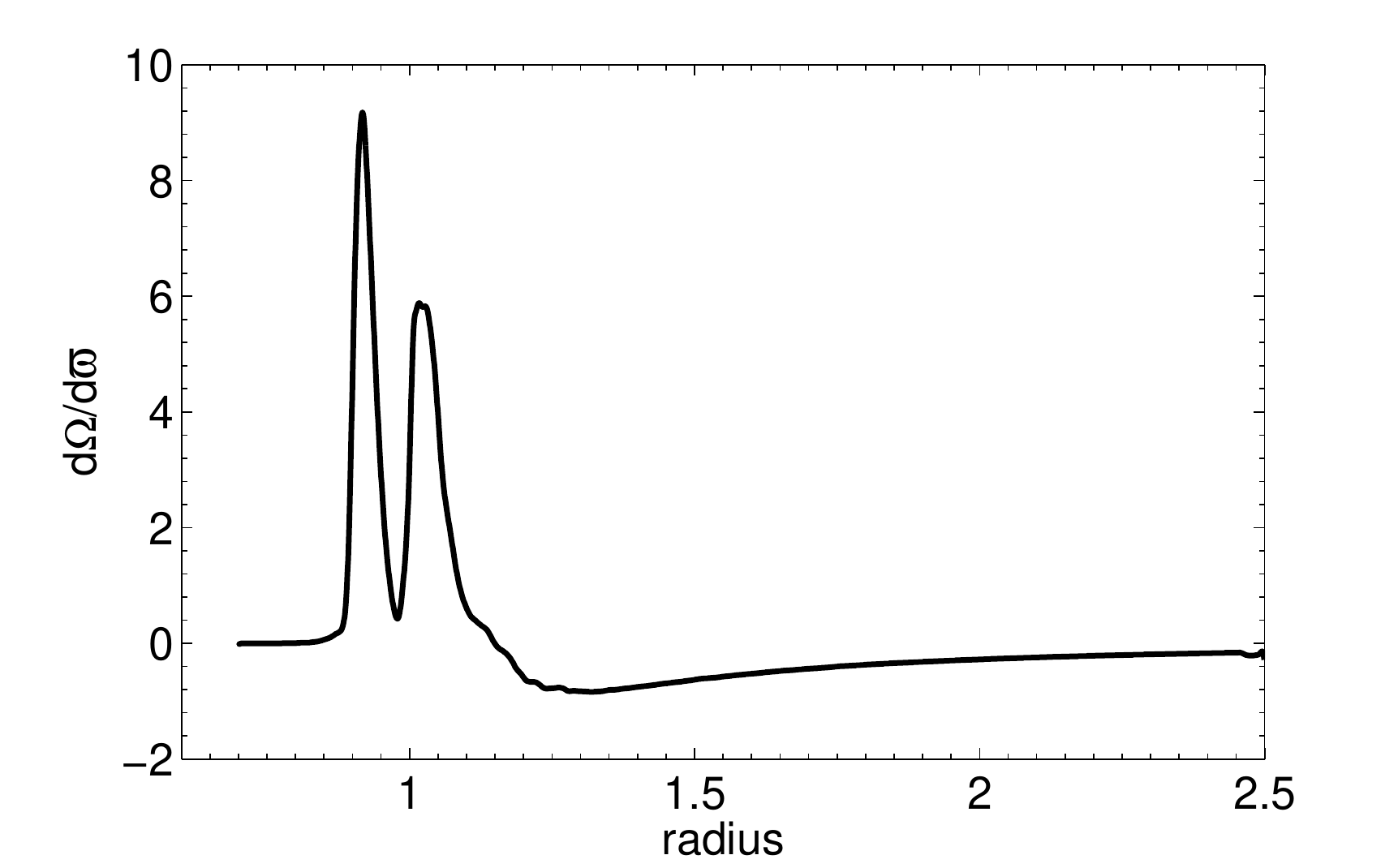}}
\caption{Plots of $\alpha_{Re}$, $\alpha_\nu$, $\ol{\Omega}$, and
  $d\ol{\Omega}/d\cp$ as functions of radii at $t=300$.}
\label{alpha300fig}
\end{figure}

\begin{figure}[!h]
\centering
  \subfigure[]{\includegraphics[width=0.49\textwidth]{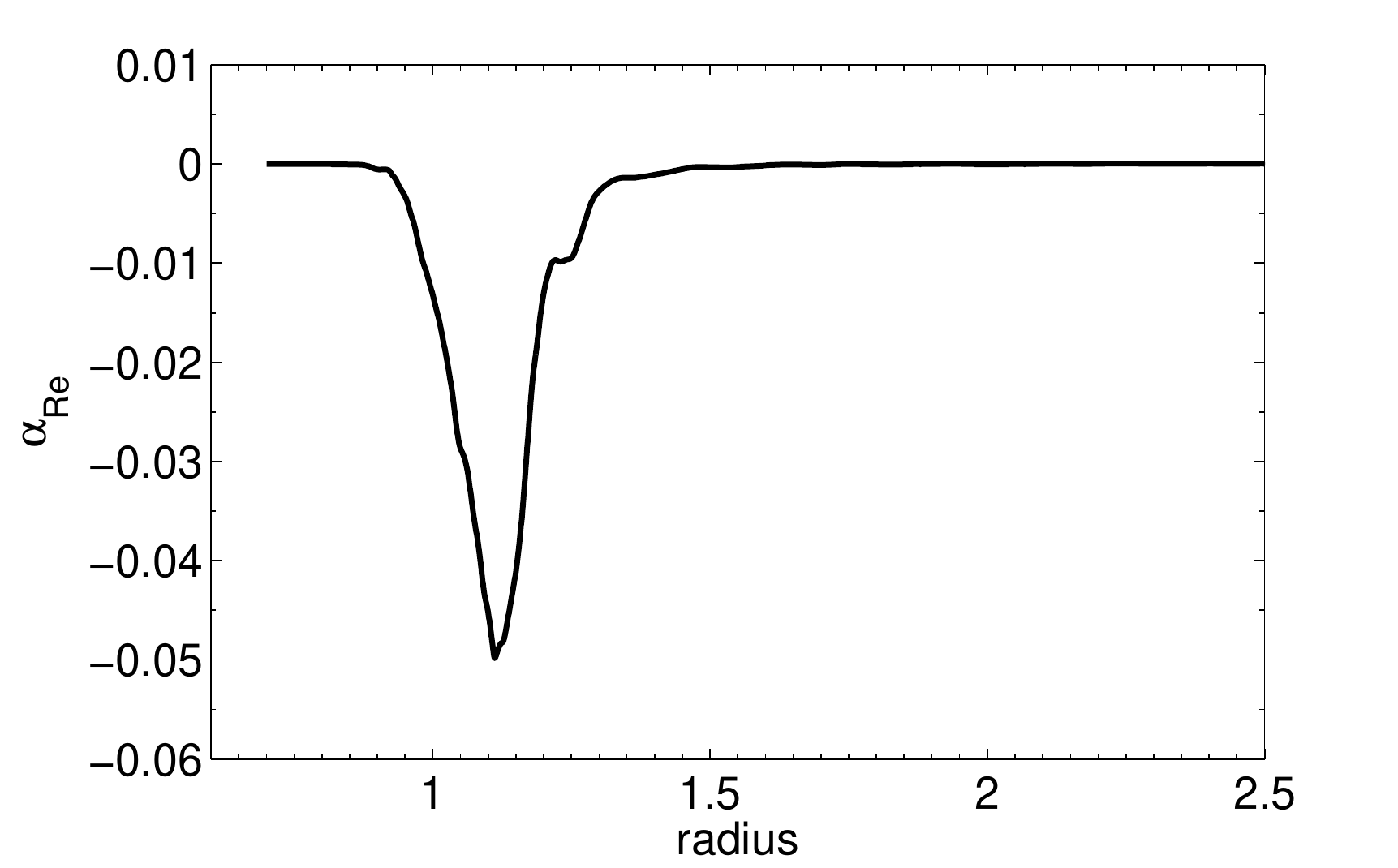}}
  \subfigure[]{\includegraphics[width=0.49\textwidth]{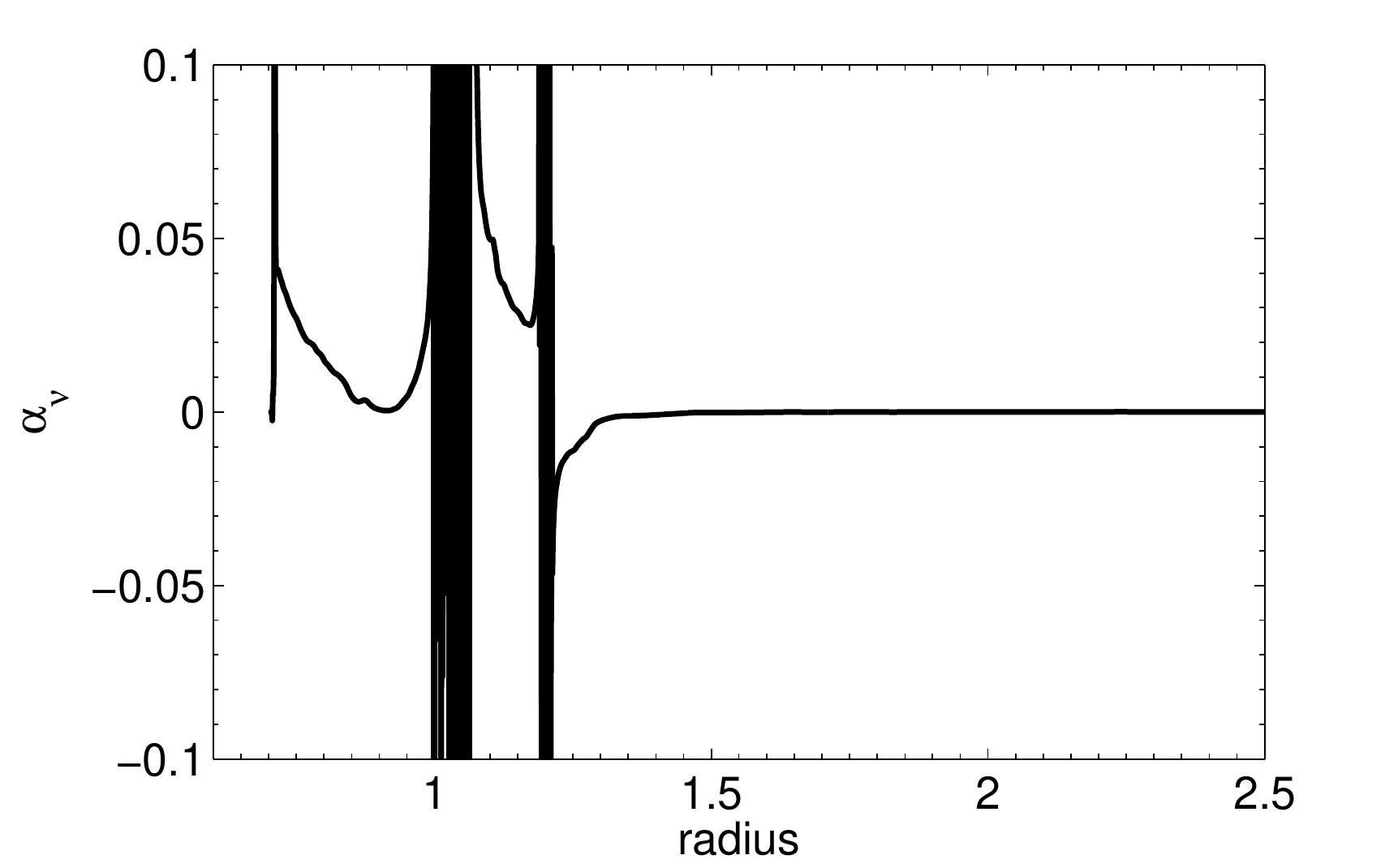}}
  \subfigure[]{\includegraphics[width=0.49\textwidth]{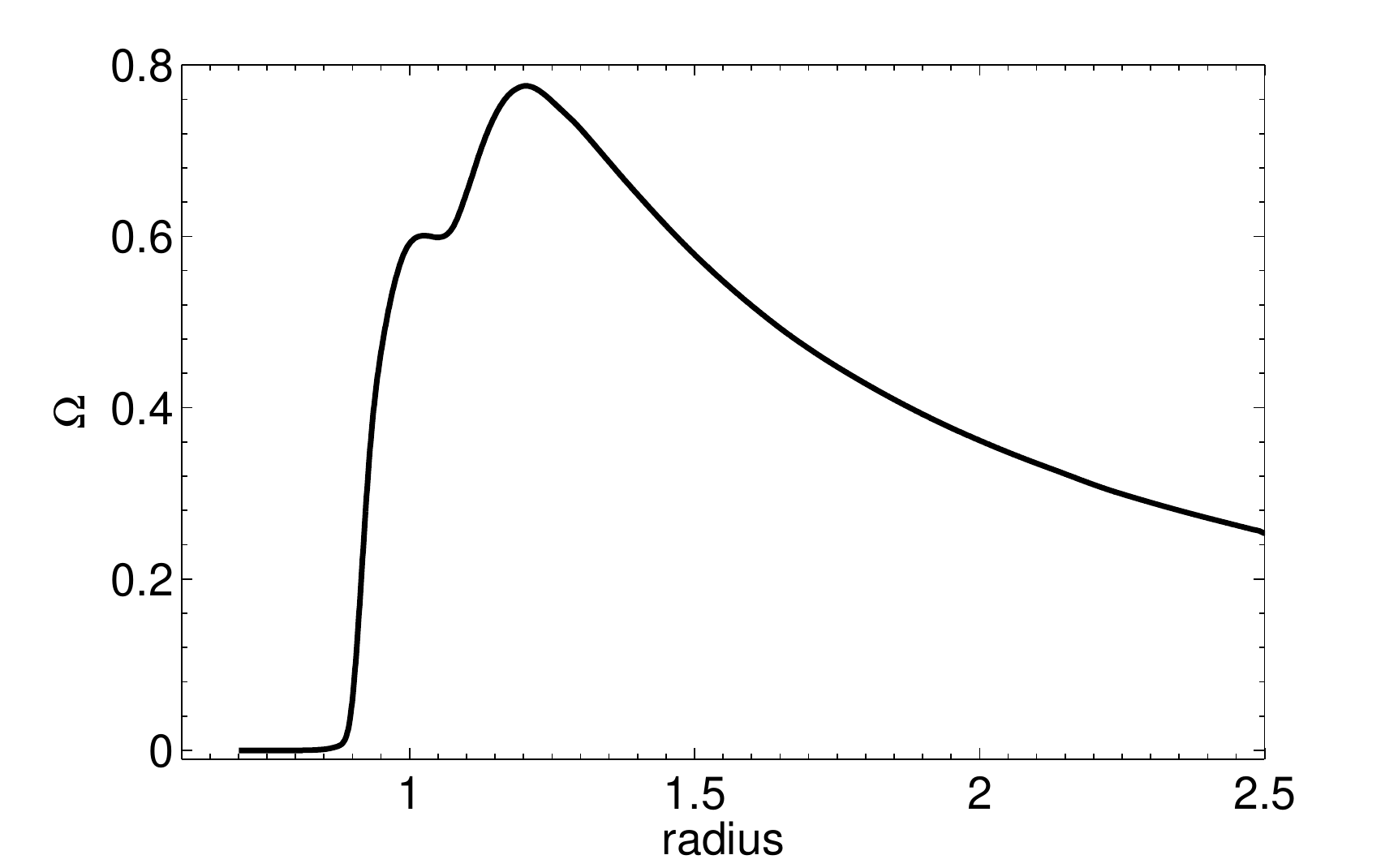}}
  \subfigure[]{\includegraphics[width=0.49\textwidth]{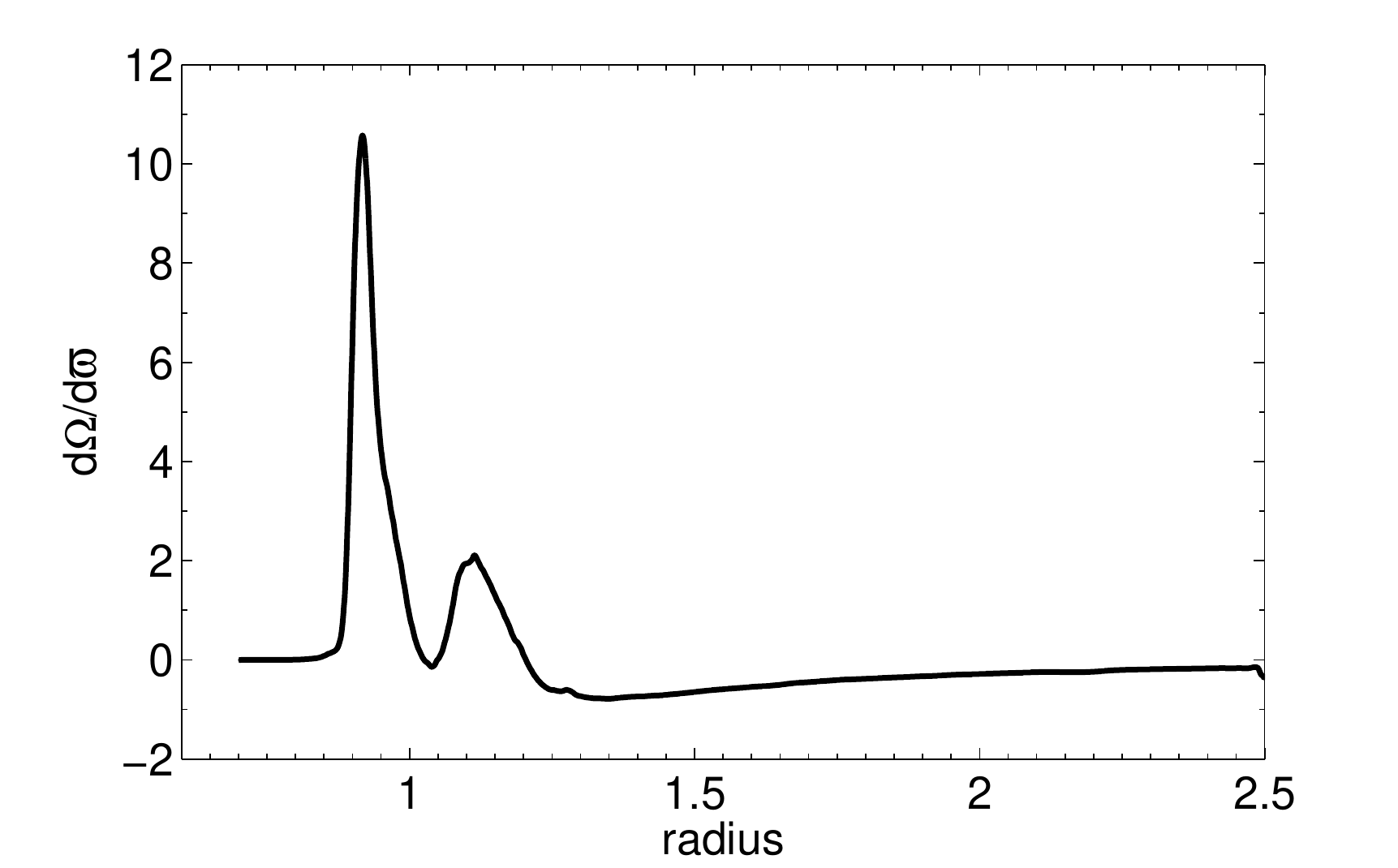}}
\caption{Plots of $\alpha_{Re}$, $\alpha_\nu$, $\ol{\Omega}$, and
  $d\ol{\Omega}/d\cp$ as functions of radii at $t=1000$.}
\label{alpha1000fig}
\end{figure}

Figures \ref{alpha300fig} and \ref{alpha1000fig} show the values of
$\alpha_{Re}$ (equation
[\ref{alphaeq}]), $\alpha_\nu$ (equation [\ref{alphanueq}]), $
\ol{\Omega}$, and $d\ol{\Omega}/d\cp$ at $t =
300$ (low state) and $t=1000$ (high state) respectively. In each
of the figures, both $\alpha_{Re}$ and $\alpha_\nu$ have been averaged in
time to reduce the level of fluctuations. 

It is clear that $\alpha_{Re}$ is negative as discussed earlier, since
the Reynolds stress is negative in the BL. On the other hand,
$\alpha_\nu$ is mostly positive in the boundary layer. However, the
condition (\ref{alphacond}) is not exactly satisfied, leading to the
large spike in Figs. \ref{alpha300fig}b and \ref{alpha1000fig}b at $\cp
\approx 1.15$ and $\cp \approx 1.2$,
respectively. Moreover, the rotation profile in the high state at
$t=1000$ is very flat between $.97 \lesssim \cp \lesssim 1.07 $
(Fig. \ref{alpha1000fig}c) and
even contains a change in sign of $d\ol{\Omega}/d\cp$
(Fig. \ref{alpha1000fig}d). This
again leads to a violation of condition (\ref{alphacond}) and leads
to the densely packed spikes between $.97 \lesssim \cp \lesssim 1.07$
in Fig. \ref{alpha1000fig}b. 

If the non-trivial nature of the rotation profile
(Figs. \ref{alpha300fig}c and \ref{alpha1000fig}c) that we observe in our
simplified simulations holds under astrophysical conditions, it seems
unlikely that the dynamics of the BL can be captured with a simple
prescription for $\alpha_\nu$. On the other hand, the dimensionless
Reynolds stress $\alpha_{Re}$ is quite smooth and well-behaved inside
the BL (Figs. \ref{alpha300fig}a and \ref{alpha1000fig}a). Thus,
$\alpha_{Re}$ could be the preferred parameter for constructing semi-analytical
models of the BL.

\subsection{High and Low Accretion States}
\label{highlow_sec}

\begin{figure}[!h]
\centering
  \includegraphics[width=0.85\textwidth]{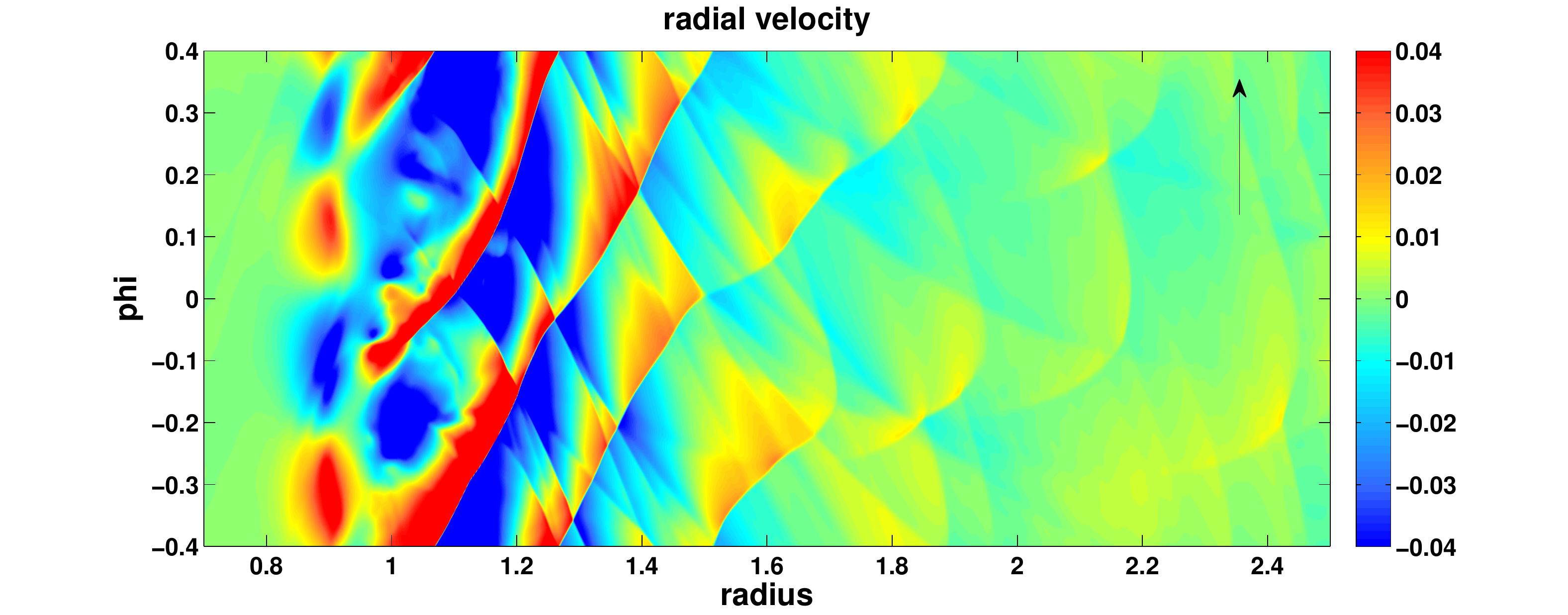}
\caption{Radial velocity during the high accretion state at $t=1000$
  for simulation A2 ($M=6$). Compare this with a snapshot of the system in
the low accretion state in Figure \ref{bump}.}
\label{highaccretionstate}
\end{figure}

From Figure \ref{M6spacetime} it is clear that our simulations are
characterized by states of high and low accretion rate. The high accretion
states are centered on $t=150$, $t=600$, $t=1000$, and $t=1400$ and
correspond to large absolute values of the Reynolds stress and of the radial
infall velocity (Fig. \ref{M6spacetime}). The low accretion state is
relatively calm and is characterized by a single dominant mode that
rotates at a constant pattern speed as
discussed in \S \ref{mode_sec}.  On the other hand, the oscillations
of the interface during the high accretion state are more violent, and
it becomes more difficult to determine a single dominant mode or a
unique pattern speed. Nevertheless, the shock structure seen very
distinctly in the low accretion state is still present to some extent
even in the high accretion state. Fig. \ref{highaccretionstate} shows a
plot of the radial
velocity in the high accretion state at $t=1000$ for simulation A2,
and can be compared with Fig. \ref{shockfig}b, which shows the radial velocity
during the low accretion state. It is clear that the high accretion
state is more violent and chaotic than the low accretion state, but
a dominant mode is still present. 

\begin{figure}[!h]
\centering
  \includegraphics[width=0.5\textwidth]{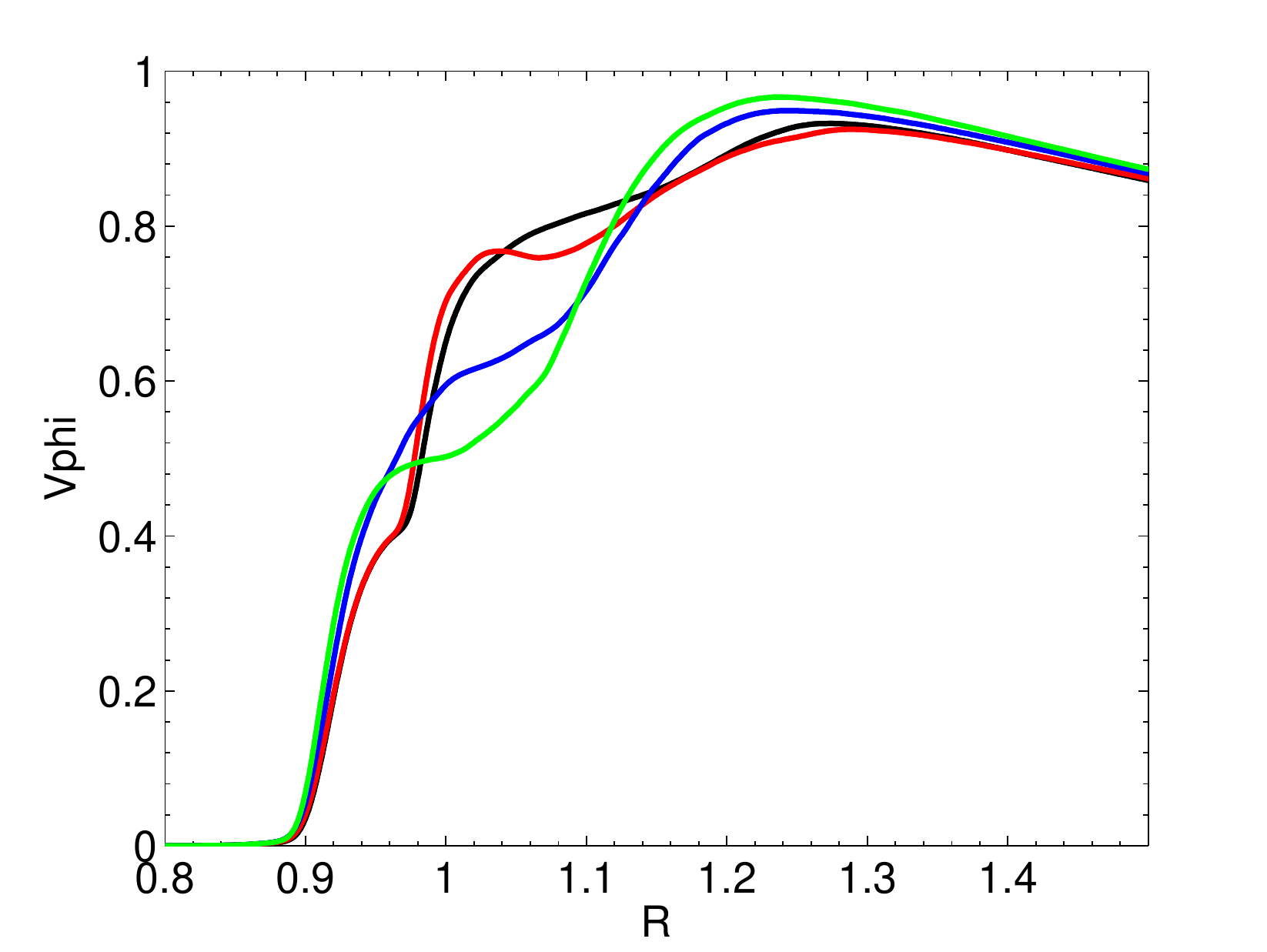}
\caption{Profiles of mean azimuthal velocity $\ol v_\phi$ at $t = 800$ 
(black), $t=900$ (red), $t=1000$ (blue), and $t=1100$ (green) for 
run A2.}
\label{vphitime}
\end{figure}

We can understand the reason for transitioning from the low accretion
state to the high accretion state by considering $\ol{v_\phi}$ as a
function of time. Fig. \ref{vphitime} shows $\ol{v_\phi}$ at
$t=800$, 900, 1000, and 1100. These times span the
duration of the third high accretion state in
Fig. \ref{M6spacetime}. We see that
between $t=800$ and $t=900$, a bump develops in the azimuthal velocity
profile. This leads to a rapid rearrangement of the velocity profile
between $t=800$ and $t=900$, which erases the bump, and by $t=1100$
the rearrangement of the velocity profile is complete.

\begin{figure}[!h]
\centering
  \includegraphics[width=0.7\textwidth]{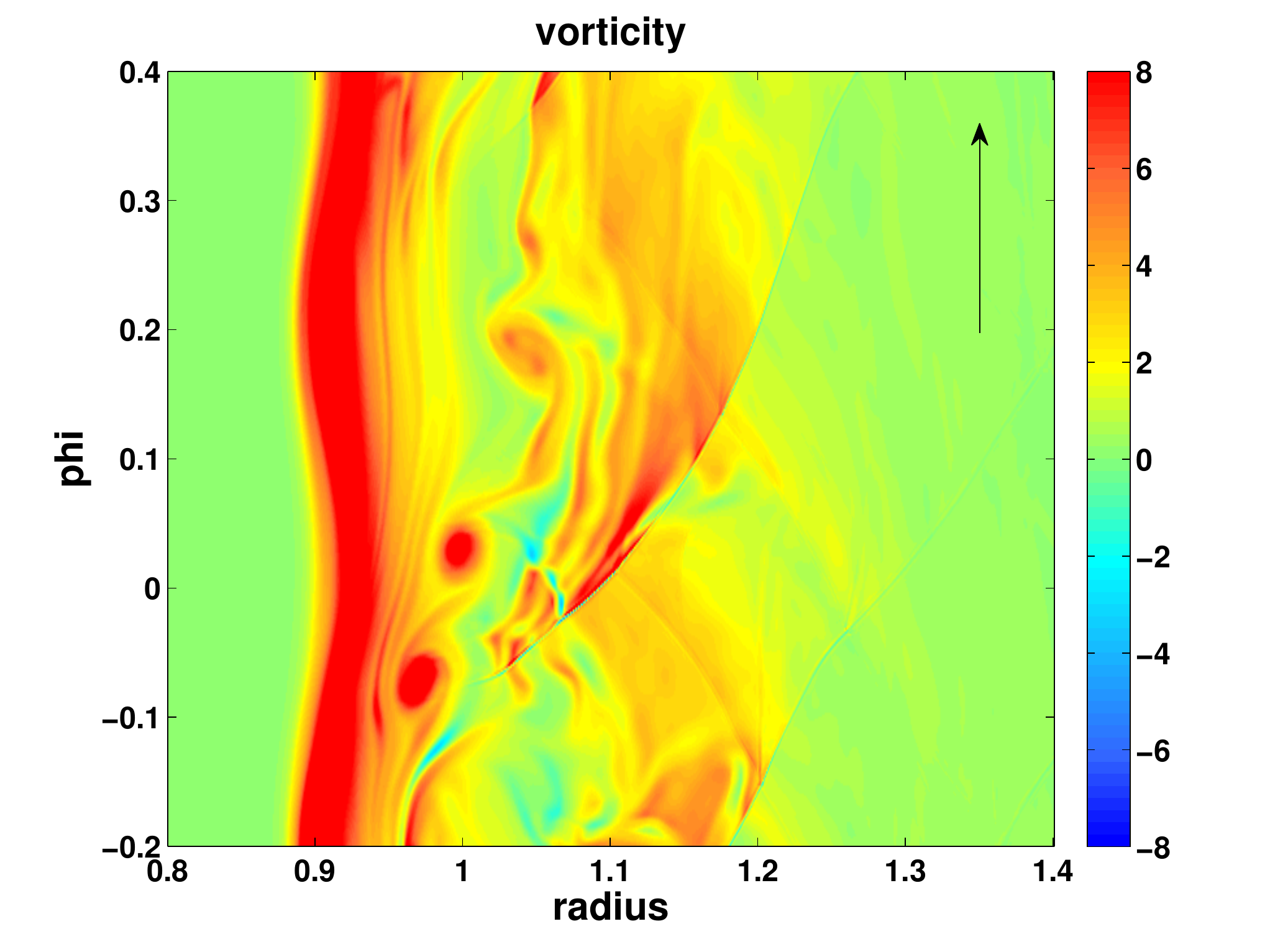}
\caption{Image of the vorticity, $\bfnabla \times \bfv$, in high 
accretion state at $t=1000$
  for simulation A2 ($M=6$). The cat's eyes are the red oval structures
  with wispy arms around $\cp = 1$ and are indicative
  of KH instability.}
\label{KHoutburst}
\end{figure}

The cause of the bump that appears in the $\ol{v_\phi}$ profile are
likely to be the oblique shocks present at the top of the BL. Each time a
fluid element at the top of the BL passes through a shock, it is
slowed down, and the collective effect of these shock passages creates
the bump. Once the bump has been created, the mechanism
for erasing it is the KH instability. KH instabilities can operate
near inflection points in the velocity profile. At
an inflection point, the radial derivative of the
vorticity is constant, making it permissible to
mix neighboring fluid elements and extract energy from the shear in
the flow. In the vicinity of the bump, the vorticity profile is very
flat, which allows the KH instability to operate
there. Fig. \ref{KHoutburst} shows a snapshot of the vorticity ($\bfnabla
\times \bfv$) in the
vicinity of the BL at $t=1000$, during the height of the
high-accretion state. One can see that around $\cp \sim 1$, the vorticity is
concentrated into
structures that resemble the cats eye pattern associated with 
classical KH instabilities. 

A curious point is that after a certain amount of time the system
stops switching between high and low states and stays in
the low state indefinitely. For instance, in
Figure \ref{M6spacetime} it is clear that after four high-accretion
states during which the
system is in the high accretion state it stays in the low accretion
state, until the end of the simulation. The cause of this behavior is
unclear, but one should keep in mind that the inner part of the disk
becomes more and more depleted of mass throughout the course of the
simulation, see \S \ref{sect:mass_accretion}. If the material in 
the inner disk were replenished, as it would be in a real disk, the 
high accretion state might not shut off and continue to operate 
periodically.

It may be tempting to associate the high and low accretion states with
the outburst and quiescent states of a CV. However, such a comparison
is superficial given the fact that we are using an isothermal
equation of state and have ignored magnetic fields. Nevertheless, the
modification of the rotation profile by shocks and the subsequent
onset of the KH instability is a very interesting phenomenon in its
own right, even if it bears no relation to the mechanism causing
outbursts in CVs.


\subsection{Mass Accretion Due to Shocks}
\label{sect:mass_accretion}

\begin{figure}[!h]
\centering
  \subfigure[]{\includegraphics[width=0.4\textwidth]{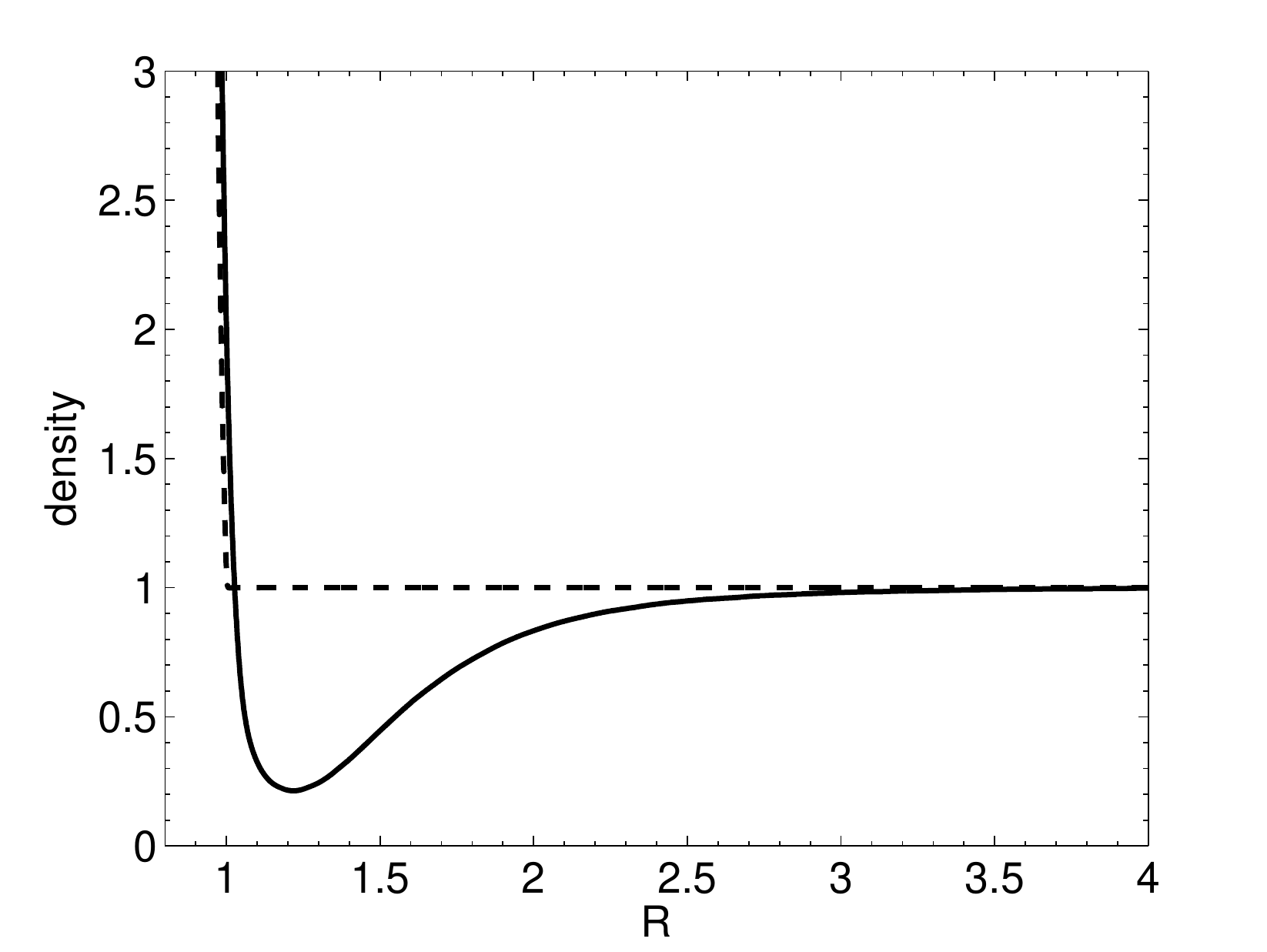}}
  \subfigure[]{\includegraphics[width=0.4\textwidth]{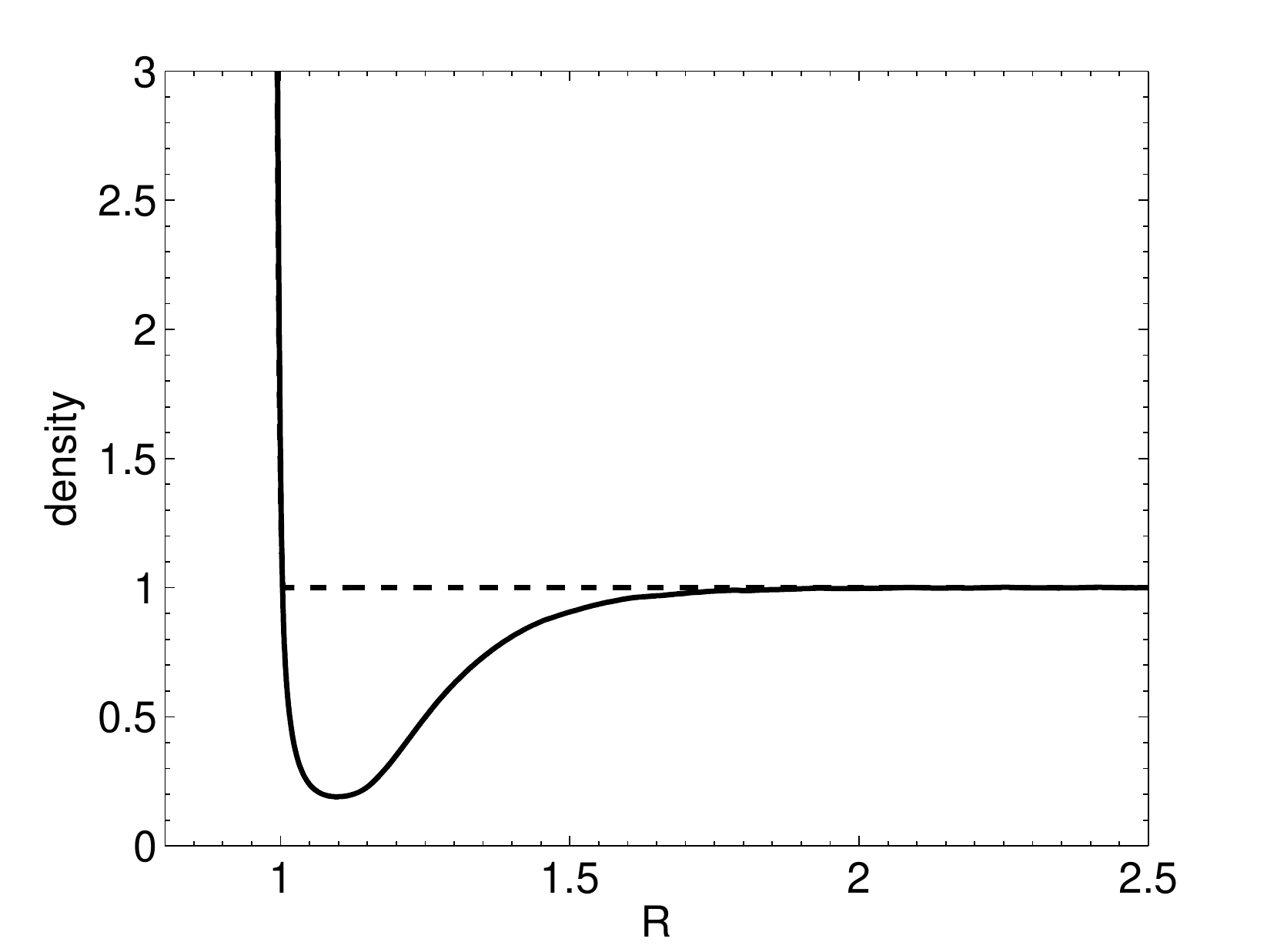}}
\caption{(a) Azimuthally-averaged density as a function of $\cp$ 
at $t=0$ (dashed line) and $t=2000$ (solid line) for run A5 ($M=6$). 
Evacuation of the part of the disk interior to the inner Lindblad 
radius (at $\cp=2.03$) is clearly visible, with mass ending up 
on the star. (b) Same as (a) but for run C1 ($M=12$).}
\label{densityfig}
\end{figure}

We confirm that even though we have purely hydrodynamical simulations,
there is mass accretion onto the star. Figure \ref{densityfig} shows the
density at $t=0$ and at $t=2000$ for runs A5 and C1. From the figure,
it is clear that the inner part of the disk has been evacuated at
$t=2000$ with material having been accreted onto the
star. In Figure \ref{densityfig}a, the innermost Lindblad radius is
at $\cp = 2.03$. One can see that the depletion of mass occurs
primarily inside of this radius, which is the region in which the
trapped modes are present. There is some depletion even beyond this
radius which is possibly due to the generation of a pressure gradient as
mass in the inner region accretes. The gain in the
radius of the star is more prominent for run A5 as opposed to run C1,
for two reasons. First, the Mach number is two times lower in run A5
as compared
to run C1 (6 vs 12), which means the radial scale height is four times
smaller and the
compression of accreted material larger in run C1. Second, more
material has accreted at $t=2000$ for run A5.  

Most of the mass accretion in the simulations occurs during the high
state, which is also when the Reynolds stress is
largest. However, during the low accretion state, it is possible to
``predict'' the mass accretion rate in the disk based on the value of $\delta
\Sigma/ \langle\Sigma\rangle$, and compare it to the ``observed'' accretion rate:
\ba
\label{massaccretobs}
\dot{M} = -2 \pi \cp \langle\Sigma\rangle \ol{v_\cp},
\ea
where the overbar represents an azimuthal density-weighted average.

To calculate the predicted mass accretion rate, we note that during the low accretion
state there is a well-defined structure of shocks that rotates at a
constant pattern speed. Given $\delta \Sigma/ \langle\Sigma\rangle$, one
can use the fact that these shocks carry angular
momentum, and because of dissipation, the angular
momentum in the shock is transferred to the bulk flow driving 
accretion. In Appendix \ref{Mdotapp}, we derive an analytical 
expression for the mass accretion rate $\dot M$ resulting from 
dissipation of weak shocks, given by equation (\ref{Mdotpred}). 
We find that the mass accretion rates measured in our simulations 
using equation (\ref{massaccretobs}) agree to order of magnitude with
the predicted mass accretion rates using equation (\ref{Mdotpred}). For
simulation A2, the mass accretion rate is $\dot{M} \sim 10^{-3}$
(in our dimensionless units), during the
low state around $t \sim 300$. This
explains why the disk can persist for hundreds of orbital periods in
the low state without being depleted. We note that during the high
accretion state,
the accretion rate can be much higher reaching values as high as $\dot{M} \sim
10^{-1} - 10^{-2}$ in the boundary layer, where shear instabilities operate.

We also point out that accretion due to numerical viscosity is 
negligible in the simulations. For instance, in simulation A2, 
numerical viscosity gives rise to a numerical accretion rate of
$\dot{M} \sim 10^{-6}$. This
estimate was made by running a simulation with the same parameters as
simulation A2, but without
perturbations. In the absence of perturbations, the sonic instability
does not set in, and we
maintain an azimuthally-symmetric flow for the duration of the
simulation, allowing us to measure the numerical mass accretion rate
directly. We point out, however, that such an azimuthally-symmetric flow is
grid-aligned in cylindrical coordinates. Thus, our
measurement of the numerical mass accretion rate likely underestimates
its true value in the science runs, where radial motions are
present. Nevertheless, the fact that the accretion rate due to
numerical viscosity in the grid-aligned case is three orders of
magnitude lower than the measured accretion rate in the low-accretion state
of simulation A2 provides convincing evidence that numerical
viscosity is negligible in the simulations.


\subsection{Stability of the Pattern Speed}
\label{pattern_sec}

\begin{figure}[!h]
\centering
\includegraphics[width=0.85\textwidth]{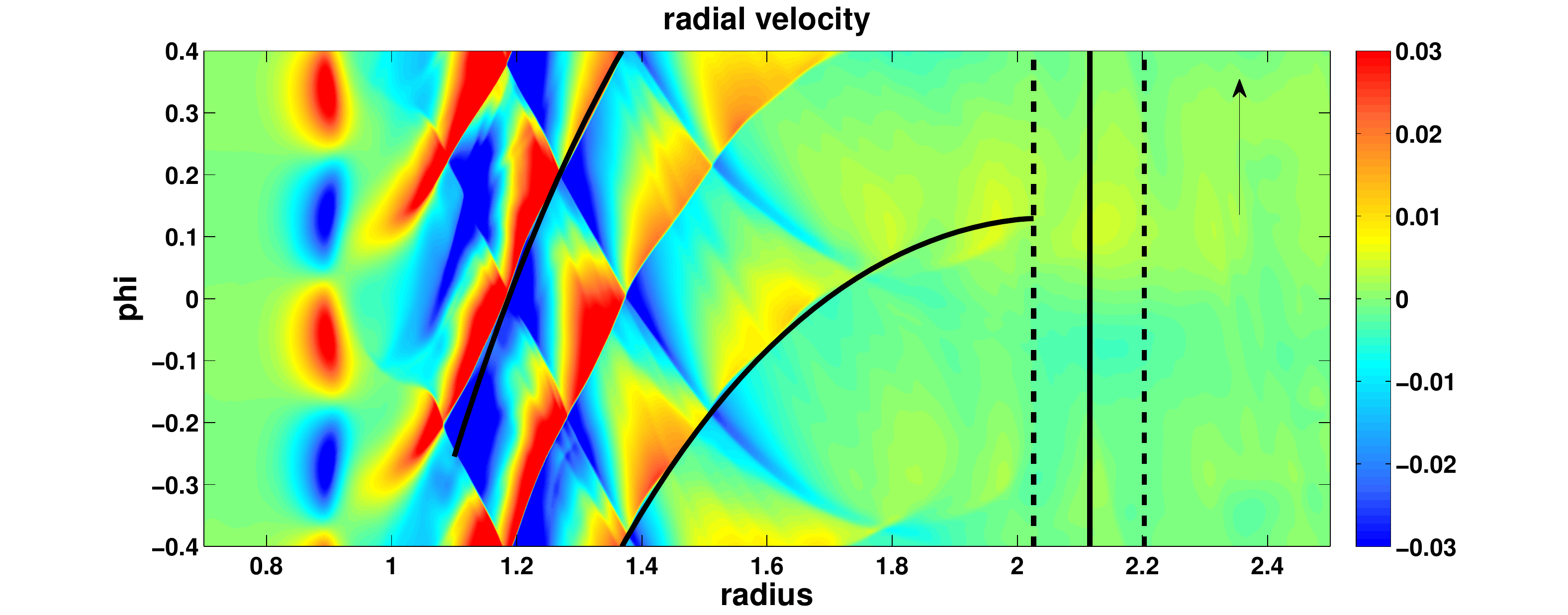}
\caption{Density plot of $v_\cp$ at $t=1284$ for simulation A2 ($M=6$). The
  black curve shows the analytic solution for the shock front from the
  dispersion relation (\ref{keq}) using $\Omega_P = 0.325$. The dashed
  vertical lines show the
  locations of the inner and outer Lindblad resonances, and the solid
  vertical line shows the location of the corotation radius.}
\label{shockfig1000}
\end{figure}

Given that the system can transition between high and low states, one
might wonder about the long-term stability of the modes
discussed in \S \ref{mode_sec}. We find that the pattern speed in
a given simulation stays constant to within a few percent over the course of
hundreds of orbital periods. This is true even if
the time interval in question
contains transitions between high and low states. Figure
\ref{shockfig1000} shows an image of the radial velocity for
simulation A2 at time $t=1284$. This image can be compared with
Figure \ref{shockfig}b which is from the same simulation but
was taken at $t=284$. The pattern speed
at $t=1284$ is approximately $\Omega_P = 0.325$, which is close
to the pattern speed at $t=284$ of $\Omega_P = 0.335$. Note that
the interval of time between $t=284$ and $t=1284$ contains two
transitions from high accretion state to low accretion state.

\subsection{Width of the BL}
\label{BLwidth_sec}  

\begin{figure}[!h]
\centering
  \includegraphics[width=0.7\textwidth]{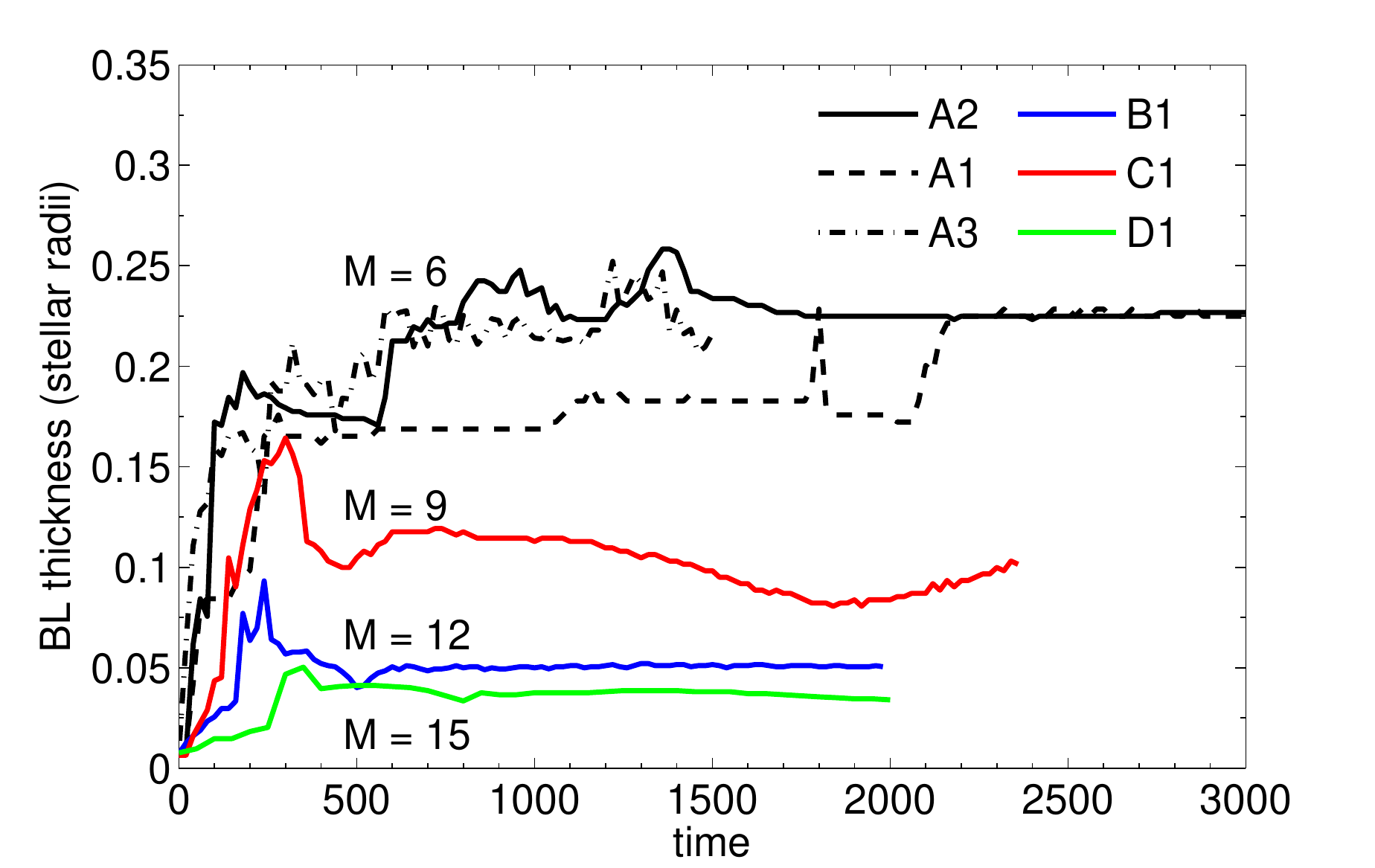}
\caption{Width of the BL as defined in equation
 (\ref{BLwidthtime}) as a function of time. Runs A1, A2, and A3 all have
  $M=6$, run B1 has $M=9$, run C1 has $M=12$, and run D1 has $M=15$.}
\label{BLwidthtime}
\end{figure}

We plot the thickness of the BL defined by equation (\ref{BLwidth}) as a
function of time in Figure \ref{BLwidthtime} for runs A1, A2, A3, B1, C1,
and D1. The thickness of the BL greatly increases in the
beginning of the simulation due to the sonic instabilities. It then
stays approximately constant during the low accretion states and
undergoes further
changes during the high accretion states, eventually settling down to an
approximately constant value. The fact that runs A1, A2, and A3
yield almost the same BL width at late times suggests that the
simulations are converged, since these three runs differ only in their
resolution. 

It follows from Figure \ref{vphitime} that after significant 
evolution has taken place --- typically after hundreds of orbital 
periods --- the region in which azimuthal velocity transitions from the 
Keplerian profile to $\ol v_\phi=0$ covers the radial interval $0.9\lesssim\cp
\lesssim 1.2$. Thus, it extends both into the star and into the disk.
This may have interesting implications for the morphology of the 
boundary layer in astrophysical systems. 

\begin{figure}[!h]
\centering
  \includegraphics[width=0.7\textwidth]{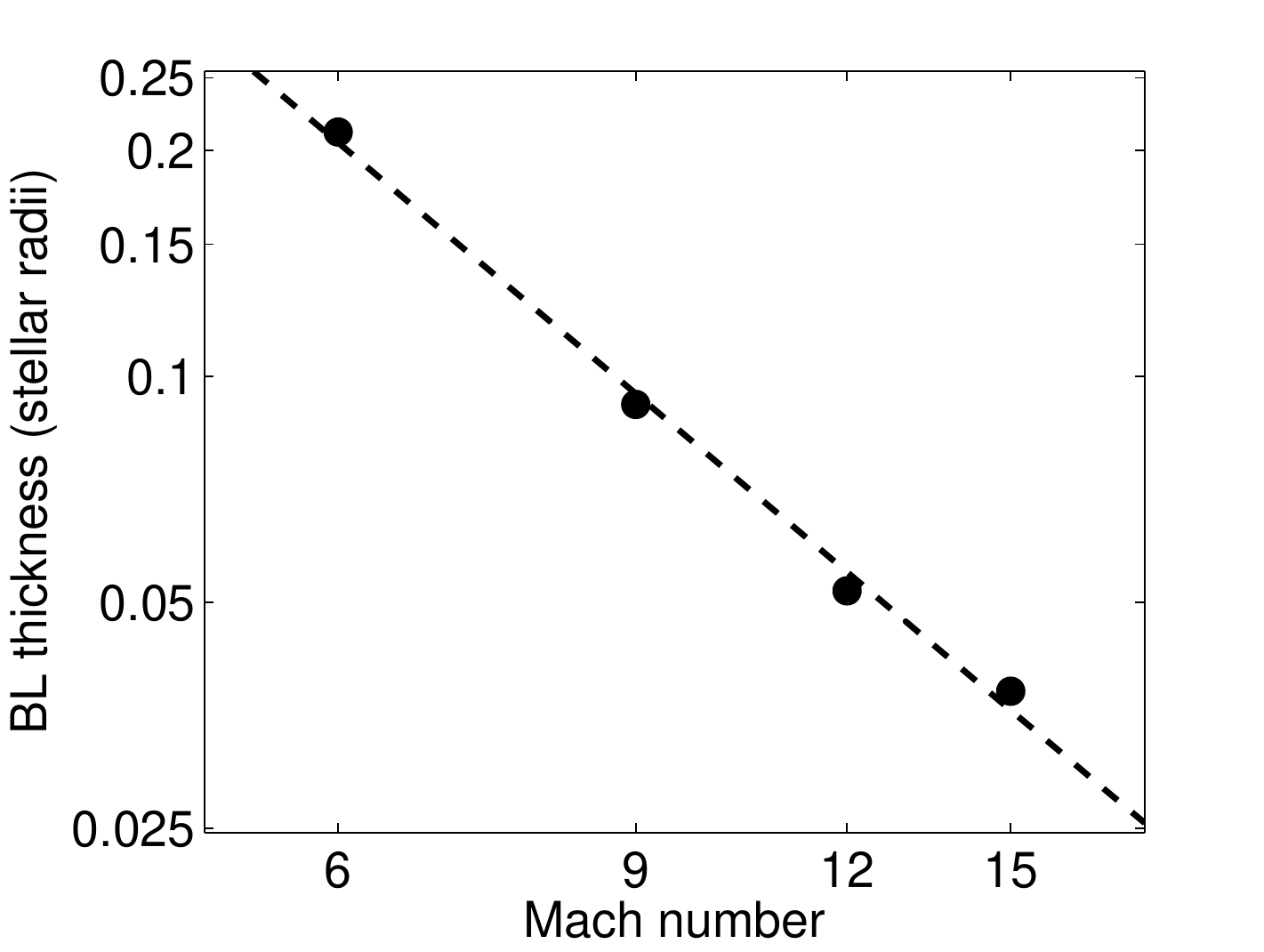}
\caption{Boundary layer width at $t=1500$ as a function of the
  Mach number in our simulations. The dashed line 
indicates the best fit line through the
  points and has a slope of $-1.9$. This is close -2, which is the
  expected value of the power law index
  for the scaling of the boundary layer width with the Mach number.}
\label{widthmach}
\end{figure}

It is clear from Figure \ref{BLwidthtime} that the boundary layer
becomes thinner with increasing Mach number. In Figure \ref{widthmach},
we plot the width of the boundary layer at $t=1500$ as a function of the Mach
number on a log-log plot. The data point for $M=6$ is a simple average of
simulations A1, A2, A3 and the points for $M=9$, $M=12$, and $M=15$
are from simulations B1, C1, and D1 respectively. The dashed line shows
the best least squares fit to the data, and from its slope we deduce that
$h_{BL} \propto M^{-1.9}$, where $h_{BL}$ is the width of the BL at
$t=1500$. Since the scale height goes as
$h_s=M^{-2}$ (equation [\ref{scheight}]), this indicates that
the deceleration of the flow in the BL occurs over an approximately
fixed number of scale heights in our simulations ($\sim 7-8$), regardless
of the Mach number.

\section{Full Disk Simulations}
\label{fulldisk_sec}

Up until now, the simulations we have discussed have had a $\phi$
extent of $0.8$ radians. In this section, we consider global disk
simulations, where the azimuthal angle runs from 0 to 2$\pi$. This is
important, since it is possible that the azimuthal wavenumber of the dominant
global mode is of order unity. Lower azimuthal wavenumbers are
inherently more interesting for explaining
physical phenomena having frequencies less than or of order the
orbital period at the surface of the star, such as DNOs. This is because the
frequency of the mode is given by $\omega = m \Omega_P$, and only if
$m$ is low enough will $\omega$ also be low enough to provide a
plausible explanation for such
phenomena. Thus, we determine the azimuthal wavenumbers and
pattern speeds of the dominant modes obtained from simulations in which
$\phi$ goes from $0$ to $2 \pi$. 

\begin{figure}[!h]
\centering
\subfigure[]{\includegraphics[width=0.49\textwidth]{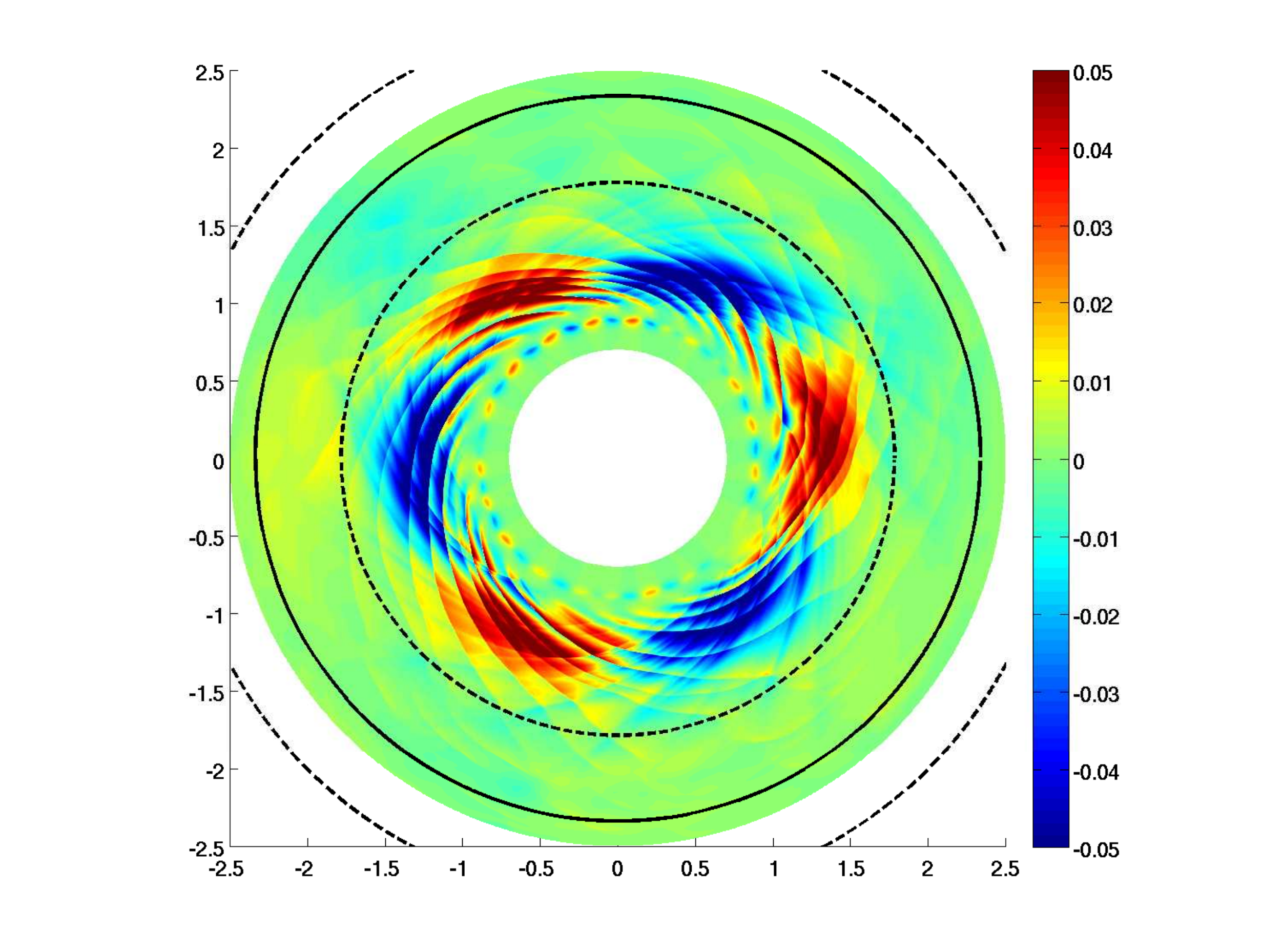}}
\subfigure[]{\includegraphics[width=0.49\textwidth]{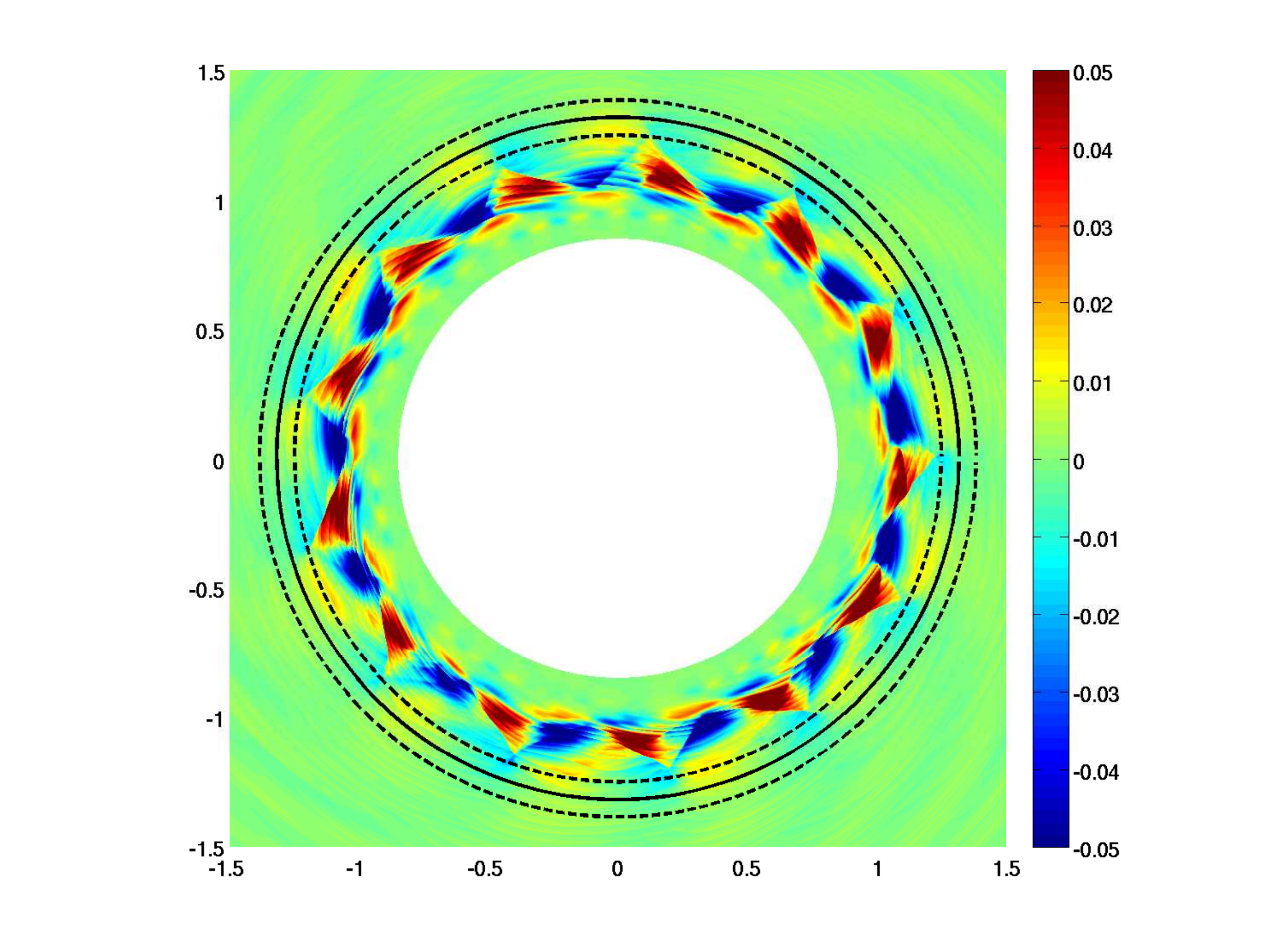}}
\subfigure[]{\includegraphics[width=0.5\textwidth]{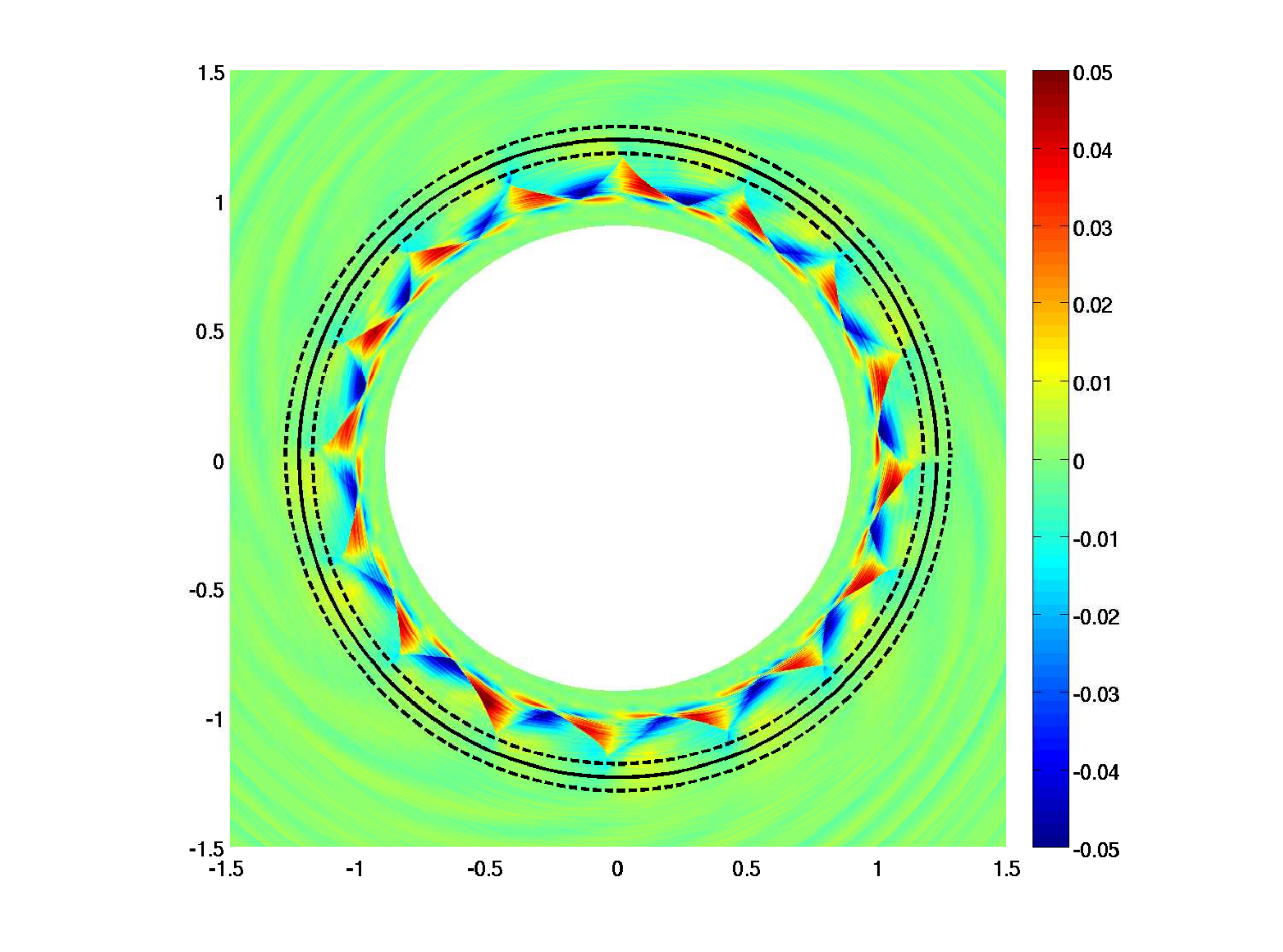}}
\subfigure[]{\includegraphics[width=0.44\textwidth]{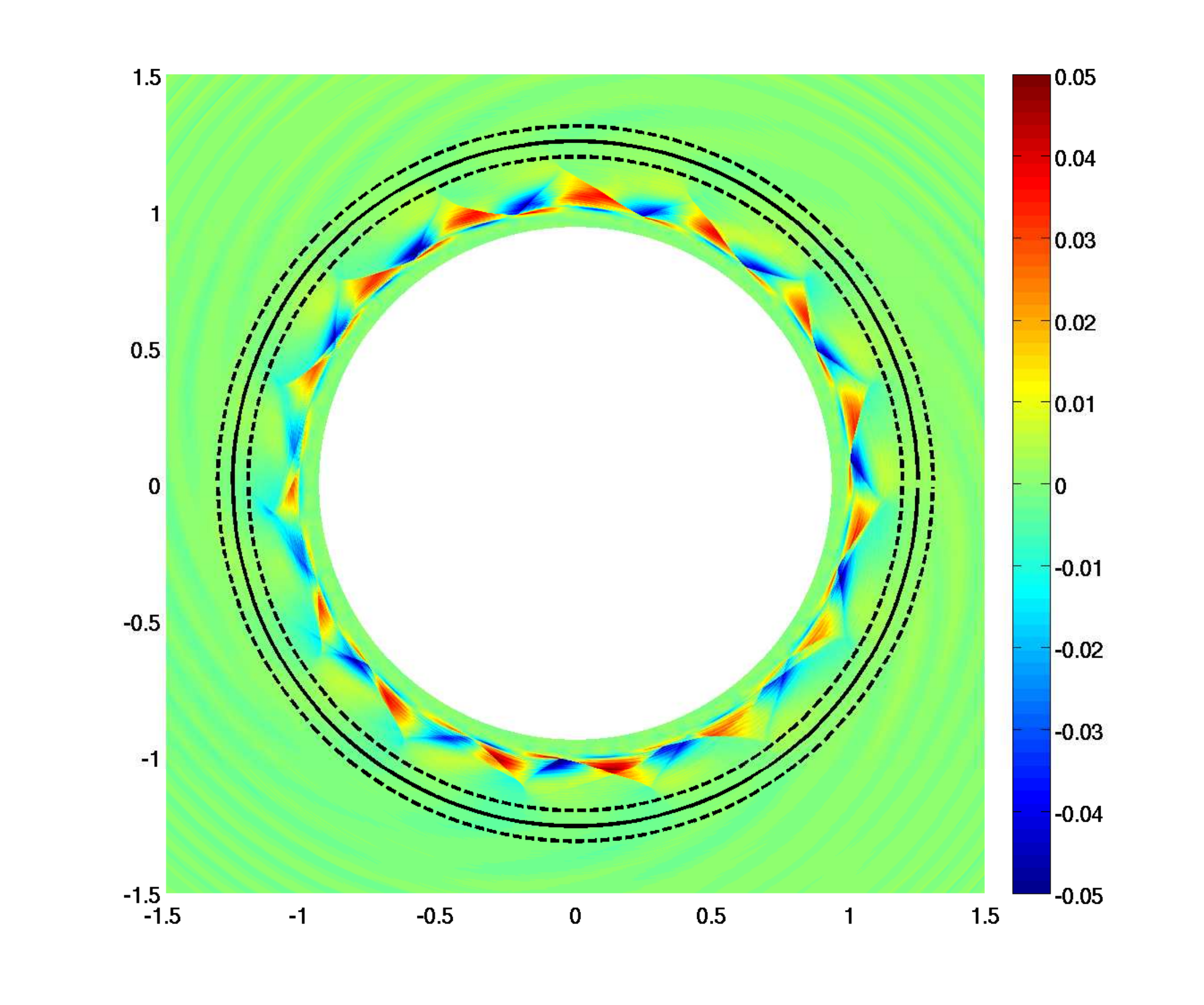}}
\caption{Images of the radial velocity for simulations A4 (panel a),
  B2 (panel b), C2 (panel c), and D2 (panel d), which have $M=6,9,12,$
  and $15$ respectively. The view has been
  zoomed in, in panels (b), (c), and (d) to give a better view of the modal
  structure in the vicinity of the BL. The $m$-numbers of the dominant
  global modes for panels (a)-(d) are $m=3$, 13, 16, and 15 respectively. 
Solid and dashed circles show the locations of the corotation and the 
inner and outer Lindblad resonances in the disk.}
\label{M6fulldiskfig}
\end{figure}

Panels (a)-(d) of Figure \ref{M6fulldiskfig} show images of the radial
velocity from simulations A4, B2, C2, and D2, which have Mach numbers
of $M=6,9,12,$ and 15 respectively. These snapshots reveal a rather 
interesting structure of fluid perturbations, which is best seen at 
low values of $M$. In particular, in panel (a), one can clearly 
discern a large-scale $m=3$ global mode along with some high frequency 
features, which are due to vortices at the base of the BL. There are more
than 20 individual vortices that reside in the BL (where $\Omega$ goes down
from its Keplerian value in the disk to 0 in the star). Each of
the vortices in the BL launches a shock into the disk, as discussed in
\S \ref{mode_sec}, but now the individual shocks are superimposed on
top of a global $m=3$ mode. 

Thus, close inspection reveals two types of periodicities in our 
simulations --- a large scale mode with relatively low wavenumber
$m$ that extends out from the BL into the disk, and a set of periodic
vortices contained in the BL with higher\footnote{Note that in 
simulations presented in \S \ref{results_sec}, which covered only 
a limited interval in $\phi$ it is generically found that $m_v=m$ 
(and typically equal to 2) since the azimuthal
extent of the domain was setting the wavenumber of the mode.}  
wavenumber $m_v$ 
($m_v\gtrsim 20$ in Figure \ref{M6fulldiskfig}a). The vortices at 
the base of the BL are only visible in panels (a) and to some extent 
$b$ of Figure \ref{M6fulldiskfig}. However, they are present as well 
in panels $c$ and $d$ and become apparent if the color scheme is 
saturated. 
The origin of the global mode is traced to a geometric 
resonance condition in \S \ref{sect:global}, but we leave the 
detailed investigation of the BL vortices to future study.

The pattern speed of the
$m=3$ global mode in panel a of Figure \ref{M6fulldiskfig} is
$\Omega_P = 0.28$. This is comparable to the
pattern speed of the shock structure for simulation A2 discussed in \S
\ref{mode_sec}, which has the same Mach number as simulation A4. The
frequency of the $m=3$ mode is
$m\Omega_P = 0.84$, and thus is lower but comparable to the orbital
frequency at the inner radius of the disk.

\subsection{Relation of Pattern Speed to Azimuthal Wavenumber}
\label{sect:global}

\begin{figure}[!h]
\centering
\subfigure[]{\includegraphics[width=0.48\textwidth]{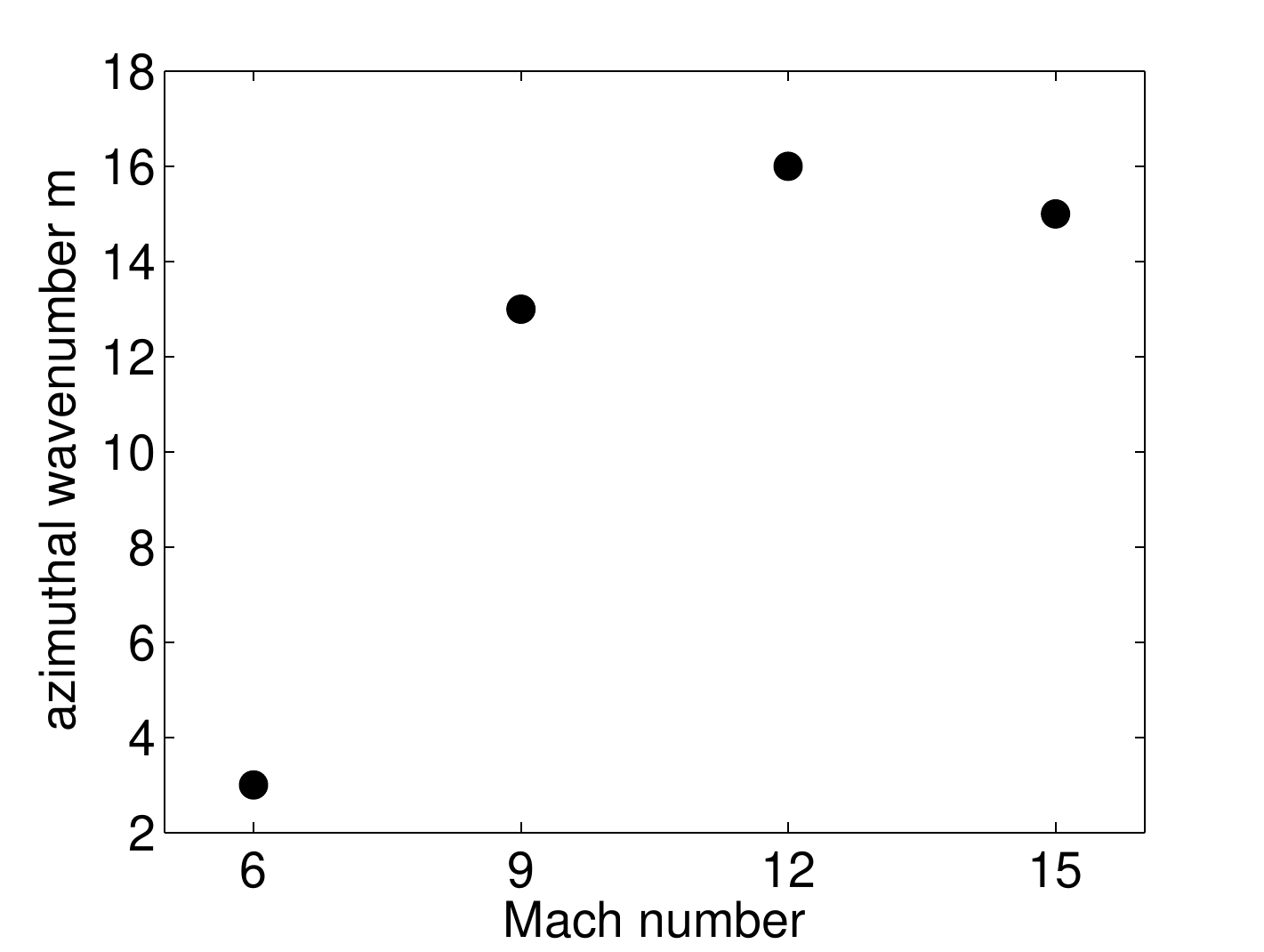}}
\subfigure[]{\includegraphics[width=0.51\textwidth]{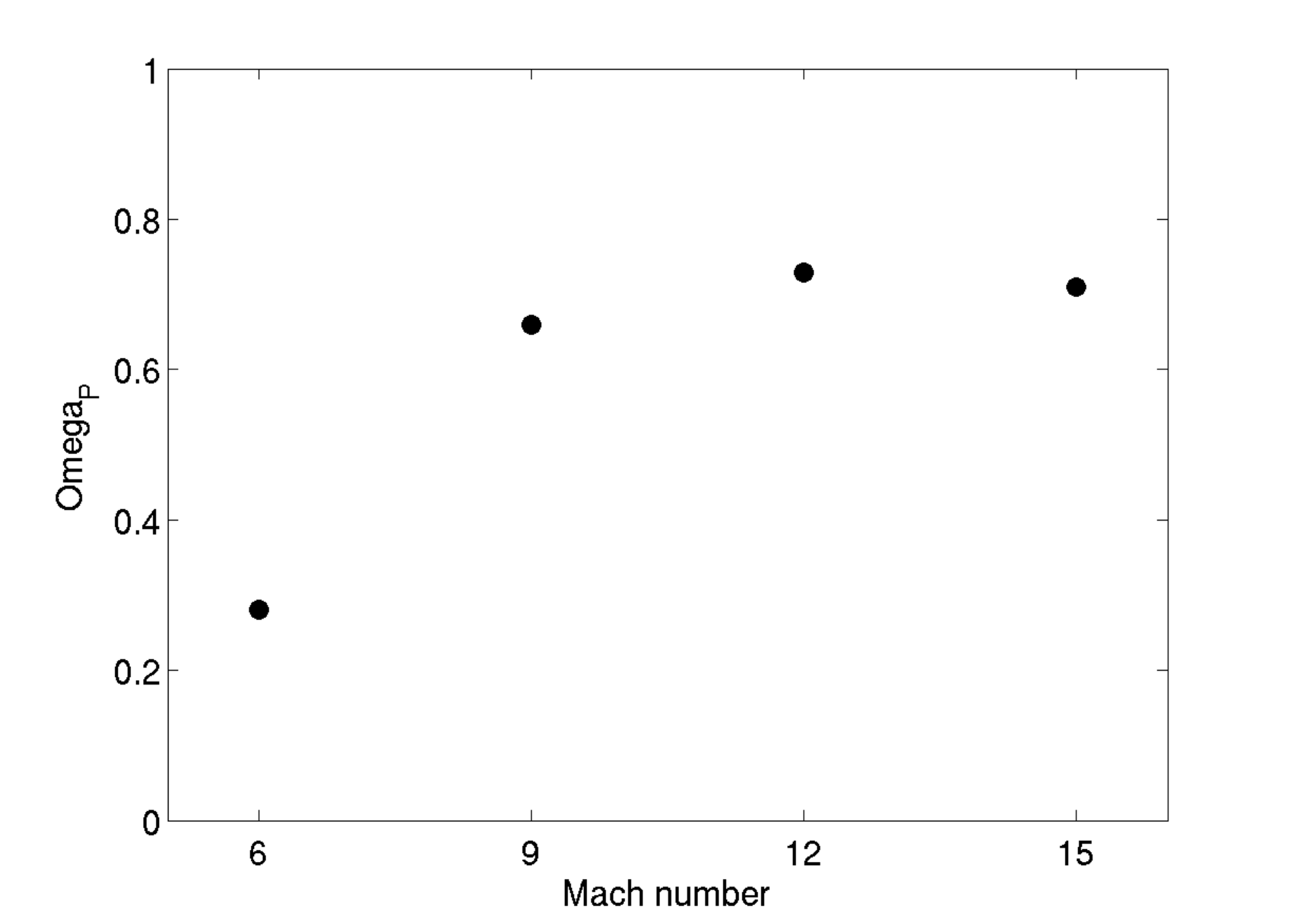}}
\caption{(a) Plot of the azimuthal wavenumber of the global mode as a
  function of Mach number. (b) Plot of the pattern speed as a function
  of Mach number for the global mode.
}
\label{mandpatternfig}
\end{figure}

As the Mach number is increased, there is a general trend for the
azimuthal wavenumber of the large scale pattern, $m$, to increase 
as well. Figure \ref{mandpatternfig}a shows
the wavenumber of the dominant global mode as a function of Mach
number. We also plot the pattern speed as a function of Mach number in
Figure \ref{mandpatternfig}b. There is again a general trend for the
pattern speed to increase with increasing Mach number.

The dependence of the pattern speed on the azimuthal wavenumber of the
global mode can be
understood if we consider the total azimuthal angle traversed by a shock
in traveling from the boundary layer to the inner Lindblad resonance
in the disk and back. This angle is given by
\ba
\label{dphieq}
\Delta \phi = \frac{2}{s} \int_{\cp_{BL}}^{\cp_{ILR}} d\cp
\sqrt{(\Omega(\cp)-\Omega_P)^2 - \left( \frac{\kappa(\cp)}{m} \right)^2},
\ea
and can be derived by using equation (\ref{keq}) together with the
relation
\ba
\frac{d \phi}{d \cp} = \frac{k(\cp)}{m},
\ea
which describes the trajectory along which the phase in equation 
(\ref{eq:WKB}) is constant. 
Here, $\cp_{BL}$ and $\cp_{ILR}$ denote the boundary layer radius and
the inner Lindblad radius in the disk, respectively. The factor of 
two out front comes from the fact that we are considering
the total change in angle for a shock that starts from the boundary
layer, is reflected at the Lindblad radius, and comes back to
the boundary layer.

Formally we need to set $\cp_{BL}$ equal to the location of the 
Lindblad resonance inside the BL. In practice, though, the shocks 
in our simulations are always observed to originate very near 
$\cp = 1$, and thus we simply set $\cp_{BL} = 1$ for the purposes 
of our calculations below.

An interesting question to ask is, given that we observe a mode with
azimuthal wavenumber $m$, what pattern speed would we predict based on
equation (\ref{dphieq}), and how well does this predicted pattern speed
match the observed pattern speed in our simulations? Assuming that the
change in azimuthal angle for one ``out and back'' trip for the shock
is given by equation (\ref{dphieq}), the condition that the shock
pattern be closed is given by
\ba
\Delta \phi = \frac{2 \pi p}{q},
\ea
where $p$ and $q$ are positive integers. It is natural to assume that $p =
1$, especially since the global modes don't appear to overlap in
Fig. \ref{M6fulldiskfig} as they would if $p > 1$. For $q$, an obvious
choice to make is $q=m$, since this naturally gives an $m$-fold
pattern. Thus, we postulate that
\ba
\label{deltaphival}
\Delta \phi = \frac{2 \pi}{m},
\ea 
where $m$ is the azimuthal wavenumber of the global mode. 

For simplicity, we assume a Keplerian rotation profile, in which case
our estimate for $\Delta \phi$ becomes
\ba
\label{phiformula}
\Delta \phi = \frac{2}{s} \int_{1}^{\cp_{ILR}} d\cp
\sqrt{(\cp^{-3/2} - \Omega_P)^2 - \left( \frac{\cp^{-3/2}}{m} \right)^2},
\ea
where $\cp_{ILR}$ is given by equation (\ref{LReq}).
Assuming a Keplerian rotation profile
gives a value of $\Delta \phi$ which is slightly larger than observed
in the simulations. For instance, see Fig. \ref{shockfig}b and note in
particular that around $\cp \sim 1$, the shock profile becomes less
steep than the analytic prediction using a Keplerian profile (black
curve), due to the fact that the angular velocity is sub-Keplerian in
that region. However, using the values of $\Omega(\cp)$ and
$\kappa(\cp)$ is complicated by the fact that the WKB approximation
breaks down in the boundary layer, (see \S \ref{mode_sec} for a more detailed
discussion).

\begin{figure}[!h]
\centering
\subfigure[]{\includegraphics[width=0.49\textwidth]{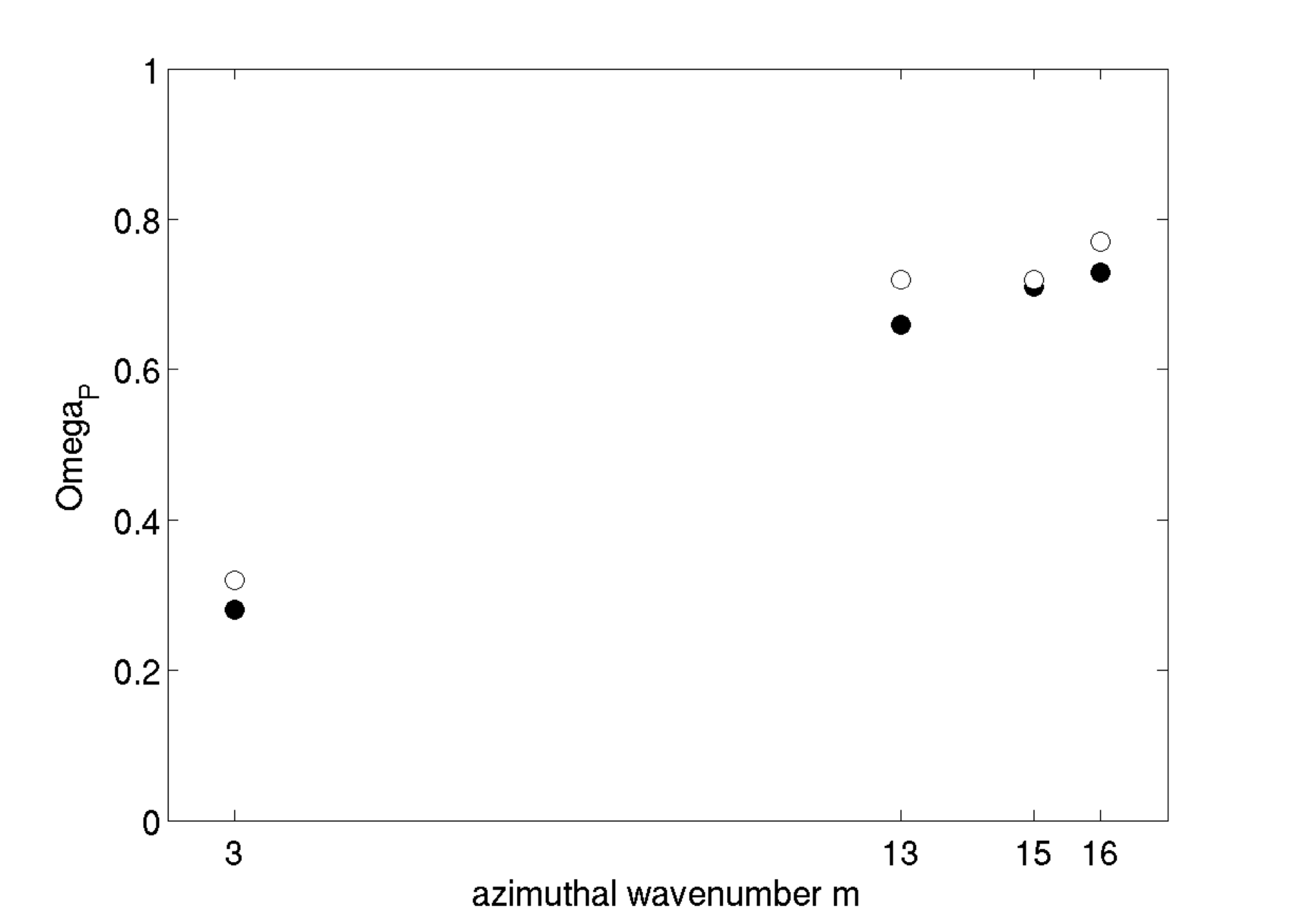}}
\caption{Plot of the pattern speed of the global mode as a
  function of the azimuthal wavenumber $m$. The closed circles show
  the pattern speed measured from the simulations and the open circles
  show the predicted pattern speed (see text after equation 
(\ref{phiformula})).}
\label{mvspattern}
\end{figure}

The open circles in Fig. \ref{mvspattern} show the value of
$\Omega_P$ that
gives the correct value of $\Delta \phi$ (equation \ref{deltaphival})
using the formula (\ref{phiformula}) for the $m$-numbers observed
in the simulations. Conversely, the solid circles in
Fig. \ref{mvspattern} show the values of $\Omega_P$ as measured from
the simulations. We find that the
predicted value of $\Omega_P$ is within $\sim 10 \%$ of the value
observed in simulations and is always larger than the observed
value. The latter point can be explained by the fact that using a
Keplerian profile results in a larger value of $\Delta \phi$ as 
explained above. A smaller value of $\Delta \phi$ would lower 
the predicted value of $\Omega_P$, bringing the analytical predictions 
into closer agreement with the simulation results.

The good agreement between the predicted and observed pattern speeds
over a range of Mach numbers
suggests that there is a direct relation between the $m$-number of the
global modes in the disk and the pattern speed. This relation is
essentially the combination of a periodicity condition and the
dispersion relation for the modes.

\section{Discussion}

The primary result of this work is the identification of 
a novel mechanism of angular momentum and mass transport 
in the boundary layer relying on non-axisymmetric acoustic modes which
are generically excited in highly supersonic shear 
layers. Dissipation of these modes in weak shocks drives 
the evolution of the system and ultimately results in the 
mass transfer from the disk into the star. Our work is 
exploratory in nature and necessarily makes a number of 
simplifying assumptions, many of which are discussed in \S 
\ref{sect:extensions}. Nevertheless we expect the main 
qualitative features of the new transport mechanism to 
persist also in more realistic settings typical for 
astrophysical boundary layers. The basic reason for this 
expectation lies in the fact that essentially the only
ingredient necessary  for the generation of acoustic 
modes is the presence of supersonic shear which is
natural in astrophysical BLs. 

Previously a number of hydrodynamical and MHD instabilities 
--- Kelvin-Helmholtz, baroclinic, Tayler-Spruit dynamo 
\citep{KT78,Fujimoto,Tayler,Spruit,PiroBildsten,InogamovSunyaev10} 
have been invoked to drive the angular momentum transport in 
the BL. However, none of these instabilities were actually 
shown to operate under the conditions typical for the BL, where
the supersonic nature of the flow is crucial for the dynamics
and can drastically change the way in which these classical 
instabilities (usually explored in incompressible limit) operate. 
\citet{Armitage} and \citet{SteinackerPapaloizou} have run 
direct 3D MHD simulations of BLs at a resolution that although state
of the art at the time is low compared to the current
simulations - their typical
resolution in the radial direction is about an order of magnitude 
lower than in our work. These authors found that magnetic field 
amplification and mass accretion took place in the BL.
However, whether the latter is due to the magnetic stresses
and not due to the hydrodynamical effects like the one discussed
here is not obvious. Transport induced by acoustic modes
does not require magnetic field to be present in the first
place, which makes this mechanism quite universal.

As part of our work we also looked at the linear stage
of the sonic instability (see \S \ref{sonic_sec}), which was 
previously explored under a more limited set of assumptions 
by \citet{BR}. In particular, \citet{BR} neglected the Coriolis 
force in their calculations and worked in Cartesian geometry. 
Despite that, their main results are in agreement with our more 
general calculations run in cylindrical geometry which fully
account for the effects of rotation. This concordance 
justifies the simplifications made by \citet{BR} in studying the sonic 
instability.


\subsection{Variability in the BL}

Because of the periodic nature of the trapped modes and their
stability over many orbital periods, one may wonder whether they can
produce a periodic modulation of the disk-star luminosity. One natural
application of such a modulation is to the 
variability observed during the dwarf novae outbursts. The leading
explanation for this variability involves
magnetically-channeled accretion, which spins
up an equatorial belt on the primary \citep{Paczynski, WarnerWoudt}. An
  alternate explanation, and one that is relevant to our results
  is in terms of modes excited on the surface of the
  WD. 

\citet{PapaloizouPringle} found that global non-axisymmetric
  modes excited in the surface layers would have the correct periods
  to account for DNOs. Later \citet{Popham} studied the possibility
  that DNOs could be caused by a bulge in the BL and
  \citet{PiroBildsten1} considered shallow water modes in a BL that
  had spread meridionally to high latitudes. However, one of the main
  impediments to such theories is identifying the physical mechanism
  that would excite such surface modes in the first place.

If the picture of sonic instabilities leading to trapped modes that we have
presented here remains valid even
when additional physics is included (e.g. 3D nature of the flow, 
realistic cooling, effects of magnetic fields), then it would 
provide a mechanism for exciting surface modes. Since the frequency of DNOs
is typically less than
the orbital frequency at the surface of the star, the most favorable
scenario for this mechanism to explain DNOs would involve a low $m$
number azimuthal mode $m
\lesssim 3$. This is because one might expect a high $m$-number mode to
produce a small signal due to integrating the light over a hemisphere
of the star. Also, the criterion that the DNO frequency be less than
the orbital frequency at the surface of the star means that $m$ cannot
be too large otherwise the criterion $\omega_{DNO} = m\Omega_P
\lesssim 1$ will not be fulfilled. 

The only one of our simulations to satisfy the constraint $m \Omega_P
\lesssim 1$ is simulation A4 ($M=6$) with the higher Mach number
simulations --- more typical for realistic astrophysical systems ---
producing values of $m \Omega_P$ that are too
high for DNOs. However, a detailed comparison to DNOs is premature at
this point given
the exploratory nature of the simulations - 2D with isothermal
equation of state.


\subsection{Extensions}
\label{sect:extensions}

We have performed 2D hydrodynamical simulations of the boundary layer
and have found excitation of non-axisymmetric surface modes by the
sonic instability. One may wonder how relevant our results are to real
3D disks with magnetic fields. We plan to address this question with
future simulations, but for now we comment on what we expect
may change in going from 2D hydro to 3D MHD simulations.

\begin{figure}[!h]
\centering
  \subfigure[]{\includegraphics[width=0.9\textwidth]{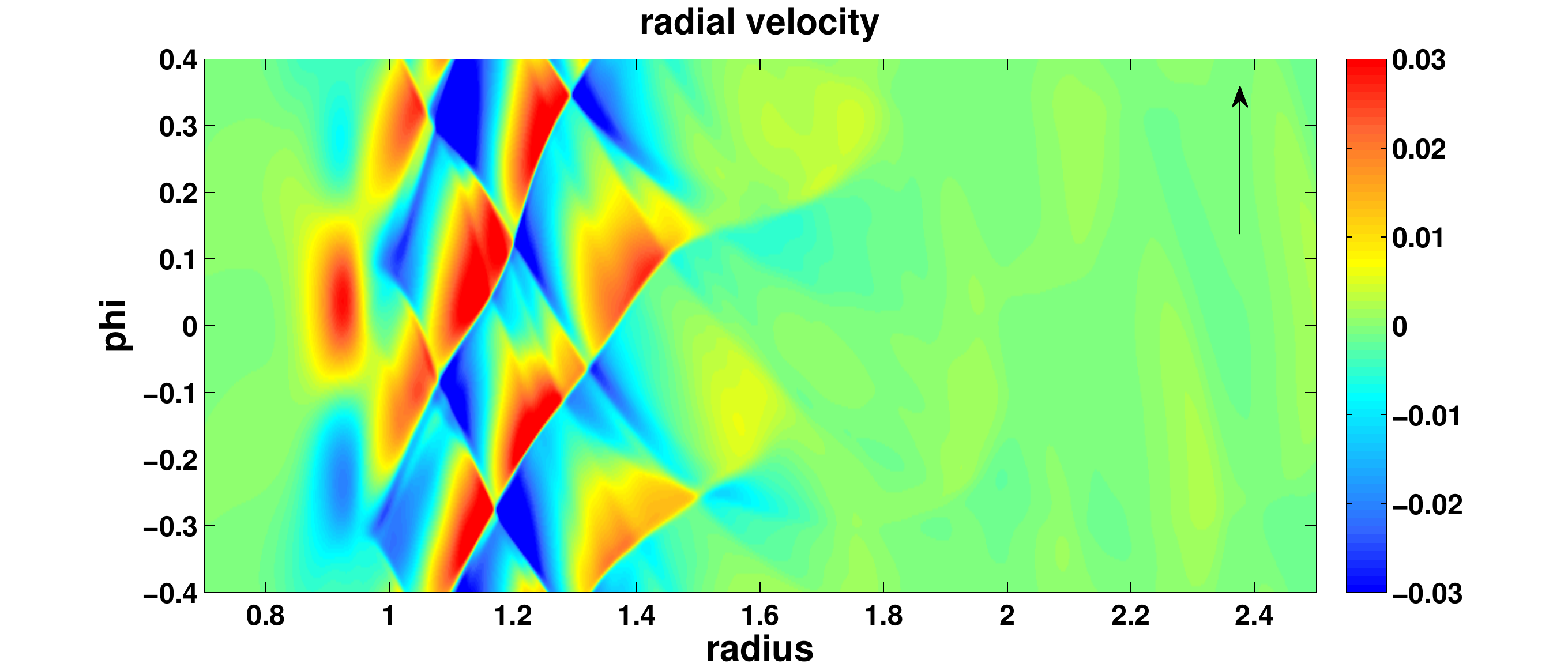}}
  \subfigure[]{\includegraphics[width=0.9\textwidth]{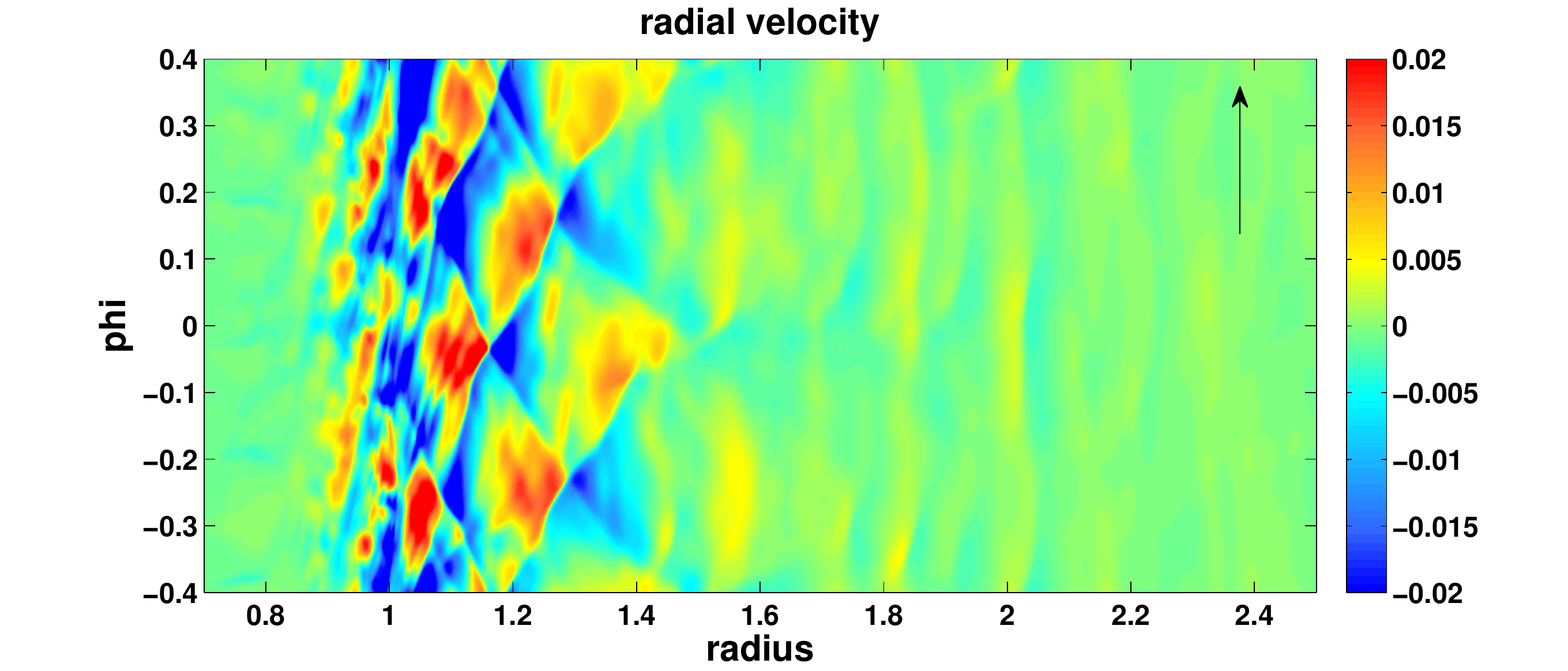}}
\caption{(a) Image of $v_\cp$ at $t=1000$ for simulation A2 ($M=6$). 
Vortices are
  clearly visible at the base of the BL. (b) Same as panel (a) but for the 3D
  simulation at $t=1000$. The base of the BL is turbulent and no clear vortices
  are present. Nevertheless, the shock structure is still
  present in the disk, although it is not as crisp.}
\label{3Dcompfig}
\end{figure}

We have carried out some preliminary calculations to 
demonstrate the fundamental difference between the 2D and 
3D cases and show the results in Figure \ref{3Dcompfig}. 
Panel (a) of Figure \ref{3Dcompfig} corresponds
to simulation A1 at time $t=1000$. Panel (b) is for a 3D simulation 
in cylindrical geometry with identical parameters except: 
the simulation domain has a 
vertical extent of $-0.167 < z <
0.167$ (i.e. $2 h_d$, where $h_d = s/\Omega$ is the disk scale
height), and there are 64 cells in the $z$-direction.
We use a cylindrical potential of the form $\Phi(\cp) =
-1/\cp$, where $\cp$ is again the cylindrical radius, so the disk is 
{\it unstratified}. The boundary condition in the $z$-direction 
is periodic for the 3D simulation.

The shock structure and the vortices at the base of the BL are clearly
visible in Figure \ref{3Dcompfig}a, but are absent or only barely 
present in Figure \ref{3Dcompfig}b. Instead, in 3D the vortices at 
the base of the BL have been replaced by turbulence. We will 
explore the appearance of vortices at the base of the BL in a 
future work, but we remark here that the appearance of vortices 
could depend on the aspect ratio of the BL. If the BL has a vertical 
extent which is small compared to its radial extent, then one might 
expect vortices to be present, whereas if the BL has a vertical 
extent which is thick compared to its radial extent, then it may 
instead be turbulent. Nevertheless, the structure of intersecting
shocks is still present in the disk even in the 3D case with
turbulence (Figure \ref{3Dcompfig}b), although it is less clear 
cut than in the 2D case with vortices. One may also expect that
the large-scale global modes discussed in \S \ref{fulldisk_sec} would
not be affected by the turbulence, since the periodicity of global 
modes is determined by the geometric resonance condition 
(\ref{deltaphival}). This does not involve the physics of what goes on
inside the BL, which occurs on scales that are small compared to the wavelength of
the global modes. 

Additional effects may emerge in 3D hydrodynamical
calculations in which vertical stratification in the disk is properly 
accounted for. \citet{LO} and \citet{Bate} have demonstrated that 
waves in thermally stratified disks tend to propagate in such a way 
that their action density is concentrated predominantly near the disk
surface, which increases wave amplitude and leads to faster nonlinear
damping. In our case this effect may influence the damping of the 
trapped modes leading to their faster dissipation. Another important 
aspect of stratified 3D simulations is that they would allow one to 
naturally explore the spreading of material on the stellar surface 
towards high altitudes, resulting in formation of a spreading
layer \citep{InogamovSunyaev,InogamovSunyaev10}.

One may also wonder how the addition of a magnetic field would modify our
results. The BL is linearly stable to the MRI, since it has a rising
rotation profile ($d \Omega^2/d \cp > 0$). Nevertheless, due to the
large shear present in the BL, a radial magnetic
field can be kinematically sheared to produce a strong toroidal
field. \citet{Armitage} and \citet{SteinackerPapaloizou} found 
such an amplification in their simulations, but the field remained 
subthermal, and the magnetic energy density was only a fraction 
($\sim 15 \%$) of the thermal energy density in the boundary layer. 
This means that even in the MHD case, most of the pressure still 
comes from the gas, so one might expect the magnetic field and 
its associated pressure to only provide a small correction to 
the dispersion relation of the trapped modes. 

Another way that MHD could affect the trapped modes is that the
velocity fluctuations in the disk due to the MRI could be much larger
than the velocity perturbations due to the shocks. \citet{HGB} found
that the velocity fluctuations due to the MRI were $\sim 10 \%$ of the
sound speed in the disk. This is comparable to the velocity
perturbation induced by the trapped modes in the disk. Thus one might
expect that the trapped modes will have their wavefronts deformed by
the MRI turbulence, but that the overall shock structure will still 
survive.


\section{Conclusion}

We have performed 2D hydrodynamic simulations of the boundary layer 
that self-consistently forms when the accretion disk is brought into 
contact with the stellar surface. Our calculations are restricted to 
the equatorial plane of the star+disk system which does not allow us 
to resolve the disk vertically but, quite importantly, captures 
the non-axisymmetric structures that emerge as a result of evolution. 
In our simulations, we used an isothermal equation of state and a 
Keplerian rotation profile inside the disk. We
varied the Mach number at the inner edge of the disk to study how it
affected the outcome of the simulations.

We find that sonic instabilities set in very rapidly in the
interfacial region of large, supersonic shear between the star and the
disk and saturate within a few orbital periods. After saturation, the
initially thin boundary layer region expands radially both into 
the star and into the disk, with its
final thickness depending on the Mach number as $\propto
M^{-1.9}$. This scaling implies that the boundary layer spans
the same number of pressure scale heights ($\sim 7-8$) in our isothermal
simulations independent of the Mach number.

After $t \sim 50$ orbital periods, a trapped mode develops between the
BL and a forbidden region in the disk, which consists of a corotation
resonance flanked by two Lindblad resonances. The trapped mode is
actually a weak shock (relative density perturbation 
$\delta \Sigma/\langle\Sigma\rangle \lesssim 1$), and one can think of it as 
a global acoustic wave that alternately reflects off the BL and 
the forbidden region in
the disk. The origin of this mode is intimately related to wave
tunneling through the forbidden region and the associated phenomenon of 
overreflection. This trapped mode can
persist for hundreds of orbital periods at a well defined pattern
speed, $\Omega_P \lesssim \Omega_K(R_\star)$. Both 
the pattern speed and azimuthal wavenumber of the the mode increase 
with increasing Mach number, and one can express the pattern speed 
in terms of the azimuthal wavenumber using a geometric resonance 
condition. 

The trapped modes and initial sonic instabilities are able to drive
accretion onto the star via dissipation in weak shocks even in 
the absence of magnetic fields. The
steady state consisting of a trapped mode rotating at a given pattern
speed can be interrupted by high-accretion states that are possibly
caused by KH
instability in the BL due to modification of the rotation profile
by shocks. The non-axisymmetric nature of the acoustic modes found
in our calculations gives rise to variability of the emission produced 
in the BL, which may be relevant for explaining phenomena such
as dwarf novae oscillations in cataclysmic variables.

\acknowledgements

We thank Jeremy Goodman for useful discussions and the referee for his
comments and suggestions. The financial support for this work is provided by 
the Sloan Foundation and NASA grant NNX08AH87G.



\appendix


\section{Angular Momentum Flux from a Superposition of Modes}
\label{totalmom}

In the WKB limit ($k \gg
m/R$), the angular
momentum flux for a normal mode with azimuthal wavenumber $m$ becomes
\ba
\label{constform}
C_{L,m}(\cp) = \pi \sgn(k)\frac{m \cp s^2 \delta \Sigma_m^2(\cp)}{\langle\Sigma\rangle |k|}
\ea
(see \citet{GT79} and  Appendix J of \citet{BinneyTremaine}).
Here $k$ is the radial wavenumber and $\sgn(k)$ determines the direction of
angular momentum transport, inward or outward. The Fourier coefficient
$\delta \Sigma_m$ in equation (\ref{constform}) is defined such that
an arbitrary density
perturbation, $\delta \Sigma(\cp, \phi, t)$, can be expressed as a sum
over Fourier components having $m \ge 0$:
\ba
\label{Fouriereq}
\delta \Sigma(\cp, \phi, t) &=& \sum_{m=0}^\infty \text{Re}\left[
  \delta \Sigma_m(\cp) e^{i(m\phi - \omega t)} \right].
\ea

Since the shocks in our simulations are not perfect normal modes, we
need to compute the angular momentum flux for an arbitrary
density profile with a given pattern speed. This can be done
by summing over the angular momentum transported by the
individual modes so we have
\ba
\label{totalCL}
C_L = \sum_{m=1}^\infty C_{L,m}.
\ea
Note that the $m=0$ term does not contribute to angular momentum
transport, which can be seen from equation (\ref{constform}). This
point allows us to perform a mathematical trick that will simplify the
analysis. Since the $m=0$ mode does not contribute to the total
angular momentum transported flux, we are free to set it to zero in
equation (\ref{Fouriereq}). Although this has an effect on the density
profile, it does not change the total angular momentum flux. 

Setting the $m=0$ component of the density profile to zero, we can
rewrite equation (\ref{Fouriereq}) as
\ba
\delta \Sigma(\cp, \phi, t) &=& \frac{1}{2} \sum_{m=-\infty}^\infty
  \delta \Sigma_m(\cp) e^{im(\phi - \Omega_P t)},
\ea
where $\Sigma_{-m}(\cp) \equiv \Sigma_m^*(\cp)$. Parseval's theorem,
which we shall shortly invoke, can then be written as
\ba
\label{Parseval}
\sum_{m=1}^{\infty} \left| \delta \Sigma_m \right|^2 =
\frac{1}{\pi}\int_0^{2 \pi} d \phi \delta \Sigma^2.
\ea

Using Parseval's theorem, we now derive an approximate formula 
for computing the
angular momentum flux that does not involve summing over the
individual Fourier coefficients as in
equation (\ref{totalCL}). We begin by making the assumption
$\Omega-\Omega_P \gg \kappa/m$ ($\kappa = \Omega$ for Keplerian
profile) so that the expression for $|k|$ (Eq. \ref{keq}) 
simplifies to
\ba
\label{ksimp}
|k| \approx \frac{m}{s}(\Omega-\Omega_P),
\ea
where $\Omega \ge \Omega_P$. Consequently, this assumption
is true if $m \gg 1$, and we are far away from any resonances so
that $\Omega - \Omega_P \sim \Omega \gg \kappa/m$.

Substituting equation (\ref{ksimp}) into equation (\ref{constform}), we
have for the angular momentum flux due to a given mode
\ba
\label{CLmsimp}
C_{L,m}(\cp) \approx \pi \sgn(k)\frac{\cp s^3 \delta \Sigma_m^2}{\langle\Sigma\rangle
  (\Omega -\Omega_P)}.
\ea
  The total angular momentum flux from equation (\ref{totalCL}) is given by
\ba
\label{CLeq}
C_{L}(\cp) = \pi \frac{\cp s^3}{\langle\Sigma\rangle
  (\Omega -\Omega_P)} \sum_{m=1}^{\infty} \sgn(k) \delta \Sigma_m^2.
\ea
We shall assume that for a given shock, all of the modes have the same sign of
  $\sgn(k)$, i.e. the shock is either incoming or outgoing. This is a
  reasonable assumption except near shock crossings, where incoming and
  outgoing shocks having opposite signs for $\sgn(k)$
  interact. For a given shock, this allows us to take $\sgn(k)$ out of
  the summation sign, assuming the individual shocks are azimuthally
  well-separated. Since $\widetilde{C_L}$ is the sum of the absolute
  values of the angular momentum flux for the individual shocks,
  assuming the shocks are well-separated, we arrive at
\ba
\label{CLtildeeq}
\widetilde{C_{L}}(\cp) = \pi \frac{\cp s^3}{\langle\Sigma\rangle
  (\Omega -\Omega_P)} \sum_{m=1}^{\infty} \delta \Sigma_m^2.
\ea
Here we have assumed $\Omega > \Omega_P$ which is true for the
  trapping region between the boundary layer and the Lindblad
  resonance in the disk. Using Parseval's theorem, equation
  (\ref{CLtildeeq}) becomes equation (\ref{finaleq}).


\section{Mass Accretion Rate Due to Weak Shocks.}  
\label{Mdotapp}

The mass accretion rate through the disk can be expressed through the
divergence of the angular momentum flux $\Fdis$ deposited in the 
disk fluid by the waves (or acoustic modes) due to their dissipation 
at the shock fronts \citep{LyndenBellPringle}
\ba
\dot M(r)=\left(\frac{dl}{d\varpi}\right)^{-1}
\frac{\partial \Fdis}{\partial r},
\label{eq:Mdot}
\ea 
where $l=\Omega(\varpi)\varpi^2$ is the specific 
angular momentum for circular orbits. The subsequent derivation 
is based in part on the work by \citet{Larson}. 

The modes are stationary in a frame rotating with angular speed
$\Omega_P$. Excitation of a wavepacket carrying angular momentum 
$\Delta J$ requires energy $\Delta E=\Omega_P\Delta J$. Dissipation
of this wavepacket at some other radial distance $\cp$ transfers 
angular momentum $\Delta J$ to the disk fluid --- a process which is 
necessarily accompanied by the mechanical work 
$\Omega(\varpi)\Delta J$ being done on the disk. The rest of energy, 
$\left[\Omega_P-\Omega(\varpi)\right]\Delta J$, is available for
heating the disk and is ultimately radiated away.
 
Thus, if the dissipation of the wave adds angular 
momentum to the disk fluid at the rate $\partial \Fdis/\partial 
\varpi$ per unit of $\varpi$ (the radial divergence of the 
wave angular momentum flux due to the wave damping), then 
the local tidal energy dissipation rate $\partial \dot 
E_{\rm w}/\partial \varpi$ is given by
\ba
\frac{\partial \dot E_{\rm w}}{\partial \varpi}=
\left[\Omega_P-\Omega(\varpi)\right]\frac{\partial \Fdis}{\partial \varpi}.
\label{eq:dEdw}
\ea 
See \citet{GoodmanRafikov} for the discussion of this issue.

It is difficult to calculate $\partial \Fdis/\partial \varpi$ 
from first principles. However, one can compute 
$\partial \dot E_{\rm w}/\partial \varpi$ and use equation 
(\ref{eq:dEdw}) to obtain $\partial \Fdis/\partial \varpi$. The 
reason behind using this approach is that dissipation along a given 
fluid element's trajectory occurs only at the shocks (unlike 
azimuthal acceleration and deceleration due to pressure gradients
that occurs also between the shocks) --- between the shocks the 
flow is adiabatic (in the absence of numerical dissipation).

For weak shocks \citet{Larson} gives the amount of energy irreversibly 
turned into heat at the shock per unit mass of the fluid crossing 
the shock as
\ba
\frac{dE}{dm}\approx s^2\frac{2}{3(\gamma+1)^2}
\frac{(M_s^2-1)^3}{M_s^4}, 
\label{eq:dE}
\ea
where \citet{LL}
\ba
M_s\approx 1+\frac{\gamma+1}{4}\frac{\Delta\Sigma}{\Sigma}
\label{eq:M_s}
\ea
is the Mach number of the weak shock, $\Delta\Sigma/\Sigma$
is the density contrast across the shock, and $s$ is the sound
speed. For an isothermal shock ($\gamma=1$) the expression 
$M_s = \sqrt{1+(\Delta\Sigma/\Sigma)}$
is exact. Thus, for $\gamma = 1$ we have
\ba
\frac{dE}{dm} = \frac{s^2}{6}\left(1 + \frac{\Delta
  \Sigma}{\Sigma} \right)^{-2}
\left(\frac{\Delta\Sigma}{\Sigma}\right)^3,
\label{eq:dE1}
\ea


Now, in an annulus of width $d\varpi$ the 
mass $dm=2m\left[\Omega_p-\Omega(\varpi)\right]\varpi\Sigma
dt d\varpi$ crosses one of the $2m$ shocks (the factor of 2 comes 
from accounting for both inward- and outward-propagating waves) 
per time $dt$, so that
\ba
\frac{\partial \dot E_{\rm w}}{\partial \varpi}=
2m\left[\Omega_p-\Omega(\varpi)\right]\varpi\Sigma\frac{dE}{dm},
\ea
meaning that $\partial \Fdis/\partial \varpi=2m\varpi\Sigma
(dE/dm)$. Finally, 
\ba
\dot M(r)=\left(\frac{dl}{d\varpi}\right)^{-1}\frac{m}{3}
\varpi\Sigma s^2
\frac{(M_s^2-1)^3}{M_s^4}
=\frac{m}{3}\frac{\Sigma s^2}{\Omega(\varpi)} \left(1 + \frac{\Delta
  \Sigma}{\Sigma} \right)^{-2}
\left(\frac{\Delta\Sigma}{\Sigma}\right)^3 \left(\frac{d\ln
  l}{d\ln\varpi}\right)^{-1}.
\label{Mdotpred}
\ea 
If we define the ``local disk mass'' to be $\Sigma\varpi^2$ then
the accretion timescale $t_{\rm acc}\equiv\Sigma\varpi^2/\dot M$
becomes (for $\Delta\Sigma/\Sigma\ll 1$)
\ba
t_{\rm acc}\approx \Omega^{-1}\frac{3}{m}\frac{M^2}{\varpi}
\left(\frac{\Delta\Sigma}{\Sigma}\right)^{-3}\frac{d\ln
  l}{d\ln\varpi},
\ea
where $M=s^{-1}$ is the Mach number of the flow.

If the disk were a $\dot M=const$ disk then 
$\dot M=3\pi\alpha_{\rm eff} s^2\Sigma/\Omega$, where 
$\alpha_{\rm eff}$ is the ``effective'' $\alpha$-viscosity
due to trapped shocks. Combining this with equation (\ref{Mdotpred})
one gets
\ba
\alpha_{\rm eff}=\frac{m}{9\pi}
\left(\frac{d\ln l}{d\ln\varpi}\right)^{-1}
\frac{(M_s^2-1)^3}{M_s^4},
\label{eq:alpha_eff}
\ea
in agreement with \citet{Larson}.

\end{document}